\documentclass[11pt,a4paper]{article}

\usepackage{fontenc}
\usepackage{inputenc}
\usepackage{amsmath,latexsym}
\usepackage{graphicx}
\usepackage{tikz}
\usepackage[nomessages]{fp}
\usepackage{jheppub}

\usepackage{accents}

\def\beq{\begin{equation}}   
\def\eeq{\end{equation}}
\def\bea{\begin{eqnarray}}  
\def\eea{\end{eqnarray}} 
\def\nn{\nonumber}
\def\r{\right} 
\def\l{\left} 
\def\f21{{}_2F_{1}}
\def\eps{\epsilon}
\def\order{\mathcal{O}}
\def\T{\boldsymbol{T}}
\def\N{\mathcal{N}}
\def\J{\mathcal{J}}
\def\M{\mathcal{M}}
\def\I{\mathcal{I}}
\def\U{\mathcal{U}}
\def\O{\mathcal{O}}

\def\d{\, \mathrm{d}}

\def\SE{\mathcal{S}}

\def\t{\widehat}
\def\mT{\mathbf{\Theta}}
\def\MI{\bold{M}}
\def\sing{\mathrm{Singular}}

\title{Geometric IR subtraction for real radiation}

\author[ab]{Franz Herzog}

\affiliation[a]{Nikhef Theory Group, Science Park 105, 1098 XG Amsterdam, The Netherlands}

\affiliation[b]{Department of Physics and Astronomy, VU University Amsterdam, De Boelelaan 1081, NL-1081 HV Amsterdam, The Netherlands}

\emailAdd{fherzog@nikhef.nl}

\abstract{
A scheme is proposed for the subtraction of soft and collinear divergences present in massless real emission phase space integrals. The scheme is based on a local slicing procedure which utilises the soft and collinear factorisation properties of amplitudes to produce universal counter-terms whose analytic integration is relatively simple. We propose that this scheme can be promoted to a fully local subtraction method. As a first application the scheme is applied to establish a general pole formula for final state real radiation at NLO and NNLO in Yang Mills theory for arbitrary multiplicities. All required counter-terms are evaluated to all orders in the dimensional regulator in terms of $\Gamma$ - and ${}_pF_q$ hypergeometric - functions. As a proof of principle the poles in the dimensional regulator of the $H\to gggg$ double real emission contribution to the $H\to gg$ decay rate are reproduced.
}

\begin{document}

\keywords{Perturbative QCD, NLO Computations, Scattering Amplitudes}  
\preprint{Nikhef 2018-018}

\maketitle

\section{Introduction}

As the LHC is entering its high precision phase the need for precise calculations of higher order perturbative corrections in QCD is becoming ever more important. A major bottleneck in such calculations is due to infrared (IR) divergences, which are related to non-integrable singularities in scattering amplitudes which arise in soft and/or collinear momentum configurations. While the KLN theorem \cite{Kinoshita:1962ur,Lee:1964is} guarantees that these singularities cancel (in the sum over real and virtual emissions for final state radiation) collinear divergences associated to partons in the initial states are handled instead by a renormalisation of the parton densities \cite{Politzer:1974fr,Georgi:1951sr,Altarelli:1977zs}. 

Since the IR divergences can be regulated dimensionally their cancellation is less problematic for analytic calculations of total inclusive quantities where one integrates over the entire phase space volume. The situation is different for differential quantities, where the domain of integration is taken over an arbitrary IR safe region of phase space. Then it may be unfeasible to carry out the integral analytically and a subtraction procedure is required to render the integrand finite. To accomplish this task has inspired the construction of a large number of different subtraction procedures.

Existing methods fall into either one of two conceptually quite different approaches. These are referred to as subtraction and slicing methods. The subtraction method is based on making the divergent integrand finite by subtracting from it a suitable counter-term whose singular behaviour matches that of the original integrand. This counter-term is subsequently added back in, analytically or numerically, integrated form.  Methods designed for dealing with real emissions at the next-to-leading order (NLO), which fall into this category, are the Catani-Seymour (CS) dipole method \cite{Catani:1996vz,Catani:1996jh}, the Frixione-Kunszt-Signer (FKS) subtraction method \cite{Frixione:1995ms,Frederix:2009yq} as well as the Nagy-Soper subtraction method \cite{Nagy:2003qn}. While all of these methods rely on the universal soft and collinear factorised limits of amplitudes to construct suitable counter-terms, they are implemented in quite different ways.

In the FKS method sets of non-overlapping divergences are separated by a partition of unity in the form of partial fractions. In each partition a suitable energy and angle parameterisation is chosen to factorise its soft and collinear divergences and subtract their singular parts via residue subtraction. The FKS method therefore defines its counter-terms not just at the level of the squared amplitude, but at the level of phase space measure times squared amplitude. 

The CS method instead constructs its counter-terms purely at the level of the squared amplitude. The counter-terms are constructed by combining together soft and collinear limits which are promoted into the full phase-space by the so-called Catani-Seymour momentum mapping. This mapping also allows to analytically integrate the counter-terms over the singular emission phase space.

Due to more complicated overlapping divergences at the next-to-next-to-leading order (NNLO) neither the FKS nor CS methods can be naively generalised. The problem for FKS-like methods based on residue subtraction is that parameterisations which completely factorise all divergences present in a given partition do not appear to exist. The most successful approaches based on residue subtraction therefore make use of sector decomposition \cite{Binoth:2004jv, Anastasiou:2003gr, Czakon:2010td, Boughezal:2011jf, Caola:2017dug} to factorise the divergences. Such approaches, especially those based on sector decomposition, have been used successfully in calculations at the current state of the art; see, e.g., \cite{Czakon:2015owf,Caola:2016trd}. While these methods can be efficiently implemented numerically they come with their own set of disadvantages: they are parameterisation dependent and the integration of the counter-terms remains, for now, numerical. 

Other approaches based on residue subtraction have been limited either to simpler applications, in which the divergences were factorisable \cite{Weinzierl:2003fx, Frixione:2004is, Anastasiou:2014nha}, or were based on topology dependent parameterisations \cite{Anastasiou:2010pw}, which is difficult to apply to more complicated final states.

Another class of subtraction methods used at NNLO are closer to the CS idea. A prominent such method is antenna subtraction \cite{GehrmannDeRidder:2005cm,GehrmannDeRidder:2007jk,Currie:2013vh}. Its counter-terms are based on physical matrix elements. Instead the colourful subtraction method relies on combining the universal soft and collinear limits \cite{Catani:1999ss} into suitable counter-terms. The advantage of the antenna method is that it leads to a comparably small number of counter-terms, whose analytic integration has been achieved. A disadvantage of the antenna method are that it is not fully local in the phase space, as certain spin correlations are ignored. This makes the phase space generation quite cumbersome. Despite this the method has been applied successfully in many state of the art NNLO calculations; see , e.g., \cite{Currie:2017eqf,Cruz-Martinez:2018rod}. In comparison the colourful subtraction method, see, e.g., \cite{Somogyi:2005xz,Somogyi:2006da,Bolzoni:2010bt}, has so far been applied only to final state radiation in, e.g., \cite{DelDuca:2016csb}. While this method is fully local it  comes with the disadvantage that the analytic integration of its counter-terms is highly non-trivial due to the appearance of Jacobians introduced by the mapping. It has thus been relying in part on numerical integration techniques for its counter-terms.

Slicing methods instead render divergent integrals finite by slicing out the singular regions. As in the subtraction method the integration over the singular region - which takes the role of the counter-term - is subsequently added back in  (analytically or numerically) integrated form. The original slicing method, implemented at NLO, was based on imposing cuts on the Mandelstam variables \cite{Giele:1991vf,Fabricius:1981sx}. These methods were never fully generalised to NNLO, although an extension was applied in mixed QCD-QED corrections in \cite{GehrmannDeRidder:1997gf}. More recent developments at NNLO include kt-subtraction \cite{Catani:2007vq} and N-jettiness subtraction \cite{Boughezal:2015aha,Gaunt:2015pea}. Both of these methods have been implemented in an impressive number of fully differential NNLO calculations; see, e.g., \cite{Catani:2011qz,Boughezal:2016dtm}. A clear advantage of these methods is the comparable ease of their implementation, since one simply implements a measurement function to cut out the singular parts of the phase space. While the kt-subtraction method is only applicable to colorless final states it has the advantage that its counter-terms are relatively simple to integrate analytically. N-jettiness subtraction can be applied to more general processes; the integration of the required soft function is however challenging and requires a numerical implementation. An advantage of both the kt-subtraction and the N-jettiness methods is that the singular limits can be obtained from general factorisation theorems. A disadvantage of these methods is that the slicing parameters must be chosen small enough for the factorisation formula to be valid; and this may lead to numerical instabilities; see e.g. \cite{Moult:2017jsg, Boughezal:2018mvf}. To address this problem one may need to add higher orders in the expansion around the slicing parameters. The challenge with this is that the structure of the sub-leading singular limits is more complicated; for instance derivatives of amplitudes will be required, which could be difficult to obtain for complicated processes.

Subtraction methods have also already been employed at next-to-NNLO (N${}^3$LO) in two incidences: an application of the projection to Born method \cite{Cacciari:2015jma} in DIS jet production \cite{Currie:2018fgr} and a novel application \cite{Dulat:2017prg} of the reverse unitarity approach \cite{Anastasiou:2002yz} in Higgs production (relying, for now, on the first two terms in an expansion around the threshold). While these are impressive calculations of so far unrivalled complexity, the methods they relied upon can not be (at least naively) applied for more general final states.

Despite the large number of existing methods, we propose - in the hope that it may overcome the shortcomings of existing methods - a new approach in this work. To accomplish maximally simple counter-terms, from the point of view of analytic integration, an FKS like residue subtraction procedure will be employed based on a Feynman diagram dependent slicing observable. By employing a slicing approach we can overcome the limitations of parameterisations which require singularities to be factorised. We will argue that this slicing approach can be promoted to a subtraction method by employing suitable phase space mappings not unlike in the CS, antenna or colourful subtraction methods. Here we will only demonstrate this idea, which was already proposed in \cite{Eynck:2001en} in the context of the conventional NLO slicing method, for a simple example in section \ref{sec:Motivation}. The purpose of this paper is instead to establish the general principles of the method. In particular we work out the combinatorics of a generalised soft-collinear subtraction formula in section \ref{sec:general} - which uniquely defines a simple prescription for the soft and/or collinear counter-terms. We apply this framework for final state radiation at NLO and NNLO in pure Yang-Mills theory for arbitrary multiplicities in section \ref{sec:CTsYM} and perform the integration of all counter-terms required. We complete the section by reproducing the poles of the gluonic double real emission correction to the gluonic Higgs decay at NNLO. Possible extensions and future developments of the method are discussed in section \ref{sec:Conclusion}.
 
\section{Notation}
In this section we introduce some of the notation used throughout the later sections.
It is convenient to define the normalisation factor
\beq
c_\Gamma =\frac{(4\pi)^{-2+\eps}}{\Gamma(1-\eps)}\,.
\eeq
Of particular  importance will be the $D=4-2\eps$ dimensional $n$-particle differential Lorentz invariant phase space measure:
\beq
\d \Phi_{1..n}(Q;m_1^2,..,m_n^2)\equiv (2\pi)^{D(1-n)+n}\; \delta^{(D)}\Big(Q-\sum_{k=1}^n p_k\Big) \prod_{k=1}^n \d^D p_i \, \delta^+(p_i^2-m_i^2)  \; .
\eeq
Here $Q$ is taken to be an off-shell time-like vector, i.e., $Q^2>0$. Final state particles are constrained to be on-shell with positive energy by the distribution 
$$
\delta^+(p^2-m^2)=\theta(p_0)\delta(p^2-m^2)\,.
$$
We mostly deal with massless vectors and masses occur in phase spaces only through the Mandelstam variables formed by squaring sums of momenta. For this purpose it is convenient to introduce the shorthand
\beq
\label{eq:momentumsum}
p^\mu_{ij..kl}=p_{i}^\mu+p_{j}^\mu +..+p_{k}^\mu+p_{l}^\mu\,.
\eeq
Mandelstam variables are defined as follows:
\bea
s_{ij}&=&2p_i.p_j,\nn\\
s_{ijk}&=&2(p_i.p_j+p_i.p_k+p_j.p_k)\,.\nn
\eea
We thus mostly suppress the dependence on masses and employ the shorthand
\beq
 \d\Phi_{1..n}(Q)=\d\Phi_{1..n}(Q;0,...,0)\,,
\eeq
for all $p_i$ massless. The total phase space volume is defined as
\beq
\Phi^{(n)}(Q;m_1^2,..,m_n^2)=\int\d \Phi_{1..n}(Q;m_1^2,..,m_n^2)\,.
\eeq
A \emph{massive sum of momenta}, e.g. $p_{12}$, with mass $s_{12}$, is indicated by bracketed notation $(12)$, e.g.,
\beq
\d\Phi_{(12)34..n}(Q;s_{12},0,..,0)=\d\Phi_{(12)34..n}(Q;s_{12})=\d\Phi_{(12)34..n}(Q)\,.
\eeq
The phase space measure satisfies the following factorisation property 
\beq
\label{eq:PSFAC}
\d \Phi_{1..n}(Q)=\frac{\d s_{12..k}}{2\pi}\; \d \Phi_{(12..k)k+1..n}(Q;s_{12..k}) \;\d\Phi_{12..k}(p_{12..k}),
\eeq
upon which much of this paper rests. Although this notation is intuitive and compact care has to be taken with identities such as $p_{1..n}=p_1+..+p_n$ which, when substituted in eq. (\ref{eq:PSFAC}), could lead to appearances of $\delta^{(D)}(0)$. Rather identites such as eq. (\ref{eq:momentumsum}) should be interpreted to arise in eq. (\ref{eq:PSFAC}) as a consequence of momentum conserving $\delta$-functions. Similar considerations apply to the Mandelstam variables $s_{ij}$.
\section{Motivation}
\label{sec:Motivation}
\begin{figure}  
\includegraphics[width=1 \textwidth,clip,trim= 0 6cm 0 4cm]{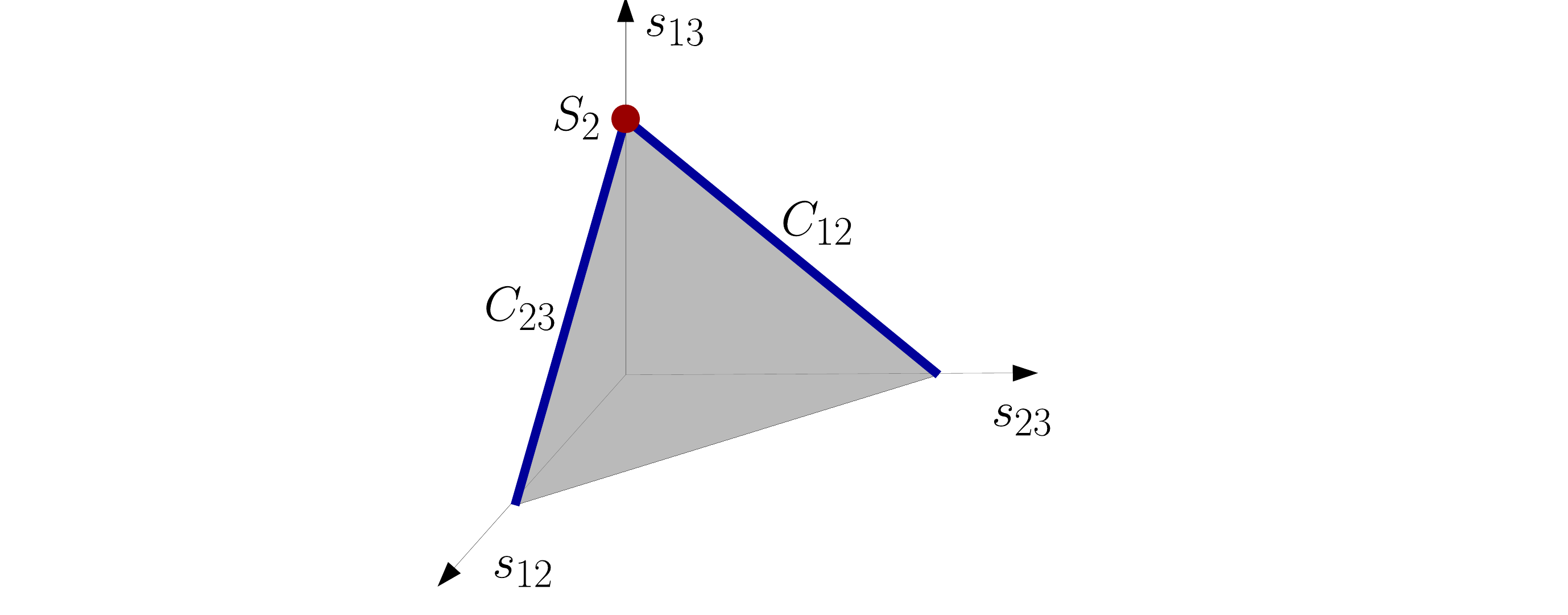}
\caption{The grey triangular surface $Q^2=s_{12}+s_{13}+s_{23}$ represents the physical phase space. The fat blue lines labelled by $C_{12}$ and $C_{23}$ show the locations of collinear singularities, while the small red circle labelled by $S_2$ shows the location of the soft singularity.}
\label{fig:Dalitz}
\end{figure}

Before describing the general method in section \ref{sec:general} we will illustrate it here in the context of a simple example of a divergent phase space integral:
\beq
\label{eq:ex1}
I(Q;D)=\int\d \Phi_{123}(Q)\; \frac{s_{13}}{s_{12}s_{23}}\,.
\eeq
This integral could appear in the real emission process $\gamma^*\to \bar q(1) q(3) g(2)$ at NLO in massless QCD. The integral is problematic in $D=4$ due to soft and collinear singularities respectively located at $2\to0$ and $1||2,2||3$. It is instructive to see where the  singularities are located in the space of Mandelstam variables. Let us express the phase space measure in terms of the variables $s_{12},s_{13}$ and $s_{23}$:
\beq
\int \d\Phi_{123}(Q)=(Q^2)^{-1+\eps} \N_3 \int_0^{Q^2} \d s_{12} \d s_{13} \d s_{23} \, \delta(Q^2-s_{12}-s_{13}-s_{23}) \,
 \, (s_{12}s_{13}s_{23})^{-\eps}\,,
\eeq
where
\beq
\N_3=\frac{1}{2}\frac{(4\pi)^{-3+2\eps}}{\Gamma(2-2\eps)}\,.
\eeq
We thus see that the 3-particle phase space can be represented as the area constrained on the surface 
\beq 
\label{eq:3simplex}
Q^2=s_{12}+s_{13}+s_{23}
\eeq 
together with $s_{ij}\ge 0$. This physical region with the locations of its singularities, embedded in the three dimensional $s_{ij}$-space, is shown in figure \ref{fig:Dalitz}. 

We now wish to construct a subtraction scheme with which to isolate the finite part of the integral in eq. (\ref{eq:ex1}). Since there exists quite some freedom how to define such a finite part it is natural to ask: how can we define the divergent part in the simplest possible way? In other words, how can the evaluation of the divergent parts be maximally simplified. A scheme that accomplishes this for the UV divergences is the minimal subtraction (MS) scheme within dimensional regularisation. This is also what is commonly used to cancel the soft and collinear divergences between real, virtual and counter-term contributions in higher order calculations. The problem with the MS-scheme is however that it does not provide us with a prescription for defining a manifestly finite integrand which can be expanded around $D=4$ before integration. This, for the practitioner of perturbative QFT, is indeed the dilemma of dimensional regularisation. Although it allows to define a unique finite part - it is not obvious how to obtain this finite part without performing the integral first in $D$-dimensions and then expanding the analytic functions around $D=4$.

For the simple example the integration is easily carried out in terms of $\Gamma$-functions, whose expansion around $D=4$ is trivial, to obtain:
\beq
\label{eq:ex1res}
I(Q;D)= (Q^2)^{-2\eps} \N_3 \frac{\Gamma(-\eps)^2\Gamma(2-\eps)}{\Gamma(2-3\eps)} 
=\frac{\Phi_3(Q^2)}{(Q^2)} \l(\frac{2}{\eps^2}- \frac{5}{\eps}+ 3+\order{(\eps)}\r)\;.
\eeq
But when increasing the perturbative order very complicated integrals appear which evaluate to complicated hypergeometric series. For such functions a Laurent expansion around $D=4$ may be difficult to obtain. Key is that one may not always be interested in integrating the integral over the entire phase-space. In fact for various reasons, such as detector efficiencies or large signal-swamping backgrounds, only part of the phase space may be experimentally accessible, or interesting. The phase space integration could then be constrained to an in principle arbitrary (infrared safe) region.

It is therefore highly desirable to have a procedure to extract the finite part of an integral which does not rely on carrying out the integration. While, as summarised in the introduction, such procedures have been developed in the past, based on subtraction, sector decomposition and phase space slicing, we believe that none of these approaches fully matches the simplicity of the divergent parts defined via the MS-scheme. Inspired by algebro-geometric schemes based on blow-ups, which have been applied in the subtraction of UV-- and IR-- divergences of Euclidean loop integrals \cite{Bloch:2008jk,Brown:2015fyf}, we intend to provide a prescription which meets this criterion as closely as possible in the following.

\subsection{Normal coordinates and phase space factorisation}

We start by identifying a set of suitable variables - we call them \textit{normal coordinates} or \textit{slicing parameters} - with which to separate the phase space into singular and finite regions. Here we shall use the word normal in the sense in which it was introduced by Sterman, see, e.g., \cite{Sterman:1978bi,Sterman:1978bj,Collins:2011zzd}; and we can identify these normal coordinates with the variables which we introduce to take soft and collinear limits. As a guide to find these variables we will use the phase space factorisation property of eq. (\ref{eq:PSFAC}) in terms of Mandelstams. It turns out that this property alone allows one to obtain suitable Lorentz invariant factorised soft and collinear limits of the phase space measure. 

Let us first see how this works for the case of collinear divergences. A choice of a variable to parameterise the collinear limit $C_{12}$ is given by $s_{12}$. The collinear limit is approached linearly as $s_{12}\to 0$. The factorisation property in eq. (\ref{eq:PSFAC}) then allows us to take this limit as follows:
\beq
\lim_{s_{12}\to0} \d\Phi_{123}(Q) = \frac{\d s_{12}}{2\pi} \,\d\Phi_{12}(s_{12})  \lim_{s_{12}\to0} \,\d\Phi_{(12)3}(Q;s_{12})\,.
\eeq
Since the 2-particle phase space measure $\d\Phi_{12}(s_{12})$ can not be simplified further, the limit operation acts only on the remaining phase space measure $\d\Phi_{(12)3}$, which has support in the limit $s_{12}\to0$. Here it is useful to introduce the Sudakov parameterisation 
\beq
p_{12}=p_{\t{12}}+\frac{s_{12}}{2p_{\t{12}}.n}n\,,\qquad p_{\t{12}}^2=0=n^2\,,
\eeq
such that we can parameterise
\bea
p_{1}&=&z_1 p_{\t{12}}+\frac{s_{12}z_2}{2p_{\t{12}}.n}n +\sqrt{s_{12}z_1z_2}e^\perp\,,\nn\\
p_{2}&=&z_2 p_{\t{12}}+\frac{s_{12}z_1}{2p_{\t{12}}.n}n -\sqrt{s_{12}z_1z_2}e^\perp \,.
\eea
with $z_1+z_2=1$ and $e^\perp$ being a space-like unit length ($|e^\perp|=1$) vector transverse to both $p_{\t{12}}$ and $n$. This allows us, at the expense of the massless reference vector $n$, to control how the off-shell vector $p_{12}$ present in $\d\Phi_{(12)3}$ approaches the massless vector $p_{\t{12}}$
\beq
\lim_{s_{12}\to0} p_{12}=p_{\t{12}} +\order{(s_{12})}\,.
\eeq
We thus obtain the following factorisation of the phase space measure in the collinear limit:
\beq
\lim_{s_{12}\to0} \d\Phi_{123}(Q) = \d\Phi_{C_{12}}  \,\d\Phi_{\t{12}3}(Q)\,,
\eeq
with the collinear phase space defined as 
\beq
 \d\Phi_{C_{12}} =\frac{\d s_{12}}{2\pi} \,\d\Phi_{12}(s_{12})\,.
\eeq
What is characteristic to this factorised limit is that the remaining $Q$ dependence is present only in the \textit{reduced} measure $\d\Phi_{\t{12}3}(Q)$. In contrast the $s_{12}$-dependence is present only in $\d\Phi_{12}(s_{12})$. This in effect means that the variable $s_{12}$ is no longer bounded from above. Since the factorisation is only valid in the limit of small $s_{12}$ it is sensible to introduce ``by hand'' a small upper cutoff for $s_{12}$ to make the integral over this measure well defined. We will describe below how to do this in a consistent manner.

We now come to the parameterisation of the soft limit. A Lorentz invariant variable suitable to parameterise the soft limit $p_2\to 0$ is
\beq 
s_{2(13)}=2p_2.p_{13}\,.
\eeq
This variable, which linearly approaches zero as $p_2\to0$, is also directly proportional to the energy of $p_2$ in the rest frame of $p_{13}$. Due to the relation eq. (\ref{eq:3simplex}) we can also identify the limit $s_{2(13)}\to 0$ with the limit $s_{13}\to Q^2$. This allows us to derive the soft phase space factorisation from the factorisation property eq. (\ref{eq:PSFAC}), as follows:
\beq
\lim_{s_{13}\to Q^2} \d\Phi_{123}(Q) =  \lim_{s_{13}\to Q^2} \frac{\d s_{13}}{2\pi} \,\d\Phi_{13}(s_{13})  \,\d\Phi_{(13)2}(Q;s_{13})
\eeq
Only the term $\d\Phi_{13}(s_{13})$ has further support in this limit, and we find:
\beq
\lim_{s_{13}\to Q^2} \d\Phi_{123}(Q) =    \d\Phi_{13}(Q^2) \, \d\Phi_{S_2}^{(1,3)}\,,
\eeq
with the soft phase space measure defined as:
\bea
\d\Phi_{S_2}^{(1,3)} &=& \frac{\d s_{2(13)}}{2\pi} \,\d\Phi_{(13)2}(Q^2;Q^2-s_{2(13)})\,,\\
&=& \d s_{2(13)}\, \frac{\d^Dp_2}{(2\pi)^{D-1}}\,\delta^{+}(p_2^2) \,\delta(s_{2(13)}-2p_{2}.p_{13})\,,\nn
\eea
where we have used $\d s_{13}=\d s_{2(13)}$. Note that the soft measure depends on the hard momenta $p_1$ and $p_3$. Thus even after integration over the soft momentum the soft limit retains a dependence on the variable $s_{13}$. In contrast the collinear limit factorises entirely and retains no dependence on any hard momenta.

\subsection{Geometry of regions}
So far we have not discussed how to construct a subtraction method which consistently combines the soft and collinear limits we introduced in the previous subsection. A way to attack this problem is to use a \textit{phase space slicing} approach. The idea here is that the phase space can be separated into a finite region ($F$) and singular, that is soft ($S_2$) and/or collinear ($C_{12},C_{13}$), regions.  To accomplish this decomposition we will associate a set of small dimensionless parameters $a_i$ for $S_{i}$ and $b_{ij}$ for $C_{ij}$ to bound the slicing parameters of each singular region from above. This procedure will naturally lead to a classification of the overlap of soft and collinear regions. We will associate: 
\begin{itemize}
   \item[$S_2$:] $\{s_{2(13)}<a_2s_{13}\}$,
   \item[$C_{12}$:] $\{s_{12}<b_{12}Q^2\}$,
   \item[$C_{23}$:] $\{s_{23}<b_{23}Q^2\}$,
   \item[$F$:]  $\{s_{2(13)}>a_2s_{13}$\,,\, $s_{12}>b_{12}Q^2$\,, \,$s_{23}>b_{23}Q^2\}$\,.
\end{itemize}
Here we have used $s_{13}$ as the scale entering the upper bound of the soft variable $s_{2(13)}$. First let us remark that for small $a_2$ and therefore also $s_{2(13)}$ we have $s_{13}\sim Q^2$, and so we could have equally well used $Q^2$ here. However $s_{13}$ is the natural hard scale appearing in the soft phase space as already commented on above, and as we will see later it will turn out important in more complicated situations to use this $s_{13}$ instead of $Q^2$. For this reason we introduce this notion already here.

\begin{figure}  
\includegraphics[width=1 \textwidth,clip,trim= 0 6cm 0 4cm]{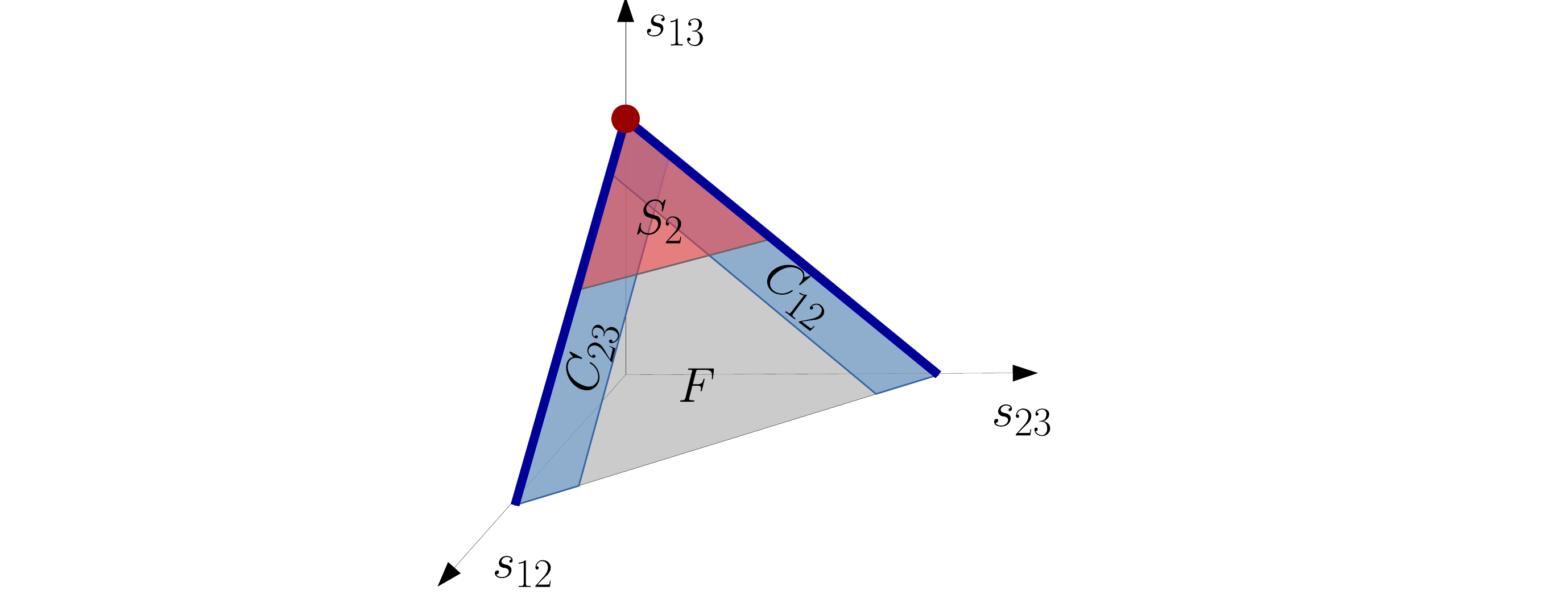}
\caption{The triangular surface $Q^2=s_{12}+s_{13}+s_{23}$ is split into singular and finite regions. The red region is the soft region $S_2$. Collinear regions $C_{12}$ and $C_{23}$ are indicated in blue. The finite region $F$ is shown in grey. The soft-collinear overlap is just visible where the blue bands intersect with the red triangular region.}
\label{fig:Desing}
\end{figure}

The decomposition of the phase space into its different finite and singular regions is best understood geometrically and is illustrated in figure \ref{fig:Desing}. 
Using the additivity of areas we arrive at the following partition of unity:
\beq
\label{eq:unityex1}
1=\Theta(F)+\Theta(S_2)+\Theta(C_{12})+\Theta(C_{23})-\Theta(C_{12}\cap S_2)-\Theta(C_{23}\cap S_2),
\eeq
with the different regions defined by:
\bea
\Theta(S_2)\;\, &=&   \Theta(s_{2(13)}<a_2s_{13})\nn\\
\Theta(C_{12}) &=&   \Theta(s_{23}<b_{23}Q^2)\nn\\
\Theta(C_{23}) &=&   \Theta(s_{12}<b_{12}Q^2)\\
\Theta(C_{23}\cap S_2) &=&   \Theta(s_{2(13)}<a_2s_{13})\Theta(s_{23}<b_{23}Q^2)\nn\\
\Theta(C_{12}\cap S_2) &=&   \Theta(s_{2(13)}<a_2s_{13})\Theta(s_{12}<b_{12}Q^2)\nn\\
\Theta(F)\;\,\, &=& \Theta(s_{2(13)}>a_2s_{13})\Theta(s_{23}>b_{23}Q^2)\Theta(s_{12}>b_{12}Q^2)\,.\nn
\eea
\subsection{Counter-term integration}
Let us now come to the evaluation of the integrals corresponding to the singular regions. We will start with the soft. A convenient parameterisation of the soft phase space measure is given by
\beq
\label{eq:softparam}
\int \d\Phi_{S_2}^{(1,3)}\Theta(S_2)=c_\Gamma s_{13}^{-1-\eps} 
\int_0^\infty \d s_{12} \d s_{23}\,  (s_{12}s_{23})^{-\eps} \,\Theta(s_{12}+s_{23}<a_2s_{13})\,,
\eeq
which allows us to obtain
\beq
I_{S1}(a_2,s_{13})=\int \d\Phi_{S_2}^{(1,3)} \frac{\Theta(S_2)s_{13}}{s_{12}s_{23}} = c_\Gamma \frac{\Gamma^2(1-\eps)}{\,\Gamma(1-2\eps)}  \frac{s_{13}^{-\eps}a_2^{-2\eps}}{\eps^2}\,.
\eeq
We continue with the evaluation of the collinear region. A convenient parameterisation for the collinear region is given by:
\beq
\label{eq:collparam}
\int \d\Phi_{C_{12}} \Theta(C_{12})= c_\Gamma \int_0^{b_{12}Q^2} \d s_{12} s_{12}^{-\eps} \int_0^1 \d z_1 \d z_2 \,\delta(1-z_1-z_2) \, (z_1z_2)^{-\eps} 
\eeq
The collinear limit of eq. (\ref{eq:ex1}) then leads us to the following integral:
\beq
I_{C_{12}}(b_{12}Q^2)=\int \d\Phi_{C_{12}} \frac{\Theta(C_{12})}{s_{12}} \frac{z_1}{z_2} =  c_\Gamma\frac{\Gamma(1-\eps)\Gamma(2-\eps)}{\Gamma(2-2\eps)}\, \frac{(b_{12}Q^2)^{-\eps}}{\eps^2}
\eeq
We continue with the evaluation of the soft-collinear overlap contribution. 
We can simplify the calculation of the overlap contribution by demanding that $b_{12}\ll a_2$.
In the limit of small $b_{12}$, which in turn forces a small $s_{12}$, the soft region then simplifies to
\beq
\lim_{s_{12}\to 0} \Theta(s_{12}+s_{23}<a_2 s_{13})=\Theta(z_2s_{\t{12}3}<a_2 s_{\t{12}3})
=\Theta(z_2<a_2)\,,
\eeq
where we also used that in the soft region $z_1\sim 1$. In other words the soft-collinear limit, in this setting, is conveniently computed by taking the limit $z_2\to 0$ in the collinear phase space. The soft-collinear phase space measure is thus given by
\beq
\label{eq:softcollparam}
\int \d\Phi_{C_{12}S_2}  \Theta(C_{12}\cap S_2)= c_\Gamma \int_0^{b_{12}Q^2} \d s_{12} s_{12}^{-\eps} \int_0^{a_2} \d z_2 \,z_2^{-\eps} \,.
\eeq
Integrating the soft-collinear limit over this measure we obtain
\beq
I_{C_{12}S_1}(b_{12}Q^2,a_2)=\int \d\Phi_{C_{12}S_2}   \frac{\Theta(C_{12}\cap S_2)}{s_{12}z_2} = c_\Gamma \frac{(a_2 b_{12} Q^2)^{-\eps}}{\eps^2}\,.
\eeq
The singular part of eq. (\ref{eq:ex1}) can now be expressed as: 
\bea
&&I_{\mathrm{Singular}}(Q;a_1,b_{12},b_{23})=\\
&&\frac{\Phi_{2}}{Q^2} \bigg[+I_{S1}(a_2,Q^2)+I_{C_{12}}(b_{12}Q^2) +I_{C_{12}}(b_{23}Q^2) -I_{C_{12}S_1}(b_{23}Q^2,a_2) -I_{C_{12}S_1}(b_{12}Q^2,a_2) \bigg]\nn
\eea
\bea
&&=\frac{\Phi_{3}}{(Q^2)^2}\bigg[   +\l(\frac{2}{\eps^2}+ \frac{-9-4\ln a_2}{\eps}+ \l(9+4\zeta_2 +18\ln a_2 +4\ln^2 a_2\r)+\order{(\eps)}\r)                   \nn\\
&&\qquad\qquad\;\,  + \l(\frac{2}{\eps^2}+ \frac{-7-2\ln b_{12}}{\eps}+ \l(4+4\zeta_2 +7\ln b_{12} +\ln^2 b_{12}\r)+\order{(\eps)}\r)        \nn \\
&&\qquad\qquad\;\,  + \l(\frac{2}{\eps^2}+ \frac{-7-2\ln b_{23}}{\eps}+ \l(4+4\zeta_2 +7\ln b_{23} +\ln^2 b_{23}\r)+\order{(\eps)}\r)        \nn \\
&&\qquad\qquad\;\, -\bigg(\frac{2}{\eps^2}+ \frac{-9-2\ln a_2-2\ln b_{12}}{\eps} + (9+6\zeta_2 +9\ln a_2+9\ln b_{12}\nn\\
&&\qquad\qquad\qquad\qquad\;\,  +2\ln a_2\ln b_{12}	 +\ln^2 a_2+\ln^2 b_{12})+\order{(\eps)}\bigg)         \nn \\
&&\qquad\qquad\;\, -\bigg(\frac{2}{\eps^2}+ \frac{-9-2\ln a_2-2\ln b_{23}}{\eps} + (9+6\zeta_2 +9\ln a_2+9\ln b_{23}\nn\\
&&\qquad\qquad\qquad\qquad\;\,  +2\ln a_2\ln b_{23}	 +\ln^2 a_2+\ln^2 b_{23})+\order{(\eps)}\bigg)       \bigg]   \\
&&=\frac{\Phi_{3}}{(Q^2)^2}\bigg[   \frac{2}{\eps^2}+ \frac{-5}{\eps}+ \l(-1 -2\ln b_{12}-2\ln b_{23}-2\ln a_2\ln b_{12}-2\ln a_2\ln b_{23}  +2\ln^2 a_2  \r)\nn\\
&&\qquad\qquad\quad+\,\order{(\eps)}\;\bigg]\nn\,.
\eea
Thus we have reproduced the correct single and double poles of eq. (\ref{eq:ex1res}). The cancellation of the logarithms at order $\eps^{-1}$ signifies a consistent treatment of the soft-collinear overlap contribution. 

\subsection{Slicing method}
\label{sec:Motslicing}
\begin{figure}  
\centering 
\begin{minipage}{.5\textwidth}
  \centering
\includegraphics[width=0.9\textwidth]{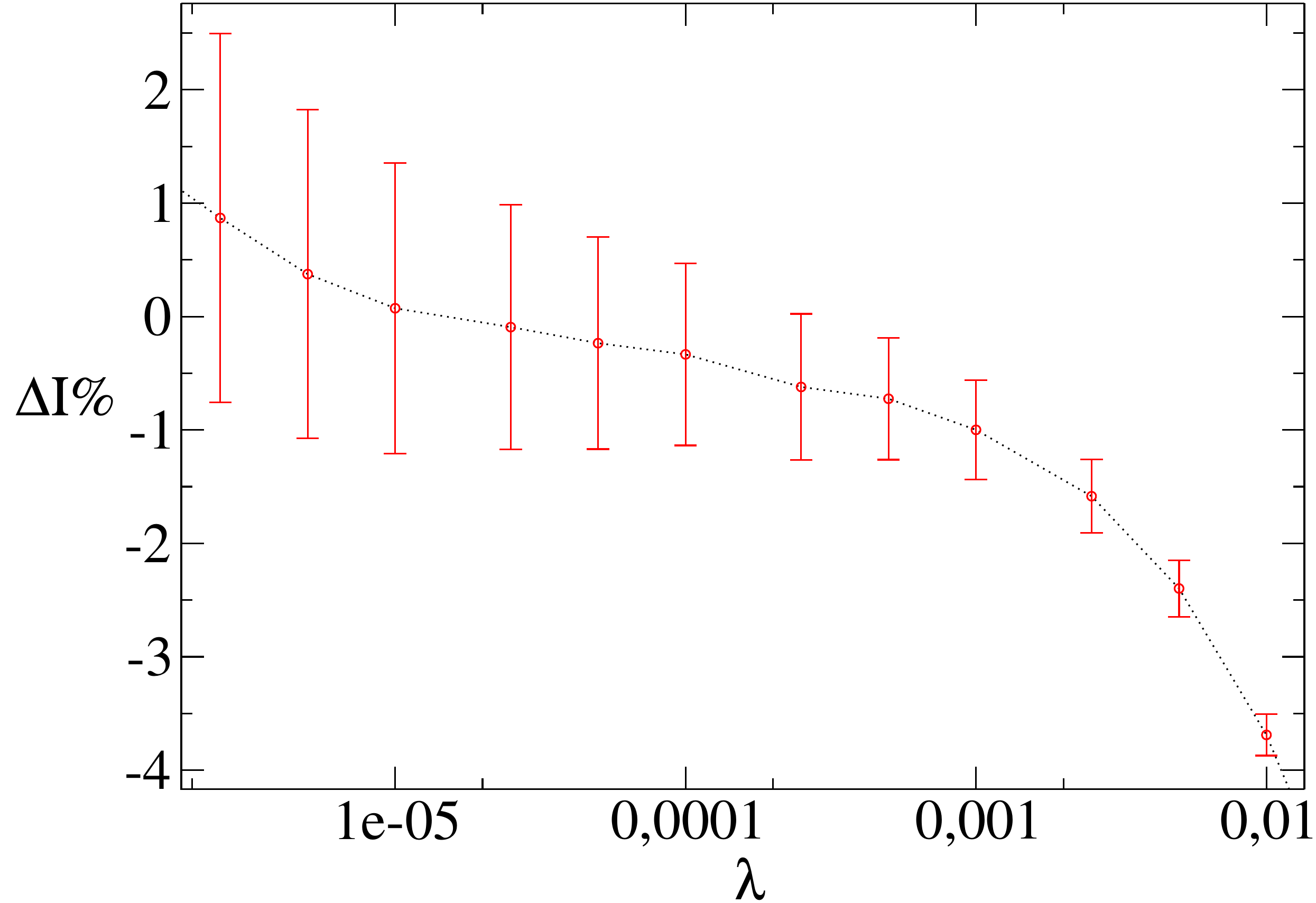}
\end{minipage}%
\begin{minipage}{.5\textwidth}
  \centering
\includegraphics[width=0.86\textwidth]{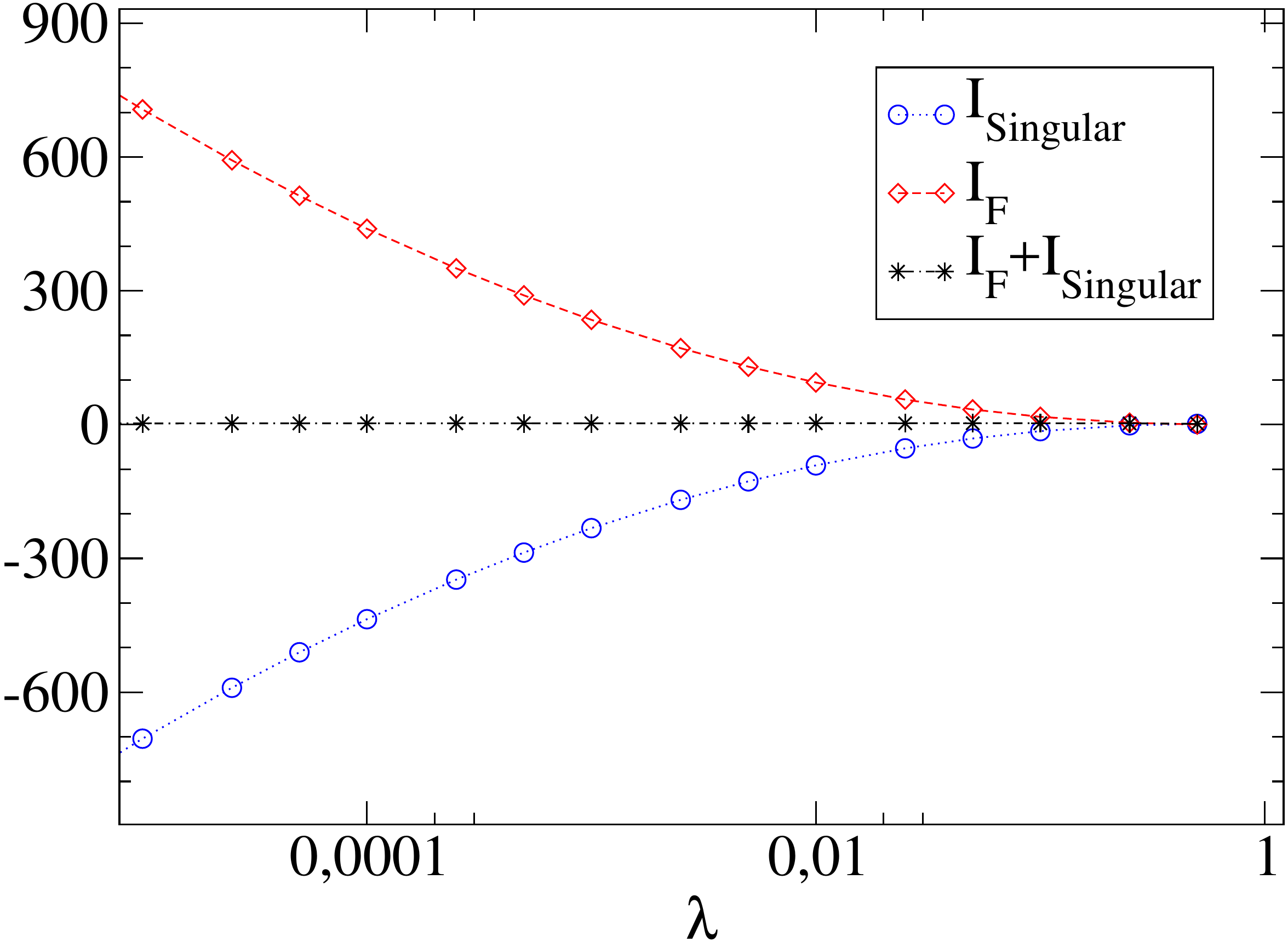}
\end{minipage}
\caption{This figure shows $\Delta I(\lambda)$ on the left and also separately the $\eps^0$ coefficients of $I_F,I_{\mathrm{Singular}}$ and their sum on the right in the slicing method. In both figures $I_F$ is evaluated numerically with the CUBA implementation of the Vegas algorithm using $10^8$ points for each value of $\lambda$.}
\label{fig:lambda}
\end{figure}

We will now test the finite part of the subtraction terms numerically using the slicing method. The advantage of the slicing approach is the simplicity with which the finite part, defined by 
\begin{equation}
\label{eq:IF}
I_F(Q;a_1,b_{12},b_{23})=\int d\Phi_{123} \,\Theta(F)\,\frac{s_{13}}{s_{12}\, s_{23}} 
\end{equation}
with 
\beq
\Theta(F)=\Theta(s_{12}>b_{12}Q^2) \Theta(s_{23}>b_{23}Q^2) \Theta(s_{2(13)}>a_{2}s_{13}) \,,
\eeq
can be implemented in a numerical simulation in $D=4$. To implement the hierarchy between soft and collinear limits we introduce a single slicing parameter $\lambda$, such that
\beq
b_{ij}=\lambda^2,\qquad a_i=\lambda\,.
\eeq
We can then define the quantity
\beq
\Delta I(\lambda)=100\cdot \frac{I_F(Q;\lambda)+I_{\mathrm{Singular}}(Q;\lambda)-I(Q)}{I(Q)}\,,
\eeq
which, since
\beq
\lim_{\lambda\to0}\Delta I(\lambda)=0\,,
\eeq 
measures the percent difference by which this quantity differs from zero for a small finite value of $\lambda$. While we know both $I_{\mathrm{Singular}}(Q)$ and $I(Q)$ analytically we can compute the finite integral $I_F(Q)$ numerically; this is plotted over a range of values of $\lambda$ in figure \ref{fig:lambda}.
\begin{figure}  
\centering 
\begin{minipage}{.5\textwidth}
  \centering
\includegraphics[width=0.95\textwidth]{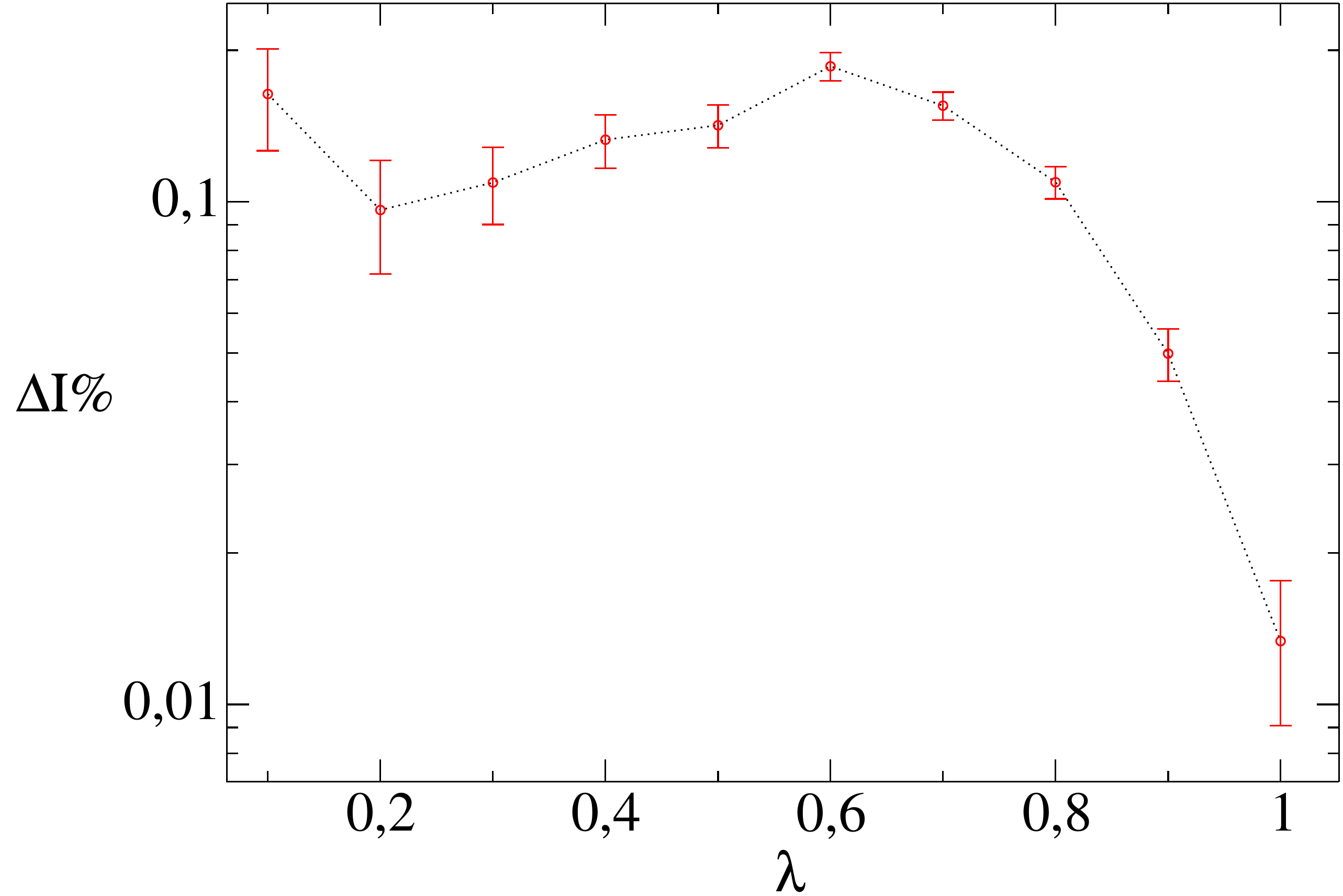}
\end{minipage}%
\begin{minipage}{.5\textwidth}
  \centering
\includegraphics[width=0.84\textwidth]{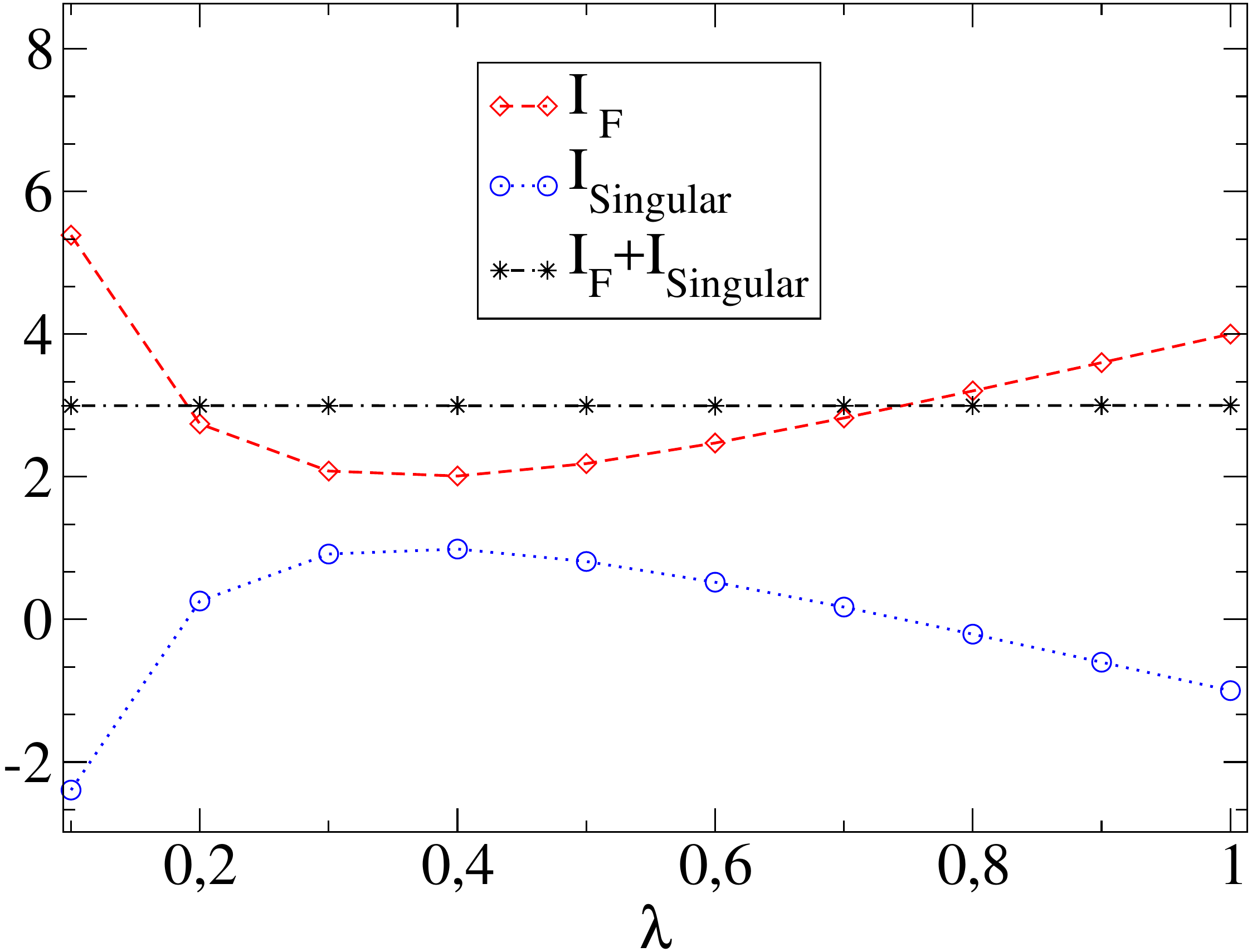}
\end{minipage}
\caption{This figure shows $\Delta I(\lambda)$ on the left and also separately the $\eps^0$ coefficients of $I_F,I_{\mathrm{Singular}}$ and their sum on the right in the subtraction method. In both figures $I_F$ is evaluated numerically with the CUBA implementation of the Vegas algorithm using $10^8$ points for each value of $\lambda$.}
\label{fig:lambda2}
\end{figure}
The figure on the left clearly shows the formation of a plateau in the range $10^{-3}< \lambda< 10^{-6}$. As is expected for a slicing scheme the numerical accuracy deteriorates as $\lambda$ is decreased and improves for larger values of $\lambda$, where however the counter-terms can not be used as a reliable approximation, and power corrections in $\lambda$ would be required. It is thus evident, given the simplicity of the example, that this approach gives rise to a rather poor numerical accuracy. Even after sampling $10^8$ points using the Cuba implementation of Vegas \cite{Hahn:2004fe} (albeit in a non-optimal phase space parameterisation) we are not able to arrive at an accuracy much better than 1\% for the most optimal values of $\lambda$.

\subsection{Subtraction method}
\label{sec:Motivation:subtraction}
An alternative to the slicing approach, presented in section \ref{sec:Motslicing}, is to rewrite the region approximants as local counter-terms. In this subsection we will illustrate - for the simple example - how a slicing method can be promoted to a fully local subtraction method. This idea is not new and was already discussed in, e.g. \cite{Gaunt:2015pea, Eynck:2001en}. 

The promotion can be accomplished in several different ways. One method would be to employ a momentum map, of the kind which has been employed by Catani and Seymour in the dipole subtraction method \cite{Catani:1996vz}. Alternatively  one can use, what we shall call, the \textit{re-weighting} approach, this mimics in some sense how the limit subtraction is embedded in the full phase space in the FKS subtraction method. In our approach we can use a similar idea, based on the particular phase space factorisation property used to derive a certain limit.

In the context of the simple example the subtraction method can be implemented very easily by noting that we can match soft and collinear measures with the full phase space measure by multiplying with the corresponding factors which were simplified to unity in the limit taking procedure. This leads us to the following relations:
\bea
\d\Phi_{S_2}^{(1,3)} \d\Phi_{13}&=&\d\Phi_{123}\Theta(s_{2(13)}<a_{2}Q^2) \l(\frac{s_{13}}{Q^2}\r)^{\eps}\,,\\
\d\Phi_{C_{12}} \d\Phi_{\t{12}3}&=&\d\Phi_{123}\Theta(s_{12}<b_{12}Q^2) \l(1-\frac{s_{12}}{Q^2}\r)^{-1+2\eps}\,,  \\
\d\Phi_{C_{12}S_2} \d\Phi_{13}&=&\d\Phi_{123} \Theta(s_{12}<b_{12}Q^2)\Theta(z_2<a_{2})  \l(1-\frac{s_{12}}{Q^2}\r)^{-1+2\eps} \l(z_{12}\r)^{\eps} \,,
\eea
where we have made explicit choices for the reference vectors $n$ in the different collinear limits. Such that the momentum fractions $z_{ij}$ become:
\bea
z_{12}+z_{21}=1,\qquad z_{12}=\frac{s_{13}}{s_{13}+s_{23}},\qquad z_{23}+z_{32}=1,\qquad z_{32}=\frac{s_{13}}{s_{13}+s_{12}}\,.
\eea
Using these relations for the measures we can derive the following alternative representation for the finite part defined in eq. (\ref{eq:IF}) in $D=4$:
\bea
\label{eq:IFS}
I_F(Q;a_1,b_{12},b_{23})=\int d\Phi_{123} \bigg[ &&\,\frac{s_{13}}{s_{12}\, s_{23}} -\frac{Q^2}{s_{12}\,s_{23}}\Theta(s_{2(13)}<a_{2}Q^2)\nn\\
&&-\frac{\big(z_{12}-\Theta(z_{21}<a_{2})\big)}{s_{12}\, z_{21}(1-s_{12}/Q^2)}\Theta(s_{12}<b_{12}Q^2)\nn\\
&&-\frac{\big(z_{32}-\Theta(z_{23}<a_{2})\big)}{s_{23}\, z_{23}(1-s_{23}/Q^2)}\Theta(s_{23}<b_{23}Q^2) \bigg]\,.\nn
\eea
A numerical evaluation of this finite part for a range of suitable values of 
\beq
\lambda=a_i=b_{ij}
\eeq
is presented in figure \ref{fig:lambda2} using again the Cuba Vegas implementation. The clear advantage of the subtraction method is that it allows to use arbitrary values for $\lambda\in (0,1]$, since the integrated counter-terms evaluate 
by construction to those used in the slicing method for any value of $a_i$ and $b_{ij}$. 

In contrast to the slicing method better convergence is observed for large values of $\lambda$. In fact it appears that $\lambda=1$, which corresponds to the counter-terms ranging over the entire phase space, is the optimal choice for this example. Notably the accuracy reached for this value of $\lambda$ is 100 times as good as that reached by the slicing method using the same number of numerical evaluations. The subtraction method - to which we have promoted the slicing method - therefore appears far superior in terms of numerical accuracy and stability when compared to its parent slicing method.

\section{General principles at NNLO and beyond}
\label{sec:general}

\subsection{Normal coordinates and phase space factorisation}
In the previous section we applied the geometric subtraction method in a simple example. Let us briefly summarise the idea behind the procedure. We introduced the variables $s_{ij}=2p_i.p_j$ to define a collinear region and the variable  $s_{i(jk)}=2p_i.p_{jk}$ to define a soft region. With these variables we then derived suitably factorised limits of the phase space measure and furthermore partitioned the phase space volume into finite and singular (soft and/or collinear) regions.

This procedure can be generalised to arbitrary perturbative order. For instance at NNLO we can use the variable $s_{ijk}$ to subtract the triple collinear limit $i||j||k$ and we can use the variable 
\beq 
s_{(ij)(kl)}=2p_{ij}.p_{kl}
\eeq 
to subtract the double soft limit $ij\to0$ sensitive to the hard momenta $k$ and $l$. The phase space factorisation property can be used to determine the corresponding factorised phase space volume limits, with which to integrate the singular limits of amplitudes. This leads us to the following phase space factorisation in the triple collinear limit:
\beq
\lim_{i||j||k}\d\Phi_{1..i..j..k..n}\to \d\Phi_{C_{ijk}} \d\Phi_{1..\t{ijk}..n}\,,
\eeq
with the triple collinear phase space measure defined by
\beq
\d\Phi_{C_{ijk}}=\frac{\d s_{ijk}}{2\pi} \d\Phi_{ijk}\,.
\eeq
To parameterise the momenta in this collinear limit we use, as before, the Sudakov parameterisation (expressed in terms of Mandelstams, rather than transverse momenta):
\bea
p_{i}&=&z_i p_{\t{ijk}}+\frac{|p_i^\perp|^2}{2z_ip_{\t{ijk}}.n}n +|p_i^\perp|e_i^\perp\,,\qquad |p_i^\perp|^2=z_i(s_{ij}+s_{ik}-z_is_{ijk})\,,\nn\\
p_{j}&=&z_j p_{\t{ijk}}+\frac{|p_j^\perp|^2}{2z_jp_{\t{ijk}}.n}n +|p_j^\perp|e_j^\perp\,,\qquad |p_j^\perp|^2=z_j(s_{ij}+s_{jk}-z_js_{ijk})\,,\\
p_{k}&=&z_k p_{\t{ijk}}+\frac{|p_k^\perp|^2}{2z_kp_{\t{ijk}}.n}n +|p_k^\perp|e_k^\perp\,,\qquad |p_k^\perp|^2=z_k(s_{ik}+s_{jk}-z_ks_{ijk})\,,\nn
\eea
with $z_i+z_j+z_k=1$ and $|p_i^\perp|e_i^\perp+|p_j^\perp|e_j^\perp+|p_k^\perp|e_k^\perp=0$ such that 
\beq 
p_{ijk}= p_{\t{ijk}}+\frac{s_{ijk}}{2p_{\t{ijk}}.n}n\,.
\eeq
As before $e_i^\perp$ are space-like unit length ($|e^\perp_i|=1$) vectors transverse to both $p_{\t{ijk}}$ and the reference vector $n$. The off-shell vector $p_{ijk}$ approaches the massless vector $p_{\t{ijk}}$ in the limit of vanishing $s_{ijk}$: 
\beq
\lim_{s_{ijk}\to0} p_{ijk}=p_{\t{ijk}} +\order{(s_{ijk})}\,.
\eeq
In the double soft limit we obtain the phase space factorisation:
\beq
\lim_{ij\to0}\d\Phi_{1..i..j..n}\to \d\Phi_{S_{ij}}^{(k,l)} \d\Phi_{1..\not{i}..\not{j}..n}\,,
\eeq
with the double soft phase space measure given by:
\beq
\d\Phi_{S_{ij}}^{(k,l)}=\frac{\d s_{(ij)(kl)}}{2\pi} \lim_{ij\to0} \d \Phi_{ij(kl)}\,.
\eeq
The pattern of these measures follows those defined at NLO. However there is subtle difference between the soft and double soft measures. The double soft measure is not simply $\d \Phi_{ij(kl)}$, since there exist further support in the limit $ij\to0$. Instead an explicit form for it is given by:
\beq
\d\Phi_{S_{ij}}^{(k,l)}=\d s_{(ij)(kl)} \,\frac{d^Dp_i}{(2\pi)^{D-1}}\delta^{+}(p_i^2)\,\frac{d^Dp_j}{(2\pi)^{D-1}}\delta^{+}(p_j^2) \,\delta(s_{(ij)(kl)}-2p_{ij}.p_{kl})\,.
\eeq
The phase space measures at yet higher order, e.g. at N${}^3$LO, can be defined similarly. For the $m$-collinear limit we would use the variable 
\beq 
s_{i_1..i_m}=(p_{i_1}+..+p_{i_m})^2
\eeq 
with the following phase space factorisation
\beq
\lim_{i_1||..||i_m}\d\Phi_{1..i_1..i_m..n}\to \d\Phi_{C_{i_1..i_m}} \d\Phi_{1..\t{i_1..i_m}..n}\,,
\eeq
and $m$-collinear phase space measure
\beq
\d\Phi_{C_{i_1..i_m}}=\frac{\d s_{i_1..i_m}}{2\pi} \d\Phi_{i_1..i_m}\,.
\eeq
Similarly we may define the $m$-soft variable 
\beq
s_{(i_1..i_m)(kl)}=2p_{i_1..i_m}.p_{kl}
\eeq
with phase space factorisation
\beq
\lim_{i_1..i_m\to 0}\d\Phi_{1..i_1..i_m..n}\to \d\Phi_{S_{i_1..i_m}}^{(k,l)} \d\Phi_{1..\not{i_1}..\not{i_m}..n}\,,
\eeq
and the $m$-soft phase space measure
\bea
\d\Phi_{S_{i_1..i_m}}^{(k,l)}&=&\d s_{(i_1..i_m)(kl)} \,
\frac{d^Dp_{i_1}}{(2\pi)^{D-1}}\delta^{+}(p_{i_1}^2)\,..\frac{d^Dp_{i_m}}{(2\pi)^{D-1}}\delta^{+}(p_{i_m}^2)\nn\\
&&\qquad\qquad\cdot  \,\delta(s_{(i_1..i_m)(kl)}-2p_{i_1..i_m}.p_{kl})\,.
\eea

\subsection{Soft and collinear forests}
From the conceptual side it remains to determine the overlap contributions. Since the phase space volume of the example given in section \ref{sec:Motivation} was simple enough, we were able to construct the partition based on the geometry of the phase space in Mandelstam variables. For higher dimensional phase spaces it will be advantageous to have at hand a formalism which allows to derive this partition in a more algebraic manner. We will employ the properties of Heaviside step functions for this purpose. 
Employing the normal coordinates defined above we associate $\Theta$-functions for each region as follows:
\bea
\Theta(S_{i_1..i_m})&=&\Theta(a_{i_1..i_m}s_{kl}\ge s_{(i_1..i_m)(kl)})\,,\\
\Theta(C_{i_1..i_m})&=&\Theta(Q^2b_{i_1..i_m}\ge s_{i_1..i_m})\,.
\eea
Here we assume that all soft regions are defined in some rest frame of $p_{kl}$. This rest frame could be chosen differently for different soft divergences. We will focus our discussion more on this point in section \ref{sec:CTsYM}, where we show how different eikonal factors can have their regions bounded in different rest frames. 

Our starting point for now is again the equation:
\beq
\label{eq:F1}
\Theta(\mathrm{Singular})+\Theta(F)=1\,,
\eeq
which follows given that finite ($F$) and singular regions are to cover the entire phase space volume. Let us now define the set $R$ as the set of all possible singular regions, excluding their overlaps, such that for the example of eq. (\ref{eq:ex1}) we would have $R=\{C_{12},C_{23},S_{2}\}$. We can then write
\beq
\label{eq:F2}
\Theta(F)=\prod_{r\in R} (1-\Theta(r))\,\,
\eeq
which combined with eq. (\ref{eq:F1}) leads to
\beq
\label{eq:Singular}
\Theta(\mathrm{Singular})= -\sum_{U\subset R} (-1)^{|U|}\prod_{r\in U} \Theta(r)\,.
\eeq
Here we sum over all nonempty subsets $U$, each of size $|U|$, of the set $R$. This expression still lacks knowledge of the geometric structure of the soft and collinear regions as well as the perturbative order. 

To get a  first feel for this equation let us study its consequences using $R=\{C_{12},C_{23},S_{2}\}$ as input. We shall use the notation $\Theta(A)\Theta(B)=\Theta(A\cap B)$
if regions $A$ and $B$ depend on common momenta. Using this notation we then find:
\bea
\Theta(\mathrm{Singular})=&&\Theta(C_{12})+\Theta(C_{23})+\Theta(S_{2})-\Theta(C_{12}\cap S_2)-\Theta(C_{23} \cap S_2)\nn\\
&&-\Theta(C_{12} \cap C_{23})+\Theta(C_{12}\cap C_{23} \cap S_2)\,,
\eea
which agrees with eq. (\ref{eq:unityex1}), if we apply the relation
\beq
\label{eq:CCnull}
\Theta(C_{12}\cap C_{23})=\Theta(C_{12}\cap C_{23}\cap S_2)\,.
\eeq
Indeed this relation follows from the geometric construction introduced in Figure \ref{fig:Desing},
since the soft region contains the intersection $C_{12}\cap C_{23}$. By demanding the hierarchy $a_{i}\gg b_{ij}$ we can thus guarantee its validity. But eq. (\ref{eq:CCnull}) would not hold for other parameter choices such as $a_{i}\ll b_{ij}$. This shows how, in a simple example, the geometry of regions plays an important role.

Let us continue with our exploration of eq. (\ref{eq:Singular}). To be able to include more complicated final states into our discussion let us introduce the measurement function $\J^{(l)}_{1..n+l}$, with $l$ denoting the maximum number of unresolved partons, which are permitted by this measurement function. In particular the measurement function obeys the relations:
\beq
\lim_{i\to0}\J^{(l)}_{1..i..n+l}=\J^{(l-1)}_{1..\not{i}..n+l}\,,\qquad 
\lim_{i||j}\J^{(l)}_{1..i..j..n+l}=\J^{(l-1)}_{1..\t{ij}..n+l}\,.
\eeq
Here the notation $\J_{..\not{i}..}$ indicates that $\J$ does no longer depend on the soft momentum $i$, while $\J_{..\t{ij}..}$ indicates that the collinear momenta $i$ and $j$ have been merged into the massless momentum $\t{ij}$. The (purely) real emission contribution to an arbitrary N${}^l$LO observable may then be defined as
\beq
\mathcal{O}_{l;1...n+l}=\int \d \Phi_{1...n+l}\, \J^{(l)}_{1..n+l}\, |\M_{1..n+l}|^2
\eeq
with $\M_{1..n+l}$ an $n+l$ parton tree-level amplitude. 

In the following we wish to define a set $\mathcal{U}^{(l)}$, whose elements are themselves sets of possible singular regions which may pass the criteria of the measurement function $\J^{(l)}$, and where we omit all those overlap regions which cancel by vitue of eq. (\ref{eq:CCnull}) and its generalisations at higher order. We can thus write 
\beq
\label{eq:singularJ}
\J^{(l)}\,\Theta(\mathrm{Singular})= -\J^{(l)} \; \sum_{U\in \U^{(l)}} (-1)^{|U|}\,\prod_{r\in U} \Theta(r) \,,
\eeq
where now cancellations among different regions have taken place and only those regions are included which can pass the criteria of the measurement function. 

Let us start by studying the case $l=1$. The only divergences allowed by $\J^{(1)}_{1..n+1}$ are collinear $i||j$ and soft $i\to0$ or their overlap with common partons. Let us recall at this point the  slight mismatch between our region definitions $C_{ij}$ and $S_i$ and the locations of singularities $i||j$ and $i\to0$ respectively. Since $C_{ij}$ is the region defined by $s_{ij}<b_{ij}Q^2$ even a soft singularity can fake the region $C_{ij}$; an apparent paradox which is resolved by subtracting the overlap contribution $C_{ij}\cap S_i$. It follows that care must be taken when considering the possible regions which may pass the criteria of the measurement function. While the NLO measurement function $\J^{(1)}$ may not allow for a singularity $i||j$ and $j||k$ which would correspond to a triple collinear singularity, the region $C_{ij} \,\cap\, C_{jk}$ can still be mimicked by a soft singularity $j\to 0$. As before we escape this apparent new region by relying on the cancellation in eq. (\ref{eq:CCnull}), as in the simple example. At NLO we thus define:
\beq
\U^{(1)}=\big\{ \{C_{ij}\} ,\{S_i\}, \{C_{ij},S_i\}  \big\} \,.
\eeq
where the notation $\{C_{ij}\}$ is not meant to indicate the set of all collinear divergences, but rather one set for each collinear region. Given for example the regions $R=\{C_{12},C_{23},S_{2}\}$ we would then have 
$$
\U^{(1)}=\big\{ \{C_{12}\} , \{C_{13}\} ,\{S_2\}, \{C_{12},S_2\}, \{C_{23},S_2\} \big\}\,.
$$
Let us now come to the definition of $\U^{(2)}$, the set of all possible singular regions which pass the criteria of the NNLO measurement function $\J^{(2)}$. To define $\U^{(2)}$ more precisely we must establish what the geometric cancellation identities are. Since these identities depend on the geometric properties of the soft and collinear regions, they depend on the parameters $a_{i..}$ and $b_{ij..}$. To make progress we choose the following order
\beq
\label{eq:aborder}
a_{ij}\gg a_{i}\gg b_{ijk}\gg b_{ij}\,,
\eeq
as it produces simple iterated phase space volumes for the counter-terms.
Further this ordering gives rise to the following region cancellation identities:
\bea
\label{eq:geometriccancellationI}
\lim_{A,B\to 0}\Theta(S_A\cap S_B)&=&\lim_{A,B\to 0}\Theta(S_A\cap S_B\cap S_{AB})\,,\\
\label{eq:geometriccancellationII}
\lim_{A||B,A||D}\Theta(C_{AB}\cap C_{AD})&=&\lim_{A||B,A||D}\Theta(C_{AB}\cap C_{AD}\cap C_{ABD})\,,
\eea
and
\bea
\label{eq:geometriccancellationIII}
\Theta(C_{A i}\cap C_{A j})&=&\Theta(S_A \cap C_{A i}\cap C_{A j})+\Theta(C_{A i j} \cap C_{A i}\cap C_{A j})   \\
&& -\Theta(S_A\cap C_{A i j} \cap C_{A i}\cap C_{A j}) \,,  \nn
\eea
which holds for
\beq
b_{Ai}\le \frac{a_Ab_{Aij}}{2}\,,
\eeq
and is therefore consistent with eq. (\ref{eq:aborder}).
Here $A,B$ and $D$ are sums of momenta, while $i$ and $j$ are single (massless) momenta. We do not claim that these are all the identities which are fulfilled, but they suffice to establish the desired region cancellations at NNLO. Pictorially these identities are illustrated in figures \ref{fig:setoverlapSABCABD} and \ref{fig:setoverlap4}. A derivation is sketched in appendix \ref{appendix:Regioncancellation}. 
\begin{figure}  
\includegraphics[width=1 \textwidth,clip,trim= 0 4cm 0 4cm]{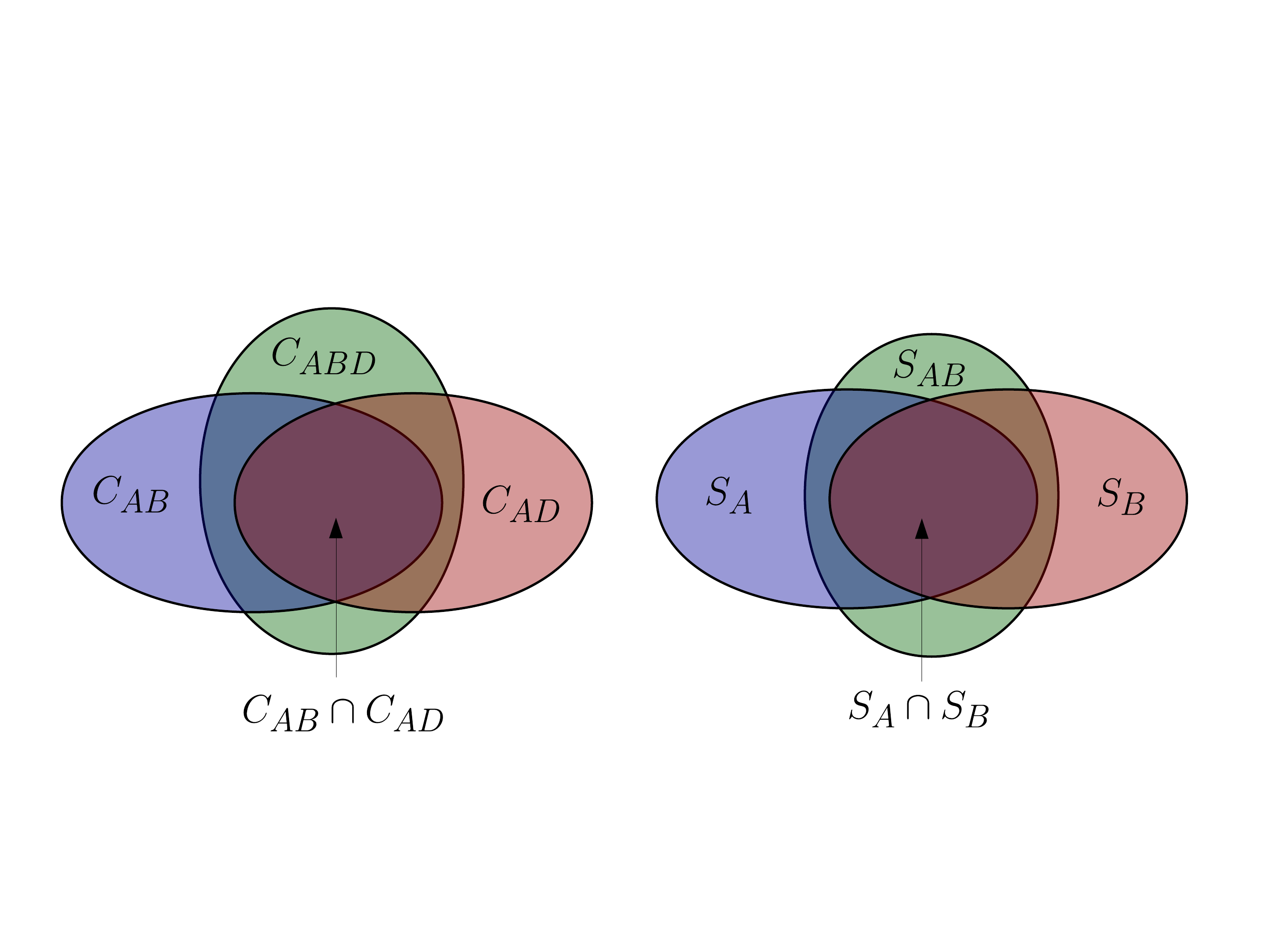}
\caption{The picture illustrates the identities $S_A\cap S_B\subset S_{AB}$ (right) and $C_{AB}\cap C_{AD}\subset C_{ABD}$ (left).}
\label{fig:setoverlapSABCABD}
\end{figure}
\begin{figure}  
\includegraphics[width=1 \textwidth,clip,trim= 0 4cm 0 4cm]{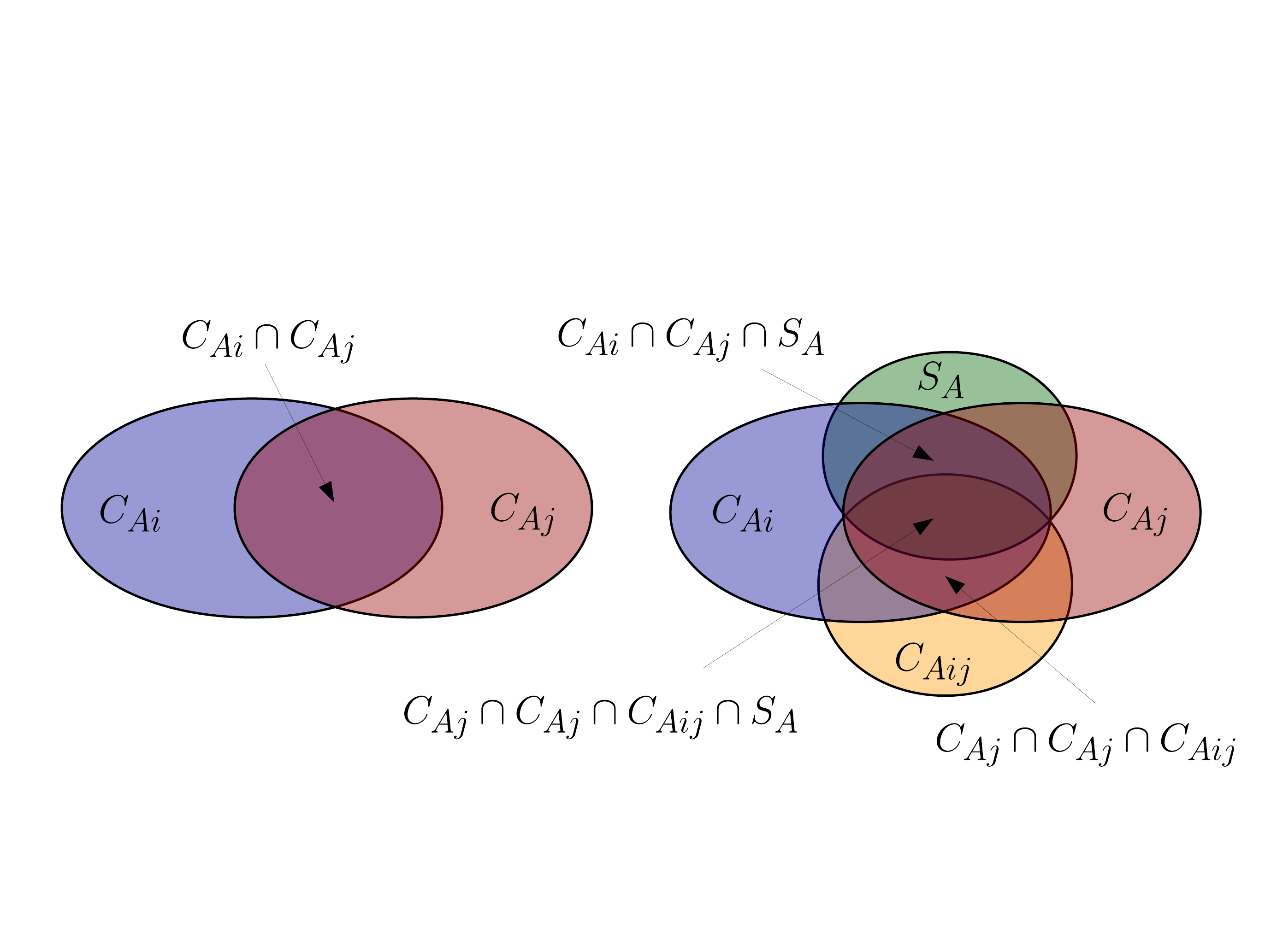}
\caption{The picture illustrates the identity $(C_{Ai}\cap C_{Aj}) \subset (C_{Aij}\cup S_{A})$.}
\label{fig:setoverlap4}
\end{figure}

The region cancellation identities are useful since they considerably reduce the number of different singular regions, or equivalently counter-terms, which are required in the subtraction. At NLO they allow us to obtain
\beq
\Theta(C_{ij}\cap C_{jk})=\Theta(C_{ij}\cap C_{jk}\cap S_j)\,,\\
\eeq
from eq. (\ref{eq:geometriccancellationIII}), since terms containing $C_{ijk}$ are rejected by the NLO measurement function. At NNLO we obtain instead the following set:
\bea
\Theta(S_i\cap S_j)&=&\Theta(S_i\cap S_j\cap S_{ij})\,,\nn\\
\Theta(C_{ijk}\cap C_{ijl})&=&\Theta(C_{ijk}\cap C_{ijl}\cap S_{ij})\,, \nn\\
\Theta(C_{ij}\cap C_{jk})&=&\Theta(C_{ij}\cap C_{jk}\cap S_j)+\Theta(C_{ij}\cap C_{jk}\cap C_{ijk})\\
&&-\Theta(C_{ij}\cap C_{jk}\cap S_j\cap C_{ijk}) \,.\nn
\eea
But furthermore these regions allow us to identify the set $\U^{(2)}$ as the Cartesian product of a set of soft forests $\U_S^{(2)}$ times a set of collinear forests $\U_C^{(2)}$ modulo measurement function constraints. We write this statement as
\beq
\label{eq:forest}
\U^{(2)}=\U_S^{(2)}\times \U_C^{(2)} \mod \J^{(2)}\,,
\eeq
with 
\beq
\U_S^{(2)}=\{\{S_{i}\},\{S_{ij}\},\{S_{i},S_{ij}\}\}\,,
\eeq
and
\beq
\U_C^{(2)}=\{\{C_{ij}\},\{C_{ijk}\},\{C_{ijk},C_{ij}\},\{C_{ij},C_{kl}\}\}\,.
\eeq
The sets $\U_S^{(2)}$ and $\U_C^{(2)}$ can be thought of as the soft and collinear analogues of Zimmermann's set of forests of UV-divergent subgraphs \cite{Zimmermann:1969jj}. This should not come as a complete surprise, since it is well known that the subtraction of both soft and collinear divergences follows similar patterns as those found in the UV; see for instance \cite{Humpert:1980xj,Smirnov:1986me,Herzog:2017jgk,Collins:1981uk} and references therein. Nevertheless it is not completely trivial that the regions defined with the particular ordering of eq. (\ref{eq:aborder}) follow this pattern as well. One may associate it to the observation that also in the context of subtracting UV-divergences an order is chosen with which to subtract singular subgraphs; such that the smaller subgraphs are subtracted before the larger. Similarly we may have chosen the relative ordering of soft and collinear divergences according to such a pattern. Regardless of this similarity eqs. (\ref{eq:geometriccancellationI}), (\ref{eq:geometriccancellationII}) and (\ref{eq:geometriccancellationIII}) are highly dependent on the relative ordering of soft and collinear regions, which has no analogue in the subtraction of UV divergences alone.

Multiplying out the product in eq. (\ref{eq:forest}) and discarding those regions which can not pass the criteria of the measurement function we arrive at:
\bea
\label{eq:U2}
\U^{(2)}&=&\big\{ \{C_{ij}\} ,\{S_i\}, \{C_{ijk}\},\{S_{ij}\},\{C_{ij},S_i\},\{C_{ij},S_{k}\} ,\{C_{ij},C_{kl}\},\{C_{ijk},C_{ij}\},\{C_{ijk},S_i\},\nn\\
&&\{C_{ijk},S_{ij}\},\{S_{ij},S_{i}\},\{C_{ik},S_{ij}\},\{C_{ij},S_{ij}\},\{C_{ij},C_{kl},S_i\},
\{C_{ij},C_{kl},S_{ik}\},\{C_{ij},S_{ij},S_i\},\nn\\
&&
\{C_{ijk},C_{ij},S_i\},\{C_{ijk},C_{ij},S_k\},\{C_{ijk},C_{ij},S_{ik}\},\{C_{ijk},C_{ij},S_{ij}\},\{C_{ijk},S_{ij},S_i\},\nn\\
&&
\{C_{ik},S_{ij},S_i\},\{C_{ik},S_{ij},S_j\},\{C_{ijk},C_{ij},S_{ik},S_i\},\{C_{ijk},C_{ij},S_{jk},S_k\},\{C_{ijk},C_{ij},S_{ij},S_i\},\nn\\
&&
\label{eq:limits}
\{C_{ij},C_{kl},S_{ik},S_i\}\big\}\,.
\eea
The size of this list shows the enormous increase of complexity which is encountered at NNLO, when compared to NLO. 

Before moving on to the definitions of the asymptotic phase space measures in the various regions let us conclude this subsection with the conjecture that eq. (\ref{eq:forest}) is valid also for the case of $l$ potentially unresolved emissions at N${}^l$LO: 
\beq
\label{eq:lforest}
\U^{(l)}=\U_S^{(l)}\times \U_C^{(l)} \mod \J^{(l)}\,,
\eeq
as long as the order
\beq
\label{eq:aborderl}
a_{i_1i_2...i_{l}}\gg..\gg a_{i_1}\gg b_{i_1i_2...i_{l+1}}\gg..\gg b_{i_1i_2}
\eeq
is chosen, with $\U_S^{(l)}$ and $\U_C^{(l)}$ sets of soft and collinear forests.

\subsection{Asymptotic phase space measures}
Having fixed the order of limits we are now in a position to determine the asymptotic measures associated to the singular regions. To compute the asymptotic measure associated to a particular region $U$ one should take the following sequence of limits of the expression:
\beq
\lim_{a_{ij}\to0}\lim_{a_{i}\to0}\lim_{b_{ijk}\to0}\lim_{b_{ij}\to0} \d\Phi \prod_{r\in U} \Theta(r)\,.
\eeq
Let us consider how this works for a few examples:
\begin{itemize}
\item $U=\{S_{ij},C_{ijk}\}$:
\bea
&&\lim_{a_{ij}\to0}\lim_{b_{ijk}\to0}\d\Phi_{..i..j..k..} \Theta(a_{ij}s_{kl}>s_{(ij)(kl)})\Theta(Q^2b_{ijk}>s_{ijk})\nn\\
&&=\lim_{a_{ij}\to0}\d\Phi_{..\t{ijk}..}\d\Phi_{C_{ijk}} \Theta(a_{ij}z_{k}>z_{ij})\Theta(Q^2b_{ijk}>s_{ijk})\\
&&=\d\Phi_{..k..}\d\Phi_{C_{ijk}S_{ij}} \Theta(a_{ij}>z_{ij})\Theta(Q^2b_{ijk}>2p_{ij}.p_k)\nn\\
&&=\d\Phi_{..k..}\d\Phi_{C_{ijk}S_{ij}}(Q^2b_{ijk},a_{ij})\nn
\eea
with 
\bea
\d\Phi_{C_{ijk}S_{ij}}&=&\d s_{(ij)k}\,\d z_{ij} \,\frac{\d^Dp_i}{(2\pi)^{D-1}}\delta^{+}(p_i^2)\,\frac{\d^Dp_j}{(2\pi)^{D-1}}\delta^{+}(p_j^2) \,\nn\\
&&\qquad\cdot\,\delta(s_{(ij)(kl)}-2p_{ij}.p_{k})\delta(z_{ij}-\frac{p_{ij}.n}{p_k.n})\,.
\eea
\item $U=\{C_{ijk},C_{ij}\}$:
\bea
&&\lim_{b_{ijk}\to0}\lim_{b_{ij}\to0}\d\Phi_{..i..j..k..} \Theta(Q^2b_{ij}>s_{ij})\Theta(Q^2b_{ijk}>s_{ijk})\nn\\
&&=\lim_{b_{ijk}\to0}\d\Phi_{..\t{ij}..k..} \d\Phi_{C_{ij}} \Theta(Q^2b_{ij}>s_{ij})
\Theta(Q^2b_{ijk}-s_{\t{ij}k})\\
&&=\d\Phi_{..\t{\t{ij}k}..} \d\Phi_{C_{\t{ij}k}}\d\Phi_{C_{ij}} \Theta(Q^2b_{ij}>s_{ij})
\Theta(Q^2b_{ijk}-s_{\t{ij}k})\nn\\
&&=\d\Phi_{..\t{\t{ij}k}..} \d\Phi_{C_{\t{ij}k}}(Q^2b_{ijk})\d\Phi_{C_{ij}}(Q^2b_{ij}) \nn
\eea
\item $U=\{S_{ij},C_{ij}\}$:
\bea
&&\lim_{a_{ij}\to0}\lim_{b_{ij}\to0}\d\Phi_{..i..j..} \Theta(a_{ij}s_{kl}>s_{(ij)(kl)})\Theta(Q^2b_{ij}>s_{ij})\nn\\
&&=\lim_{a_{ij}\to0}\d\Phi_{..\t{ij}..}\d\Phi_{C_{ij}} \Theta(a_{ij}s_{kl}>s_{\t{ij}(kl)})\Theta(Q^2b_{ij}>s_{ij})\\
&&=\d\Phi_{..\not{\t{ij}}..}\d\Phi_{S_{\t{ij}}}^{(k,l)}\d\Phi_{C_{ij}} 
\Theta(a_{ij}s_{kl}>s_{\t{ij}(kl)})\Theta(Q^2b_{ij}>s_{ij})\nn\\
&&=\d\Phi_{..\not{\t{ij}}..}\d\Phi_{S_{\t{ij}}}^{(k,l)}(s_{kl},a_{ij})\d\Phi_{C_{ij}} (Q^2b_{ij})\nn
\eea
\item $U=\{S_{ij},S_{i}\}$:
There are two different cases, of which the more complicated one is:
\bea
&&\lim_{a_{ij}\to0}\lim_{a_{i}\to0}\d\Phi_{..i..j..} \Theta(a_{ij}s_{kl}>s_{(ij)(kl)})\Theta(a_{i}s_{jk}>s_{i(jk)})\nn\\
&&=\lim_{a_{ij}\to0}\d\Phi_{..i..\not{j}..}\d\Phi_{S_{i}}^{(j,k)} \Theta(a_{ij}s_{kl}>s_{j(kl)}) \Theta(a_{i}s_{jk}>s_{i(jk)})\\
&&=\d\Phi_{..\not{i}..\not{j}..}\d\Phi_{S_{i}}^{(j,k)}\d\Phi_{S_{j}}^{(l,k)} 
\Theta(a_{ij}s_{kl}>s_{j(kl)}) \Theta(a_{i}s_{jk}>s_{i(jk)})\nn\\
&&=\d\Phi_{..\not{i}..\not{j}..}
\d\Phi_{S_{i}}^{(j,k)}(s_{jk},a_i) 
\d\Phi_{S_{j}}^{(l,k)}(s_{lk},a_{ij}) \,.\nn
\eea
There exists a subtlety when taking the limit $a_{ij}\to 0$ in the third line, which forces the single soft limit $j\to 0$, since the argument of the factor $\Theta(a_{i}s_{jk}>s_{i(jk)})$ does not simplify in this limit. To understand this better let us study the scaling with $a_{ij}$ which the different factors portray:
\bea
\label{eq:itersoftconstraint}
a_{i}s_{jk} &>&\;\; s_{ij}\;\;\;\,+\;\;s_{ik}\\
\order(a_{ij}) &>& \order(a_{ij})+\order(1)\nn
\eea
The two terms on the right hand side are thus (at least naively) not of the same size; their exist however no support to further simplify this expression. The only plausible interpretation is that the upper bound on the left forces both Mandelstams on the right, i.e. $s_{ij}$ and $s_{ik}$ to scale as $\sim a_{i}a_{ij}$. Given that both $i$ and $j$ are sufficiently soft it thus appears that $i$ is forced to become collinear to $k$ for $s_{ik}$ to be of similarly size as $s_{ij}$.

A different way to come to a similar conclusion comes from the result of the integral: 
\beq
\int \d\Phi_{S_{i}}^{(j,k)}(s_{jk},a_i)\,\SE_i^{(j,k)}\sim (s_{jk})^{-\eps}\,,
\eeq 
where $\SE_i^{(j,k)}$ corresponds to the singular limit of an amplitude. It is thus clear that this integral is homogeneous in $p_j$, although the constraint eq. (\ref{eq:itersoftconstraint}) does no appear to be so.
\end{itemize}
We will leave it as an exercise for the reader to work out the asymptotic forms of the measures corresponding to other regions. Compact expressions for the resulting measures for all required regions can be found in appendix \ref{sec:Integrated counterterms}. In particular we find that at NLO and NNLO all regions can be written using the following \textit{primitive} phase space measures:
\bea
\d \Phi_{C_{ij}}(Q^2b_{ij})&=&\d \Phi_{C_{ij}}\,\Theta(s_{ij}<Q^2b_{ij})\\
\d \Phi_{C_{ijk}}(Q^2b_{ijk})&=&\d \Phi_{C_{ijk}}\,\Theta(s_{ijk}<Q^2b_{ijk})\\
\d \Phi_{S_i}^{(j,k)}(s_{jk},a_i)& = & \d \Phi_{S_i}^{(j,k)}\,\Theta(s_{i(jk)}<a_is_{jk})\\
\d \Phi_{S_{ij}}^{(l,k)}(s_{kl},a_{ij})& =&  \d \Phi_{S_{ij}}^{(k,l)}\,\Theta(s_{(ij)(kl)}<a_{ij}s_{kl})\\
\d \Phi_{C_{ij}S_i}(Q^2b_{ij},a_i)&=&\d \Phi_{C_{ij}S_i}\,\Theta(s_{ij}<Q^2b_{ij})\Theta(z_i<a_i)\\
\d \Phi_{C_{ijk}S_{ij}}(Q^2b_{ijk},a_{ij})&=&\d \Phi_{C_{ijk}S_{ij}}\,\Theta(s_{(ij)k}<Q^2b_{ijk})\Theta(z_{ij}<a_{ij})
\eea
It is left to future work to show that an analogous statement will hold also at higher orders, i.e. that one requires a certain set of new primitive measures corresponding to, e.g., $\{C_{ijkl}\}$, $\{S_{ijk}\}$ and $\{C_{ijkl},S_{ijk}\}$ at N$^{3}$LO, while all other regions will factorise into the measures already present at lower orders.

\subsection{Soft and collinear master integrals at NNLO}
\label{sec:masters}
The integration of singular limits of amplitudes over the primitive NNLO phase spaces ($\{C_{ijk}\}$, $\{S_{ij}\}$ and $\{C_{ijk},S_{ij}\}$) is more complicated than the integration over NLO primitive phase spaces or their iterated limits which occur at NNLO. While the latter evaluate to $\Gamma$-functions, the primitive NNLO limits can lead  to hypergeometric functions of type ${}_pF_q$. 

In order to minimise the total number of integrals it is advantageous to express them in terms of a basis of \textit{master integrals}. This basis can be constructed by solving the relevant IBP identities using the LaPorta algorithm. Using reverse unitarity \cite{Anastasiou:2002yz,Anastasiou:2013srw} to treat Dirac $\delta$-functions which appear in the definition of their associated phase space measures, it is a well defined task to use public codes to set up the IBP reduction for these limits. To accomplish this task we have employed the public softwares FIRE \cite{Smirnov:2008iw,Smirnov:2014hma} and AIR ~\cite{Anastasiou:2004vj}. 

For the double soft this procedure leads to exactly two master integrals, they can be easily extracted from the two independent double soft Master integrals which were presented and evaluated in \cite{Anastasiou:2012kq}, where they appeared in the threshold limit of the Higgs boson cross section in gluon fusion. In our conventions this leads to:
\bea
\MI^{(2;1)}_{S}(s_{12},a_{34})&=&\int\d\Phi_{S_{34}}^{(1,2)}(s_{12},a_{34}) 
\frac{(s_{12})^2}{(s_{(12)(34)})^4}\\
&=&-c_\Gamma^2 \frac{(s_{12})^{-2\eps}(a_{34})^{-4\eps}}{4\eps}\frac{\Gamma^4(1-\eps)}{\Gamma(4-4\eps)}\,,\nn\\
\MI^{(2;2)}_{S}(s_{12},a_{34})&=&\int\d\Phi_{S_{34}}^{(1,2)}(s_{12},a_{34})\; 
\frac{s_{12}}{s_{34}s_{13}s_{24}}\\
&=&\MI^{(2;1)}_{S}(s_{12},a_{34})\, {}_3F_2(1,1,-\eps;1-\eps,1-2\eps;1)\,.\nn
\eea
The double soft-triple collinear master integrals are in one-to-one correspondence to those in the double soft limit, and results for these slightly simpler integrals can be read off from their double soft counter parts using the phase space parameterisation which was presented in section 5.2 of \cite{Anastasiou:2015yha}. 
\bea
\MI^{(2,2;1)}_{SC}(Q^2b_{134},a_{34})&=&\int\d\Phi_{C_{134}S_{34}}(Q^2b_{134},a_{34}) 
\frac{1}{(s_{1(34)})^2(z_{34})^2} \\
&=&c_\Gamma^2 \frac{(Q^2a_{34}b_{134})^{-2\eps}}{4\eps^2}\frac{\Gamma^4(1-\eps)}{\Gamma^2(2-2\eps)}\,,\nn\\
\MI^{(2,2;2)}_{SC}(Q^2b_{134},a_{34})&=&\int\d\Phi_{C_{134}S_{34}}(Q^2b_{134},a_{34}) \; \frac{1}{s_{34}s_{13}z_4}\\
&=&\MI^{(2,2;1)}_{SC}(Q^2b_{134},a_{34})\, {}_3F_2(1,1,-\eps;1-\eps,1-2\eps;1)\,.\nn
\eea
In the triple collinear limit we will require four master integrals. These, among two other simpler ones, were presented and evaluated in \cite{Ritzmann:2014mka}, where they appeared in the context of jet fragmentation. In our conventions these results can be expressed as follows
\bea
\MI^{(2;1)}_{C}(Q^2b_{123})&=&\int\d\Phi_{C_{123}}(Q^2b_{123})\frac{1}{s_{123}^2} \qquad\qquad\qquad\qquad\qquad\qquad\qquad\qquad\qquad\\
&=&   -c_\Gamma^2 \frac{(Q^2b_{123})^{-2\eps}}{2\eps}\frac{\Gamma^5(1-\eps)}{\Gamma(2-2\eps)\Gamma(3-3\eps)}\,,\nn
\eea
\bea
\MI^{(2;2)}_{C}(Q^2b_{123})&=&\int\d\Phi_{C_{123}}(Q^2b_{123}) \; \frac{1}{s_{123}s_{12}z_{23}} \\
&=&   -\frac{2-3\eps}{\eps}\;  \MI^{(2;1)}_{C}(Q^2b_{123})\; {}_3F_2(1, 1 - 2\eps, 1 - \eps; 2 - 3\eps, 2 - 2\eps; 1)\,,\nn
\eea
\bea
&&\MI^{(2;3)}_{C}(Q^2b_{123})=\int\d\Phi_{C_{123}}(Q^2b_{123}) \; \frac{1}{s_{12}s_{13}z_{13}z_{12}} \\
&&\qquad =   c_\Gamma^2 \frac{(Q^2b_{123})^{-2\eps}}{2\eps}\frac{\Gamma^4(1-\eps)}{\Gamma(1-4\eps)} 
\; {}_4F_3 (1 -\eps, -2\eps, -2\eps, -2\eps; 1 - 2\eps, 1 - 2\eps, -4\eps; 1)\,,
\nn
\eea
\bea
&&\MI^{(2;4)}_{C}(Q^2b_{123})=\int\d\Phi_{C_{123}}(Q^2b_{123}) \; \frac{1}{s_{12}s_{13}z_2z_3} \\
&&\quad=c_\Gamma^2 (Q^2b_{123})^{-2\eps}\bigg[
\;3\frac{\Gamma(1-\eps)^5}{\eps^4\Gamma(1-2\eps)\Gamma(1-3\eps)}\nn\\
&&\qquad\qquad-\frac{\Gamma(1-2\eps)\Gamma(1-\eps)^3\Gamma(1+\eps)}{2\eps^4\Gamma(1-4\eps)}
{}_3F_2 (-2\eps, -2\eps, -2\eps; 1 - 2\eps, -4\eps; 1)\nn\\
&&\qquad+\frac{\Gamma(1-\eps)^5}{\eps^2(1-\eps)(1+\eps)\Gamma(1-3\eps)\Gamma(1-2\eps)}
{}_4F_3(1, 1 - \eps, 1 - \eps, 1 - \eps; 1 - 3\eps, 2 - \eps, 2 + \eps; 1) \bigg]\,.\nn
\eea
The expansions around $\eps=0$ of the hypergeometric functions, can be obtained with the HypExp package \cite{Huber:2005yg}.

\subsection{Example at NNLO}
\begin{figure}  
\centering 
\begin{minipage}{.5\textwidth}
  \centering
\includegraphics[width=0.87\textwidth]{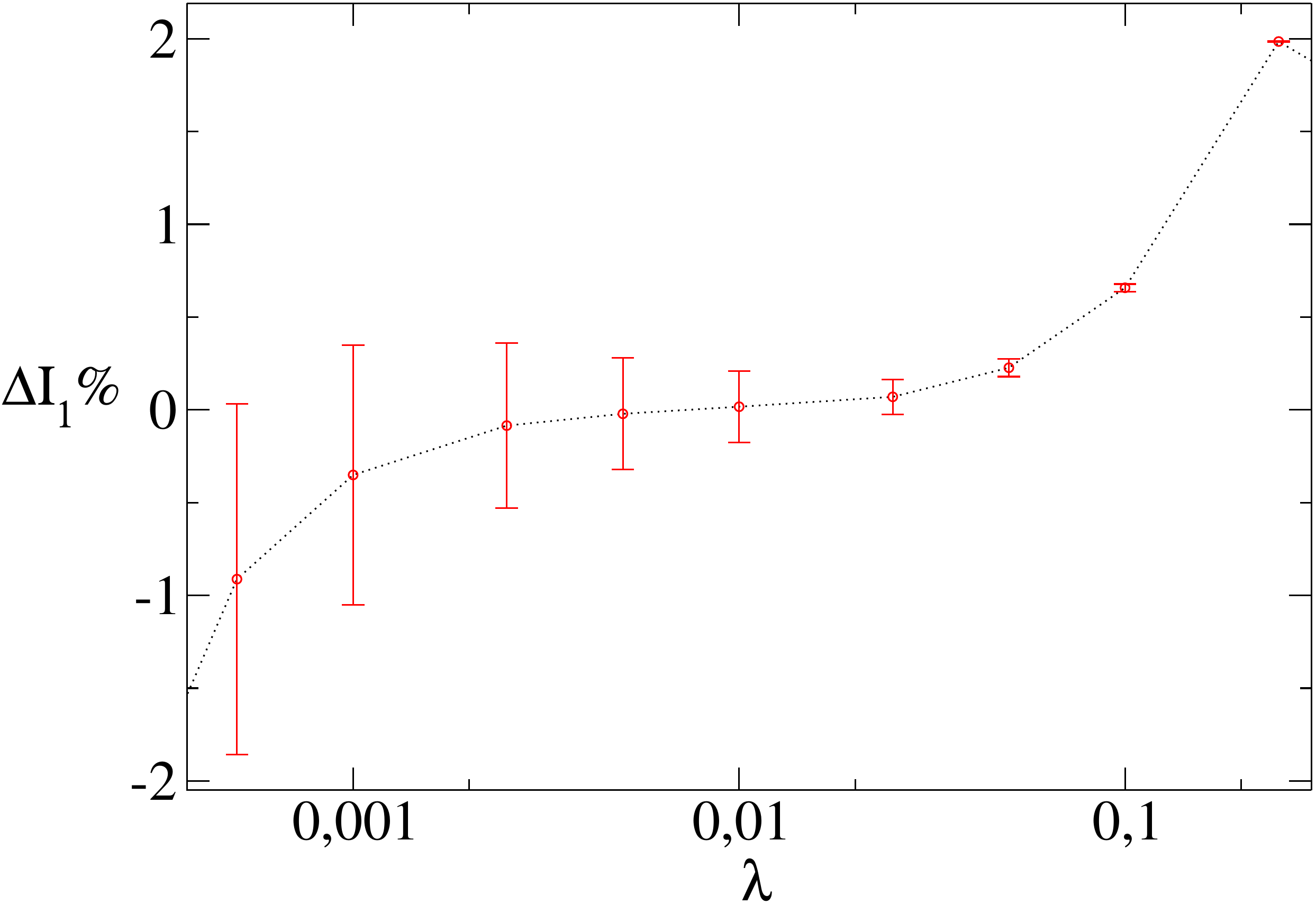}
\end{minipage}%
\begin{minipage}{.5\textwidth}
  \centering
\includegraphics[width=0.84\textwidth]{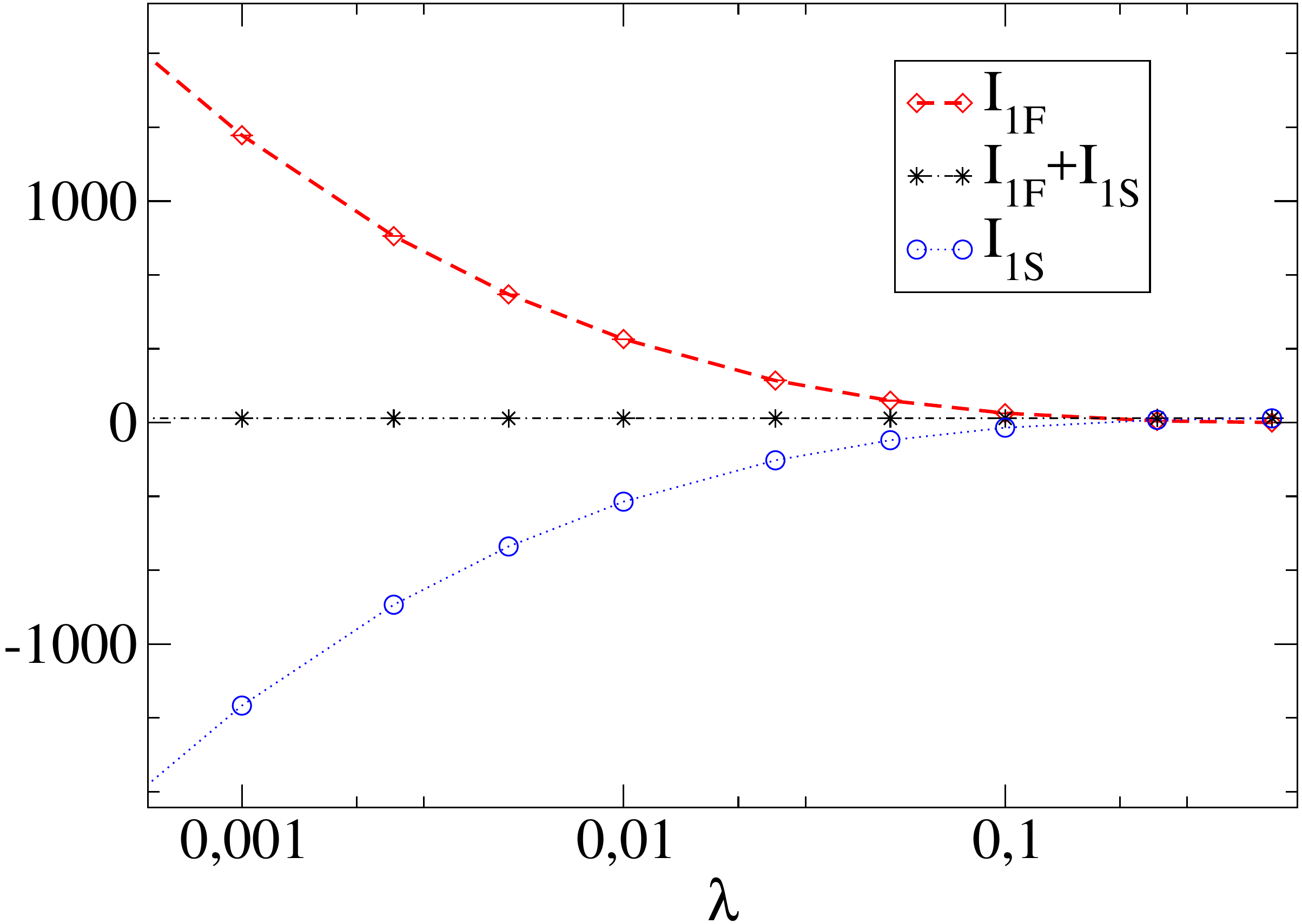}
\end{minipage}
\caption{This figure shows $\Delta I_1(\lambda)$ on the left and also separately the $\eps^0$ coefficients of $I_{1F},I_{1S}$ and their sum on the right in the subtraction method. In both figures $I_{1F}$ is evaluated numerically using $10^8$ points for each value of $\lambda$.}
\label{fig:lambda3}
\end{figure}

Let us now consider an example where we can apply the ideas we developed in the last section in a simple setting. The example we shall consider is 
\beq
\label{eq:NNLOEx1}
I_1=\int  \frac{\d\Phi_{1234} }{s_{34}s_{134}s_{234}}=S_\Gamma\l[ -\frac{1}{4\epsilon^3}-\frac{1}{2\epsilon^2}+\Big(\frac{5}{2}\zeta_2-1\Big)\frac{1}{\epsilon}+5\zeta_2+11\zeta_3-2 +\order{(\eps^0)}\r]
\eeq
where we have set $Q^2=1$ and 
\beq
S_\Gamma=\Phi_2\;(c_\Gamma)^2.
\eeq
We have analytically evaluated this integral applying again reverse unitarity, IBP reduction as well as the results for the master integrals given in \cite{Gehrmann-DeRidder:2003pne}. This integral contains the following set of singular regions:
\beq
R=\{C_{34},S_{34},C_{134},C_{234}\}\,,
\eeq
which leads to the following set of regions:
\bea
\mathcal{U}=&&\qquad\qquad\qquad\big\{\{C_{34}\},\{S_{34}\},\{C_{134}\},\{C_{234}\},\nn\\
&&
\;\;\{C_{34},S_{34}\},\{C_{34},C_{134}\},\{C_{34},C_{234}\},\{S_{34},C_{134}\},\{S_{34},C_{234}\},\\
&&
\;\;\qquad\qquad\quad\{S_{34},C_{134},C_{34}\},\{S_{34},C_{234},C_{34}\}\big\}\nn\,.
\eea
Making use of permutation invariance the number of different counter-terms is reduced to seven. Using IBP reduction we can evaluate them using the master integrals given in section \ref{sec:masters}. We list analytic results in the following:
\begin{itemize}
   \item $\{C_{34}\}$:
   \beq
   \int   \frac{\d\Phi_{12\t{34}} }{s_{1\t{34}}s_{2\t{34}}}\,\int \frac{\d\Phi_{b_{34}}(b_{34})}{s_{34}} 
   = -S_\Gamma \frac{(b_{34})^{-\eps}}{\eps^3}\frac{(1-3\eps)(2-3\eps)\Gamma^5(1-\eps)}{\Gamma(3-3\eps)\Gamma(2-2\eps)}  
   \eeq   
   \item $\{S_{34}\}$:
   \beq
   \int   \d\Phi_{12} \,\int \frac{\d\Phi_{S_{34}}^{(1,2)}(s_{12},a_{34})}{s_{34}s_{1(34)}s_{2(34)}} 
   = -S_\Gamma \frac{(a_{34}^4)^{-\eps}}{2\eps^3}\frac{(1-4\eps)(3-4\eps)\Gamma^4(1-\eps)}{\Gamma(4-4\eps)}
   \eeq
   \item $\{C_{134}\}$:
   \beq
   \int  \frac{ \d\Phi_{\t{134}2}}{s_{\t{134}2}} \,\int \frac{\d\Phi_{C_{134}}(b_{134})}{s_{34}s_{134}z_{34}} 
   = -S_\Gamma \frac{(b_{134}^2)^{-\eps}}{4\eps^3}\frac{(1-3\eps)(2-3\eps)\Gamma^5(1-\eps)}{\Gamma(3-3\eps)\Gamma(2-2\eps)}
   \eeq
   \item $\{C_{134}, S_{34}\}$:
   \beq
   \int  \frac{ \d\Phi_{\t{134}2}}{s_{\t{134}2}} \,\int \frac{\d\Phi_{C_{134}S_{34}}(b_{134},a_{34})}{s_{34}s_{1(34)}z_{34}} 
    =-S_\Gamma \frac{(a_{34}^2b_{134}^2)^{-\eps}}{4\eps^3}\frac{(1-2\eps)\Gamma^4(1-\eps)}{\Gamma^2(2-2\eps)}
   \eeq
   \item $\{C_{34}, S_{34}\}$:
   \beq
    \int   \d\Phi_{12} \int \frac{\d\Phi_{S_{\t{34}}}^{(1,2)}(s_{12},a_{34})}{s_{1\t{34}}s_{2\t{34}}}\,\int \frac{\d\Phi_{C_{34}}(b_{34})}{s_{34}} 
    =  -S_\Gamma \frac{(a_{34}^2b_{34})^{-\eps}}{\eps^3}\frac{(1-2\eps)\Gamma^4(1-\eps)}{\Gamma^2(2-2\eps)}                 
    \eeq   
   \item $\{C_{34}, C_{134}\}$:
   \beq
   \int  \frac{ \d\Phi_{\t{134}2}}{s_{\t{134}2}} \,\int \frac{\d\Phi_{C_{1\t{34}}}(b_{134})}{s_{1\t{34}}z_{\t{34}}} \,\int \frac{\d\Phi_{C_{34}}(b_{34})}{s_{34}}=  -S_\Gamma \frac{(b_{34}b_{134})^{-\eps}}{\eps^3}\frac{(1-2\eps)\Gamma^4(1-\eps)}{\Gamma^2(2-2\eps)}                                               
   \eeq 
   \item $\{S_{34}, C_{234}, C_{34}\}$:
   \beq
    \int  \frac{ \d\Phi_{\t{134}2}}{s_{\t{134}2}} \,\int \frac{\d\Phi_{C_{1\t{34}}S_{\t{34}}}(b_{134},a_{34})}{s_{1\t{34}}z_{\t{34}}} \,\int \frac{\d\Phi_{C_{34}}(b_{34})}{s_{34}} = -S_\Gamma \frac{(a_{34}b_{34}b_{134})^{-\eps}}{\eps^3}\frac{\Gamma^2(1-\eps)}{\Gamma(2-2\eps)}                            
    \eeq
\end{itemize}
Summing the counter-terms, weighted with the appropriate signs, we obtain:
\bea
I_{1S}=&&\int  \frac{\d\Phi_{1234} }{s_{34}s_{134}s_{234}}\Theta(\mathrm{singular})\\
=&&S_\Gamma\Big[-\frac{1}{4\epsilon^3}-\frac{1}{2\epsilon^2}
 +\Big(-1+\frac{5}{2}\zeta_2\Big)\frac{1}{\epsilon}\nn\\
&&\qquad
-2+5\zeta_2+10\zeta_3-2\zeta_2L_{a_{34}} +L_{a_{34}}^3-\frac{1}{2}L_{a_{34}}^2(L_{b_{134}}+L_{b_{234}})\nn\\
&&\qquad
-L_{b_{34}}L_{a_{34}}^2-\frac{1}{2}L_{a_{34}}(L_{b_{134}}^2+L_{b_{134}}^2)+L_{a_{34}}L_{b_{34}}(L_{b_{134}}+L_{b_{234}})+\order{(\eps^1)}\Big]\nn
\eea
where we use the notation 
\beq
L_z^n=\log^n z\,.
\eeq 
The singular contribution thus correctly reproduces the poles of eq. (\ref{eq:NNLOEx1}). We continue with a numerical check of the finite part of eq. (\ref{eq:NNLOEx1}), which is defined by:
\beq
I_{1F}=\int  \frac{\d\Phi_{1234} }{s_{34}s_{134}s_{234}}\Theta(F)\,,
\eeq
with
\beq
\Theta(F)=\Theta(s_{34}>b_{34})\Theta(s_{(12)(34)}>a_{34}s_{12})\Theta(s_{134}>b_{134})\Theta(s_{234}>b_{234})\,.
\eeq
A numerical evaluation of $I_{1F}$ using the slicing method for a range of values of a parameter $\lambda$, which fixes 
the cut-off parameters via
\beq
a_{34}=\lambda,\qquad b_{134}=b_{234}=\lambda^2,\qquad b_{34}=\lambda^3\,,
\eeq
is shown in figure \ref{fig:lambda3}. The strict hierarchy which the slicing cut offs must satisfy unfortunately limits the range of possible values of $\lambda$ which can be chosen without risking loss of numerical stability. Nevertheless good numerical convergence is observed in the range $\lambda\in[0.1,0.001]$ for this particular example. In general we may not expect such good convergence in the range $\lambda\sim 0.1$ which is likely due to the trivial numerator. 

\section{Counter terms for final state real emissions in Yang Mills theory}
\label{sec:CTsYM}
In the following section we will employ the geometric subtraction formalism to construct suitable counter-terms for tree-level Yang Mills amplitudes; that is QCD amplitudes without quarks, which we ignore here for simplicity. While amplitudes factorise completely in collinear limits, soft limits are color correlated. To make the counter-terms as simple as possible we will employ tailor-made soft volumes for the different color correlated eikonal factors, which make up a particular soft limit. 

To accomplish this task we correlate the sum over singular regions with individual interference terms contributing to the squared amplitude:
\beq
\mathcal{O}_{l;1..n+l}^{\mathrm{Singular}}=\int \d\Phi_{1..n+l}\,\mathcal{J}_{1..n+l}^{(l)}\,\mT(\mathrm{Singular}) \ast |\mathbf{\M}_{1..n+l}|^2  \,,
\eeq
with
\beq
\mT(\mathrm{Singular})\ast |\mathbf{\M}_{1..n+l}|^2 =\sum_{k,m} \,
(\M_k^*)_{1..n+l}\,
(\M_m)_{1..n+l}\,
\Theta(\mathrm{Singular}(k,m))\,,
\eeq
where the sum over $k,m$ labels different color projected (sets of) Feynman diagrams contributing to the matrix element $\M$, such that $\M=\sum_k \M_k$. We can then use eq. (\ref{eq:singularJ}) individually for each component of $\mT(\sing)$. In order to ensure gauge invariance in the counter-terms it is sufficient that sets of Feynman diagrams multiplying a particular entry of $\mT(\sing)$ conspire to a singular limits which are gauge invariant, such as the Altarelli-Parisi splitting function or the eikonal factor.

In the following we shall present how to define $\mT(\sing)$ explicitly at NLO and at NNLO. Further more we will provide  all integrated counter-terms for final state emissions at these first two orders.

\subsection{Counterterms at NLO}
At NLO we define the sum over singular regions as
\bea
\label{eq:CataniU}
&&\mathcal{O}_{1;1..n+1}^{\mathrm{Singular}}=-\lim_{a_i\to 0}\,\lim_{b_{ij}\to0} \\
&&\qquad\qquad  \sum_{U\in \U^{(1)}} (-1)^{|U|} \int \d\Phi_{1..n+1}\,\mathcal{J}_{1..n+1}^{(1)}\, 
\prod_{r\in U}  \mT(r)\ast|\mathbf{\M}_{1..n+1}|^2 ,\nn
\eea
with the set of regions defined by
\beq
\U^{(1)}=\{\{C_{ij}\},\{S_i\},\{C_{ij},S_i\}\}\,.
\eeq
To define a suitable $\mT$-matrix it is sufficient to define how it behaves in the singular limit. In the soft limit we require it to behave as follows:
\beq
\lim_{a_{k}\to0} \mT(S_{k})\ast |\mathbf{\M}_{1..n+1}|^2 
=\sum_{ij}\,|\M^{(i,j)}_{1..\not{k}..n+1}|^2\, \SE_{k}^{(i,j)}\,\Theta(a_{k}s_{ij}-s_{k(ij)})\,
\eeq
where the eikonal factor is given by
\beq
\SE_{k}^{(i,j)}=2\,\frac{s_{ij}}{s_{ik}s_{jk}}\,,
\eeq
and $|\M^{(i,j)}_{1..n}|^2$ denotes the color correlated squared (Born) amplitude:
\beq
|\M^{(i,j)}_{1..n}|^2=\langle\M_{1..n} |\T_i.\T_j |\M_{1..n}\rangle\,.
\eeq
Here $\T_i$ denote the (by now standard) color charge operator; see, e.g., \cite{Catani:1999ss}. In the collinear limit we require the $\mT$-matrix to  factorise completely: 
\beq
\lim_{b_{ij}\to0}\, \mT(C_{ij}) \ast|\mathbf{\M}_{..i..j..}|^2 =
\frac{2}{s_{ij}}\, (P_{ij})_{\mu_1\mu_2}\,|\mathbf{\M^{\mu_1\mu_2}}_{..\t{ij}..}|^2\, 
\Theta(b_{ij}Q^2-s_{ij})\,.
\eeq
Here $(P_{ij})_{\mu_1\mu_2}$ is the standard spin correlated gluonic Altarelli-Parisi splitting function defined in, e.g., eq. (12) of \cite{Catani:1999ss} and $|\mathbf{\M^{\mu_1\mu_2}}_{..\t{ij}..}|^2$ is the spin correlated squared matrix element, denoted $\mathcal{T}^{\mu\nu}$ in \cite{Catani:1999ss}. Let us remark also that the different soft volumes $\Theta(a_{n+1}s_{ij}-s_{n+1(ij)})$ all collapse to the same limit in the soft-collinear limit; since 
\beq
\lim_{i||k}\Theta(a_{k}s_{ij}-s_{k(ij)})=\Theta(a_{k}-z_k)\,.
\eeq
To write down the integrated counter-term it is convenient to define the functions: 
\bea
\I_{g}^S(s_{kl},a_i)&=&\int \d\Phi_{S_{i}}^{(k,l)}(s_{kl},a_{i})\, \SE_{i}^{(k,l)} =   2c_\Gamma \frac{(a_i^2s_{kl})^{-\eps}}{\eps^2}\frac{\Gamma(1-\eps)^2}{\Gamma(2-2\eps)}\,,
\eea
\bea
\I_{gg}^C(Q^2b_{ij})&=&\int \d\Phi_{C_{ij}}(Q^2b_{ij})\,\frac{2}{s_{ij}}\, \langle P_{gg}(z_i) \rangle \\
&=& 6C_Ac_\Gamma  \frac{(Q^2b_{ij})^{-\eps}}{\eps^2} \frac{(1-\eps)(4-3\eps)}{(3-2\eps)} \frac{\Gamma(1-\eps)^2}{\Gamma(2-2\eps)} \,,\nn
\eea
\bea
\I_{gg}^{SC}(Q^2b_{ij},a_i)&=&\int \d\Phi_{C_{ij}S_{i}}(Q^2b_{ij},a_{i})\,\frac{2}{s_{ij}}\, \langle P_{gg}(z_i) \rangle \Big|_{z_i\to0}\\
&=&
 4C_Ac_\Gamma \frac{(Q^2b_{ij}a_i)^{-\eps}}{\eps^2}\,,\nn
\eea
as well as the following linear combination: 
\beq
\I_{ab}^{\t{C}}(Q^2b_{ij},a_i,a_j)=
\I_{ab}^C(Q^2b_{ij})-\I_{ab}^{SC}(Q^2b_{ij},a_i)-\I_{ab}^{SC}(Q^2b_{ij},a_j)\,.
\eeq
Here $\langle P_{gg}(z_i) \rangle$ denotes the spin averaged Altarelli-Parisi splitting function. In terms of these functions one can write down a compact formula for the quantity $\O_{1;1..n+1}^{\mathrm{Singular}}$:
\bea
\label{eq:nlopoles}
\O_{1;1..n+1}^\sing&=&\sum_{i>j} \I_{ij}^{\t{C}}(Q^2b_{ij},a_i,a_j) \O_{0;1..\t{ij}..n+1} \,\\
&+&\sum_{i}\sum_{k,l\neq i} \int \d\O_{0;1..\not{i}..n+1}^{(k,l)}   \I_{g_i}^S(s_{kl},a_i)\,,\nn
\eea
with
\beq
\d\O_{l;1..n+l}^{(i,j)}= \d\Phi_{1..n+l}\, |\M_{1..n+l}^{(i,j)}|^2 \J^{(l)}_{1..n+l}\,.
\eeq
It is straight forward to show that eq. (\ref{eq:nlopoles}) agrees with the corresponding one-loop 
pole operator given by Catani in, e.g., \cite{Catani:1998bh}.

\subsection{Counter terms at NNLO}
At NNLO we define the sum over singular regions similarly as
\bea
\label{eq:CataniU}
&&\O_{2;1..n+2}^\sing=-\lim_{a_{ij}\to 0}\,\lim_{a_i\to 0}\;\lim_{b_{ijk}\to 0}\,\lim_{b_{ij}\to0} \\
&&\qquad\qquad  \sum_{U\in \U^{(2)}} (-1)^{|U|} \int d\Phi_{1..n+2}\,\mathcal{J}_{1..n+2}^{(2)}\, \prod_{r\in U}  \mT(r) \ast |\mathbf{\M}_{1..n+2}|^2 \;,\nn
\eea
with $\U^{(2)}$ defined similarly, although not identically due to the more elaborate soft structure, as in eq. (\ref{eq:U2}). To define the limits $\mT(C_{ij})$ and $\mT(S_{i})$ we simply use the NLO definitions. The triple collinear limit is defined similarly to the double collinear:
\bea
\lim_{b_{ijk}\to0}\,\mT(C_{ijk}) \ast |\mathbf{\M}_{..i..j..k..}|^2 &=&\frac{4}{(s_{ijk})^2}\, (P_{ijk})_{\mu_1\mu_2}\,|\mathbf{\M^{\mu_1\mu_2}}_{..\t{ijk}..}|^2\, \Theta(Q^2b_{ijk}-s_{ijk})\,,\nn\\
\eea
with $(P_{ijk})_{\mu_1\mu_2}$ the triple collinear Altarelli-Parisi splitting function, defined, e.g., in eq. (66) of \cite{Catani:1999ss}. 

The double soft limit receives two independent color-correlated contributions:
\bea
\label{eq:SS}
\lim_{k,l\to0} |\mathbf{\M}_{1..n+2}|^2 &=&\frac{1}{2}\sum_{i,j,r,t=0}^n\,|\M^{(i,j)(r,t)}_{1..\not{k}..\not{l}..n}|^2\, \SE^{(i,j)}_{k}\, \SE_{l}^{(r,t)}\,\,\\
&-&\frac{1}{2}C_A\sum_{i>j=1}^n\,|\M^{(i,j)}_{1..\not{k}..\not{l}..n}|^2\, \Big(2\,\SE_{kl}^{(i,j)}-\SE_{kl}^{(i,i)}-\SE_{kl}^{(j,j)}\Big)\,,\nn
\eea
with $\SE_{kl}^{(i,j)}$ the double-soft eikonal function defined in eq. (109) of \cite{Catani:1999ss},
\beq
|\M^{(i,j)(r,t)}_{1..\not{k}..\not{l}..n}|^2=
\langle\M_{1..\not{k}..\not{l}..n+2}| \{\T_i.\T_j,\T_k.\T_l\} |\M_{1..\not{k}..\not{l}..n+2} \rangle\,
\eeq
and where we have used the color conservation identity
\beq
\sum_i\T_i|\M\rangle=0\,,
\eeq
to shift the color diagonal terms  $\SE_{kl}^{(i,i)}$ into the color off-diagonal terms. 

The double soft limit thus contains two terms. The first term factorises over the soft momenta and contains color-kinematic correlations with up to four hard partons (Wilson lines). Instead the second term contains kinematic correlations of the two soft momenta and color-kinematic correlations with up to two hard partons. It is interesting to note that while both terms are singular in triple collinear regions $i||l||k$ and $j||l||k$ only the second term contributes to the limit containing $k||l$ and only the first term contributes to limits containing $i||k$, $i||l$, $j||k$ and $j||l$.

A natural measure for the second term is $\d\Phi_{S_{kl}}^{(i,j)}(s_{ij},a_{kl})$ which we introduced earlier. Instead we shall treat each of the eikonal factors in the first term with a single soft phase space measure, i.e. with $\d\Phi_{S_{k}}^{(i,j)}(s_{ij},a_{k})\d\Phi_{S_{l}}^{(r,t)}(s_{rt},a_{l})$. The ``true'' double soft measure will thus be associated only to the second term, while the first term is naturally associated to the product of two single soft limits. These considerations lead us to define:
\bea
&&\lim_{a_{kl}\to0} \mT(S_{kl})\ast|\mathbf{\M}_{1..n+2}|^2 =\\
&&\qquad-\frac{1}{2}C_A\sum_{i,j=1\neq k,l}^{n+2}\,|\M^{(i,j)}_{1..\not{k}..\not{l}..n+2}|^2\, (2\SE_{kl}^{(i,j)}-\SE_{kl}^{(i,i)}-\SE_{kl}^{(j,j)})\,\Theta(a_{kl}s_{ij}-s_{(kl)(ij)})\,,\nn
\eea
and
\bea
&&\lim_{a_{kl}\to0} \lim_{(a_k,a_l)\to0} (1-\mT(S_{kl}))\mT(S_{k})\mT(S_{l})\ast|\mathbf{\M}_{1..n+2}|^2 =\\
&&\qquad+\frac{1}{2}\sum_{i,j,r,t\neq k,l}\,|\M^{(i,j)(r,t)}_{1..\not{k}..\not{l}..n+2}|^2\, \SE^{(i,j)}_{k}\, \SE_{l}^{(r,t)}\,\,\Theta(a_{k}s_{rt}-s_{k(rt)})\,\Theta(a_{l}s_{ij}-s_{l(ij)})\,.\nn
\eea
With this distribution of the theta functions it follows that the double soft limit $k,l\to0$ is not entirely controlled by the limit $a_{kl}\to 0$, instead also $a_k,a_l\to0$ is required for both terms in eq. (\ref{eq:SS}) to diverge. The region cancellation between the regions $\Theta(S_{ij})\Theta(S_{i})\Theta(S_{j})$ and $\Theta(S_{i})\Theta(S_{j})$, which was given in eq. (\ref{eq:geometriccancellationI}), therefore only holds for the second term in eq. (\ref{eq:SS}), since the first does not contribute to the region $\Theta(S_{ij})$. 

Let us now consider what happens in the strongly ordered double soft limits corresponding to $\{S_{ij},S_{i}\}$. One can show, by taking the successive soft limits, that this limit becomes:
\bea
&&\lim_{a_{kl}\to0}\lim_{a_{k}\to0}  \mT(S_{k})\mT(S_{kl}) \ast|\mathbf{\M}_{1..n+2}|^2 =\nn\\&&\qquad\qquad\qquad
-C_A\sum_{i,j\neq k,l}^{n+2}\,|\M^{(i,j)}_{1..\not{k}..\not{l}..n+2}|^2\,\SE_{l}^{(i,j)}\,\Theta(a_{kl}s_{ij}-s_{l(ij)})\\
&&\qquad\qquad\qquad\cdot
\Big(\SE_{k}^{(l,j)}\Theta(a_{k}s_{lj}-s_{k(lj)}) +\SE_{k}^{(l,i)}\Theta(a_{k}s_{li}-s_{k(li)}) -\SE_{k}^{(i,j)}\Theta(a_{k}s_{ij}-s_{k(ij)})\Big)\,.
\nn
\eea
Thus the different single eikonal factors which contribute to the strongly ordered limit of the non-Abelian double soft limit come with their distinct single soft phase spaces. 
A caveat of the method is that in the strongly ordered soft limit certain collinear limits such as $\{S_{kl},S_k,C_{il}\}$, which would usually not survive in the non-Abelian double soft factor, e.g. $\{S_{kl},C_{ik}\}$ is not singular, now survive:
\bea
\label{eq:CikSijcaveat}
&&\lim_{a_{jk}\to0}\lim_{a_{k}\to0} \lim_{b_{il}\to0}\mT(C_{il})\mT(S_{k})\mT(S_{kl}) \ast|\mathbf{\M}_{1..n+2}|^2 
=\\&&\qquad\qquad\qquad\; 
 -C_A\sum_{j\neq k,l}^{n+2}\,|\M^{(i,j)}_{1..\not{k}..\not{l}..n+2}|^2\, 
\frac{2}{s_{il}} \langle P_{il}(z_l)\rangle\Big|_{z_l\to0}\,\Theta(a_{kl}-z_l)\,\Theta(b_{kl}Q^2-s_{il})\nn\\
&&\qquad\qquad\qquad\qquad\cdot\;
\SE_{k}^{(\t{il},j)}\Big(
\Theta(a_{k}z_ls_{\t{il}j}-z_ls_{k\t{il}}-s_{kj}) -
\Theta(a_{k}s_{\t{il}j}-s_{k\t{il}}-s_{kj})
\Big)\,.\nn
\eea
The re-scaling invariance of the eikonal factor,
\beq
\SE_{k}^{(\t{il},j)}=\SE_{k}^{(z_l\t{il},j)}\,,
\eeq
ensures that the last two terms in eq. (\ref{eq:CikSijcaveat}) would cancel, if it was not for the differing $\Theta$-functions which break the re-scaling invariance upon which the cancellation mechanism relies.

The chosen distribution of single and double soft $\Theta$-functions similarly splits the various overlapping soft-collinear limits. For instance the triple collinear double soft limit splits  into a non-Abelian part:
\bea
&&\lim_{a_{kl}\to0}\lim_{b_{ikl}\to0} \mT(S_{kl})\mT(C_{ikl})\ast|\mathbf{\M}_{1..n+2}|^2 =\\
&&\qquad C_A^2\,|\M_{1..\not{ikl}..n+2}|^2\, (2\SE_{kl}^{(i,n)}-\SE_{kl}^{(i,i)}-\SE_{kl}^{(n,n)})\Big|_{z_i\to1}\,
\Theta(a_{kl}-z_k-z_l) \Theta(Q^2b_{ikl}-s_{il}-s_{ik})\,,\nn
\eea
and an Abelian part:
\bea
&&\lim_{a_{kl}\to0}\lim_{(a_k,a_l)\to0} \lim_{b_{ikl}\to0} (1-\mT(S_{kl}))\mT(S_{k})\mT(S_{l}))\mT(C_{ikl})\ast|\mathbf{\M}_{1..n+2}|^2 =\\
&&\qquad+4 C_A^2\,|\M_{1..\t{ikl}..n+2}|^2\, \SE^{(i,n)}_{k} \SE^{(i,n)}_{l}\Big|_{z_i\to1}\,\,\Theta(a_{k}-z_k)\,\Theta(a_{l}-z_l)\Theta(Q^2b_{ikl}-s_{il}-s_{ik})\,,\nn
\eea
where $n$ denotes the collinear reference vector such that, e.g.,
\beq
\SE^{(i,n)}_{k}=2\frac{z_i}{s_{ik}z_k}\,.
\eeq
The single soft triple collinear limits instead are split into three different eikonal factors:
\bea
\label{eq:C123S1}
&&\lim_{a_{k}\to0}\lim_{b_{ijk}\to0} \mT(S_{k})\mT(C_{ikl})\ast|\mathbf{\M}_{1..n+2}|^2 =\nn\\&&\qquad 
+C_A\,|\M_{1..\t{ikl}..n+2}|^2\, \frac{2}{s_{ij}} P_{ij}(z_i) 
\Big( \SE_{k}^{(i,j)} \Theta(a_{k}s_{ij}-s_{i(kl)}) \Theta(Q^2b_{ikl}-s_{ij})\\&&\qquad
+\SE_{k}^{(\t{ij},n)}\Theta(Q^2b_{ikl}-s_{ij}-s_{k\t{ij}})
\big(\Theta(a_{k}z_i-z_k) +\Theta(a_{k}z_j-z_k) \big)
\Big),
\nn
\eea
with 
\beq
\SE_{k}^{(\t{ij},n)}=\SE_{k}^{(z_i\t{ij},n)}=\SE_{k}^{(z_j\t{ij},n)}=2\frac{1}{s_{\t{ij}k}z_k}\,.
\eeq
It is interesting to note how the different eikonals contributing to this limit come with their individual phase space volume constraints. The first term can be interpreted as a degeneration of the limit $\{C_{ijk},S_k\}$ into something like an ordered $\{C_{ij},S_k\}$ with the soft scale smaller then the collinear one. The second term on the other hand takes the form of a soft collinear limit where the soft and collinear scales are of the same order. Let us briefly analyse the kinematics of this second limit. Taking the limit $a_k\to 0$ in $\Theta(a_{k}z_i-z_k)$  forces the momentum fraction $z_k$ to vanish; in turn this means that 	
\beq
p_k^\mu\to \frac{p_k.p_{\t{ij}}}{n.p_{\t{ij}}} n^\mu \quad \Rightarrow \quad 2p_k.p_{ij} \to \frac{p_k.p_{\t{ij}}}{n.p_{\t{ij}}} 2n.p_{ij}=2p_k.p_{\t{ij}}
\eeq
and so $s_{ijk}\to s_{ij}+s_{k\t{ij}}$. It appears almost as something of a miracle that the triple collinear splitting function produces a factor $(s_{ij}+s_{k\t{ij}})^2$ in the numerator which precisely cancels the overall denominator $1/s_{ijk}^2$. A feature which leads to a welcome simplification for the integrated counter-term associated to this limit.

The strongly ordered double soft limit receives contributions only from the non-Abelian double soft limit of the triple collinear region and yields 
\bea
&&\lim_{a_{ik}\to0}\lim_{a_{k}\to0}\lim_{b_{ijk}\to0} \mT(S_{k})\mT(S_{ik})\mT(C_{ikl})\ast|\mathbf{\M}_{1..n+2}|^2 =\\&&\qquad 
+C_A\,|\M_{1..\t{ikl}..n+2}|^2\, \frac{2}{s_{ij}} P_{ij}(z_i)\Big|_{z_i\to0} \Theta(a_{ik}-z_i)  
\Big( \SE_{k}^{(i,j)} \Theta(a_{k}s_{ij}-s_{i(kl)}) \Theta(Q^2b_{ikl}-s_{ij})\nn\\&&\qquad \qquad\qquad\qquad\qquad\qquad+\Theta(Q^2b_{ikl}-s_{ij}-s_{k\t{ij}})\SE_{k}^{(\t{ij},n)}\big(\Theta(a_{k}z_i-z_k) -\Theta(a_{k}-z_k) \big)
\Big),
\nn
\eea
Other limits can be worked out similarly starting from these expressions and using the soft and collinear limits of amplitudes. All in all, with this choice of the single and double soft color correlated phase space boundaries, the NNLO set of regions which enters eq. (\ref{eq:CataniU}) is given by:
\bea
\U^{(2)}=
&&\Big\{\{S_{i}\},\{S_{ij}\},\{C_{ij}\},\{C_{ijk}\},\{C_{ijk}, C_{ij}\}, \{C_{ijk}, S_{ij}\}, \{C_{ijk}, S_{i}\}, \{C_{ij}, C_{kl}\},\nn\\&&\;\;
\{C_{ij}, S_{ij}\}, \{C_{ij}, S_{i}\}, \{C_{ij}, S_{k}\}, \{S_{ij}, S_{i}\}, \{S_{i}, S_{j}\}, \{S_{i}, S_{j},S_{ij}\},\{C_{ijk}, C_{ij}, S_{ij}\},\nn\\&&\;\;
\{C_{ijk}, C_{ij}, S_{i}\}, \{C_{ijk}, C_{ij}, S_{k}\}, \{C_{ijk}, S_{ij}, S_{i}\},\{C_{ijk}, S_{i}, S_{j}\},\{C_{ijk}, S_{i}, S_{j}, S_{ij}\},\nn\\&&\;\;
\{C_{ij}, C_{kl}, S_{i}\}, \{C_{ij}, S_{ij}, S_{i}\}, \{C_{ij}, S_{i}, S_{k}\}, \{C_{ij}, S_{i}, S_{k}, S_{ik}\}, \{C_{jk}, S_{ij}, S_{i}\},\nn\\&&\;\;
\{C_{ijk}, C_{ij}, S_{ij}, S_{i}\}, \{C_{ijk}, C_{ij}, S_{ik}, S_{k}\}, 
\{C_{ijk}, C_{ij}, S_{i}, S_{k}\},\{C_{ijk}, C_{ij}, S_{i}, S_{k},S_{ik}\},\nn\\&&\;\;
\{C_{ij}, C_{kl}, S_{i}, S_{k}\}, \{C_{ij}, C_{kl}, S_{i}, S_{k},S_{ik}\}\Big\}\,.
\eea
It is convenient to re-organise the sum over regions in eq. (\ref{eq:CataniU}) by introducing the sub-divergence subtracted regions $\bar{C}_{ijk}$, $\hat{S}_{ij}$, and  $\bar{C}_{ij}$, as certain subsets of $\U^{(2)}$. The region $\bar{C}_{ijk}$ is defined as the set of all regions which contain $C_{ijk}$. The region $\hat{S}_{ij}$ is defined as the set of all regions 
containing $S_{ij}$ apart from those also containing $C_{ijk}$; and the region $\bar{C}_{ij}$ includes $C_{ij}$ and its overlaps with the regions $S_i$ and $S_j$.

Using $\mT$s we can also define these regions as follows:
\bea
\mT(\bar{C}_{12})&=&\mT(C_{12})\,\Big(1-\mT(S_1)-\mT(S_2)\Big)\,,\\
\mT\big(\hat S_{12})&=&\mT(S_{12})\Big[\big(1-\mT(S_1)-\mT(S_2)\big)\big(1-\mT(C_{12})\big)\nn\\
\label{eq:Shat}
&&  \qquad\qquad  + \mT(S_1)\sum_{k\neq 1,2}\mT(C_{2k}) +\mT(S_2)\sum_{k\neq 1,2}\mT(C_{1k})  \Big]\\
&&-\mT(S_1)\mT(S_2)(1-\mT(S_{12}))\,, \nn\\\
\mT(\bar{C}_{123})&=&\mT(C_{123})\,\Big[
\big(1-\sum_{k=1}^3 \mT(S_k)\big)\big(1-\sum_{i>j=1}^3 \mT(C_{ij})\big)\nn \\
 &&
+\sum_{i>j=1}^3 \sum_{k=1\neq i,j}^3  \big(1-\mT(S_{ij})\big) \mT(S_{i}) \mT(S_{j}) \big(1-\mT(C_{ik})-\mT(C_{jk})\big)\nn \\
 &&
+\sum_{i>j=1}^3 \sum_{k=1\neq i,j}^3  \mT(S_{ij})\Big( \big(1- \mT(S_{i}) -\mT(S_{j})\big)\big(1-\mT(C_{ij})\big) \nn \\
&&\qquad\qquad\qquad +\mT(S_{j})\mT(C_{ik})+\mT(S_{i})\mT(C_{jk}) \Big)
\Big]\,.
\eea
Note in particular that the term $\mT(S_1)\mT(S_2)(1-\mT(S_{12}))$  in eq. (\ref{eq:Shat}) only receives contributions from the first term in eq. (\ref{eq:SS}), instead all other terms in eq. (\ref{eq:Shat}) receive contributions only from the respective second term in eq. (\ref{eq:SS}).

Associated to these $\Theta$-functions we define the integrals 
\bea
&&\lim_{a_{ij}\to 0}\,\lim_{a_i\to 0}\;\lim_{b_{ijk}\to 0}\,\lim_{b_{ij}\to0}\int \mT(\bar{C}_{123})\ast\d\O_{2:123..n+2}=\\
&&\qquad\qquad\qquad\I_{g_1g_2g_3}^{\bar{C}}(t_{123},t_{12},t_{13},t_{23},a_{12},a_{13},a_{23},a_1,a_2,a_3)\,\int \d\O_{0;\t{123}..n+2}\,\nn
\eea
and
\bea
&&\lim_{a_{ij}\to 0}\,\lim_{a_i\to 0}\,\lim_{b_{ij}\to0}\int \mT(\hat{S}_{12})\ast\d\O_{2:123..n+2}=\\
&&\qquad\qquad\qquad-\frac{C_A}{2}\sum_{i,j\neq 1,2}\int \d\O^{(i,j)}_{0;3..n+2}\,\I_{g_1g_2}^{\hat{S}}(s_{ij},a_{12},a_1,a_2,t_{12},t_{1i},t_{1j},t_{2i},t_{2j})\nn\\
&&\qquad\qquad\qquad+\sum_{i,j,k,l\neq 1,2}\int \d\O^{(i,j)(k,l)}_{0;3..n+2}\,\I_{g_1}^{S}(s_{ij},a_{1})\,
\I_{g_2}^{S}(s_{kl},a_{2})\,,\nn
\eea
where $t_{ij..}=Q^2b_{ij..}$ and
\bea
&&\d\O_{0;1..\not{k}..\not{l}..n+2}^{(i,j)(k,l)}=\d\Phi_{1..\not{k}..\not{l}..n+2}\, |\M^{(i,j)(r,t)}_{1..\not{k}..\not{l}..n}|^2\,\J^{(0)}_{1..\not{k}..\not{l}..n+2}\,.
\eea
In terms of these combinations we can then write out the sum in eq. (\ref{eq:CataniU}) as follows:
\bea
&&\O_{2;1..n+2}^\sing=\lim_{a_{ij}\to 0}\,\lim_{a_i\to 0}\;\lim_{b_{ijk}\to 0}\,\lim_{b_{ij}\to0} \int d\Phi_{1..n+2}\,\mathcal{J}_{1..n+2}^{(2)}\,\nn\\
&&\qquad\qquad\quad\Big( \sum_{i} \mT(S_{i})
+\sum_{i>j} \mT(\bar C_{ij})
+\sum_{i>j} \mT(\hat S_{ij})
+\sum_{i>j>k}\, \mT(\bar C_{ijk})\\
&&\qquad\qquad\qquad\qquad
-\sum_{i>j>k>l} \mT(\bar C_{ij}) \mT(\bar C_{kl})
-\sum_{i>j}\sum_{k\neq i,j} \mT(\bar C_{ij})\mT(S_{k})
\Big)\ast  |\mathbf{\M}_{1..n+2}|^2 \,.\nn
\eea
An explicit representation for the pole part in terms of the different region approximants 
can then be written as follows:
\bea
\label{eq:ngluonpoles}
\O_{2;1..n+2}^\sing&=&\sum_{i>j} \I_{g_ig_j}^{\bar{C}}(t_{ij},a_i,a_j)\,\O_{1;1..\t{ij}..n+2}\nn\\
&-&\sum_{k}\sum_{i,j\neq k} \int \d\O_{1;1..\not{k}..n+2}^{(i,j)}\, \I_{g_k}^{S}(s_{ij},a_k)\nn\\
&-&\sum_{i>j>k>l} \I_{g_ig_j}^{\bar{C}}(t_{ij},a_i,a_j)\,\I_{g_kg_l}^{\bar{C}}(t_{kl},a_k,a_l)\,\O_{0;1..\t{ij}..\t{kl}..n+2}\nn\\
&+&\sum_{i>j>k} \I_{g_ig_jg_k}^{\bar{C}}(t_{ijk},t_{ij},t_{ik},t_{jk},a_{ij},a_{ik},a_{jk},a_i,a_j,a_k)\,\O_{0;1..\t{ijk}..n+2}\\
&+&\sum_{i>j}\sum_{k\neq i,j}\sum_{l,m \in \{1,..,\t{ij},..,\not{k},..n+2\}}\I^{\bar{C}}_{g_ig_j}(t_{ij},a_i,a_j)\,\int\d\O^{(l,m)}_{0;1..\t{ij}..\not{k}..n+2}\,\I^S_{g_k}(s_{lm},a_{k})\nn\\
&+&\sum_{k>l}\sum_{i,j,m,n\neq k,l} \int \d\O_{0;1..\not{k}..\not{l}..n+2}^{(i,j)(m,n)}\, 
\I_{g_k}^{S}(s_{ij},a_k)\,\I_{g_l}^{S}(s_{mn},a_l)\nn\\
&-&\frac{C_A}{2}\sum_{k>l}\sum_{i,j\neq k,l} \int \d\O_{0;1..\not{k}..\not{l}..n+2}^{(i,j)}\, 
\I_{g_kg_l}^{\hat{S}}(s_{ij},a_{kl},a_k,a_l,t_{kl},t_{ik},t_{jk},t_{il},t_{jl})\nn
\eea
Let us remark here that the poles of the observable $\O_{1..n+2}$ do of course not depend on the parameters $a_{i..}$ and $b_{ij..}$. To get a simpler expression independent of these parameters one can alternatively set all the parameters to unity, i.e., $b_{ij..}=1,a_{i..}=1$. However to explicitly verify the cancellation of these parameters in the pole parts constitutes a welcome cross-check for its validity for a given process.

Given the results of all the integrated counter-terms, for which we present simple integral representations in Appendix \ref{sec:Integrated counterterms}, one can assemble the functions $\I_{g_ig_jg_k}^{\bar{C}}$ and $\I_{g_ig_j}^{\hat{S}}$ which make up the basic new building blocks needed at NNLO to construct the integrated counter-terms for arbitrary multiplicities. We give these functions as expansions in $\eps$ including terms up to $\order(\eps^0)$; although the results of this paper allow to construct them to arbitrary order if needed. Since these functions are lengthy, due to the many parameters on which they depend, we provide them in computer readable format with this publication. We nevertheless provide them here for the useful case where one sets 
\beq 
\label{eq:paramchoice}
Q^2b_{ijk}=\beta_2,\quad Q^2b_{ij}=\beta_1,\quad a_{ij}=\alpha_2,\quad a_{i}=\alpha_1
\eeq
for all $i,j,k$:
\bea
&&\I_{gg}^{\hat{S}}(s_{ij},\alpha_2,\alpha_1,\alpha_1,\beta_1,\beta_1,\beta_1,\beta_1,\beta_1)= 12(c_\Gamma)^2\nn\\
&&\quad\; \cdot\Bigg\{
\;\frac{1}{\eps^4}+\frac{1}{\eps^3}\Big[-\frac{2}{3}L_{s_{ij}}-\frac{4}{3}L_{\beta_1}+\frac{11}{6} \Big] 
+\frac{1}{\eps^2}\Big[
\frac{4}{3}L_{s_{ij}}L_{\beta_1}-\frac{2}{3}L_{\alpha_2}^2-\frac{8}{3}L_{\alpha_1}L_{\alpha_2}\nn\\
&&\qquad\qquad 
+\frac{4}{3}L_{\alpha_1}^2+\frac{2}{3}L_{\beta_1}^2-\frac{11}{9}L_{s_{ij}}-\frac{22}{9}L_{\alpha_2}
-\frac{22}{9}L_{\beta_1}-3\zeta_2+\frac{67}{18}
\Big]\nn\\
&&\qquad 
+\frac{1}{\eps}\Big[
\Big(2L_{s_{ij}}+\frac{8}{3}L_{\alpha_2}+\frac{8}{3}L_{\alpha_1}+4L_{\beta_1}-\frac{11}{9}\Big)\zeta_2
+\frac{2}{9}L_{s_{ij}}^3-\frac{2}{3}L_{s_{ij}}^2L_{\beta_1}-\frac{4}{3}L_{\alpha_1}^2L_{s_{ij}}\nn\\
&&\qquad \qquad 
-\frac{2}{3}L_{s_{ij}}L_{\beta_1}^2+\frac{2}{3}L_{\alpha_2}^3+\frac{4}{3}L_{\alpha_1}L_{\alpha_2}^2+\frac{4}{3}L_{\alpha_2}^2L_{\beta_1}+\frac{16}{3}L_{\beta_1}L_{\alpha_1}L_{\alpha_2}-\frac{4}{3}L_{\alpha_1}^3\nn\\
&&\qquad \qquad 
-\frac{4}{3}L_{\alpha_1}^2L_{\beta_1}-\frac{2}{9}L_{\beta_1}^3+\frac{22}{9}L_{s_{ij}}L_{\beta_1}+\frac{44}{9}L_{\beta_1}L_{\alpha_2}+\frac{11}{9}L_{\beta_1}^2-\frac{67}{27}L_{s_{ij}}\nn\\
&&\qquad \qquad 
-\frac{134}{27}L_{\alpha_2}-\frac{134}{27}L_{\beta_1}-5\zeta_3+\frac{202}{27}
\Big]\nn\\
&&\qquad 
+\Big[
-\frac{8}{3}L_{\beta_1}L_{\alpha_1}^2L_{\alpha_2}+\frac{4}{3}L_{\alpha_2}^2L_{\alpha_1}^2+\frac{134}{27}L_{s_{ij}}L_{\beta_1}
+\frac{2}{9}L_{s_{ij}}L_{\beta_1}^3+\frac{1}{3}L_{s_{ij}}^2L_{\beta_1}^2\nn\\
&&\qquad \quad\;\; 
+\frac{2}{9}L_{s_{ij}}^3L_{\beta_1}+\frac{4}{3}L_{\alpha_1}^3L_{s_{ij}}+\frac{2}{3}L_{s_{ij}}^2L_{\alpha_1}^2+\frac{2}{3}L_{s_{ij}}^2L_{\alpha_2}^2-\frac{44}{9}L_{\beta_1}L_{\alpha_2}L_{s_{ij}}\nn\\
&&\qquad \quad\;\; 
-\frac{4}{3}L_{s_{ij}}L_{\beta_1}L_{\alpha_2}^2+\frac{8}{9}L_{\alpha_1}^3L_{\alpha_2}+\frac{20}{9}L_{\alpha_1}L_{\alpha_2}^3+\frac{268}{27}L_{\beta_1}L_{\alpha_2}+\frac{2}{3}L_{\beta_1}^2L_{\alpha_1}^2\nn\\
&&\qquad \quad\;\; 
+\frac{4}{3}L_{\beta_1}L_{\alpha_1}^3-\frac{8}{3}L_{\beta_1}^2L_{\alpha_1}L_{\alpha_2}+\Big(8L_{\beta_1}+\frac{32}{3}L_{\alpha_1}-\frac{4}{3}L_{\alpha_2}+2L_{s_{ij}}+\frac{22}{9}\Big)\zeta_3\nn\\
&&\qquad \quad\;\; 
+\Big(-\frac{16}{3}L_{\beta_1}L_{\alpha_2}-8L_{\alpha_1}L_{\alpha_2}-\frac{16}{3}L_{\alpha_1}L_{s_{ij}}-4L_{s_{ij}}L_{\beta_1}+\frac{44}{9}L_{\beta_1}-\frac{44}{9}L_{\alpha_2}\nn\\
&&\qquad \quad\;\;  
-\frac{22}{9}L_{s_{ij}}-4L_{\alpha_1}^2+2L_{\alpha_2}^2-2L_{\beta_1}^2-\frac{67}{27}\Big)\zeta_2+\frac{44}{9}L_{s_{ij}}L_{\alpha_2}^2+\frac{22}{9}L_{s_{ij}}^2L_{\alpha_2}\nn\\
&&\qquad \quad\;\; 
+\frac{1214}{81}-\frac{404}{81}L_{s_{ij}}-\frac{9}{2}\zeta_4-\frac{1}{6}L_{s_{ij}}^4-\frac{808}{81}L_{\alpha_2}-\frac{13}{18}L_{\alpha_2}^4-\frac{808}{81}L_{\beta_1}-\frac{11}{27}L_{\beta_1}^3\nn\\
&&\qquad \quad\;\; 
+\frac{4}{3}L_{\alpha_1}^2L_{\beta_1}L_{s_{ij}}+\frac{8}{3}L_{\alpha_1}^2L_{\alpha_2}L_{s_{ij}}+\frac{8}{3}L_{s_{ij}}^2L_{\alpha_1}L_{\alpha_2}+\frac{16}{3}L_{\alpha_1}L_{\alpha_2}^2L_{s_{ij}}\nn\\
&&\qquad \quad\;\; 
-\frac{16}{3}L_{\alpha_1}L_{\beta_1}L_{\alpha_2}L_{s_{ij}}
-8L_{\beta_1}L_{\alpha_1}L_{\alpha_2}^2-\frac{11}{9}L_{s_{ij}}L_{\beta_1}^2-\frac{11}{9}L_{s_{ij}}^2L_{\beta_1}+\frac{1}{18}L_{\beta_1}^4\nn\\
&&\qquad \quad\;\;  
+\frac{11}{27}L_{s_{ij}}^3+\frac{88}{27}L_{\alpha_2}^3+\frac{67}{27}L_{\beta_1}^2+\frac{7}{9}L_{\alpha_1}^4-\frac{44}{9}L_{\alpha_2}^2L_{\beta_1}-\frac{22}{9}L_{\beta_1}^2L_{\alpha_2}\nn\\
&&\qquad \quad\;\;  
-\frac{4}{3}L_{\beta_1}L_{\alpha_2}^3-\frac{2}{3}L_{\beta_1}^2L_{\alpha_2}^2
\Big]+\order(\eps)\Bigg\}\,,
\eea
and
\bea
&& \I_{g_ig_jg_k}^{\bar{C}}(\beta_2,\beta_1,\beta_1,\beta_1,\alpha_2,\alpha_2,\alpha_2,\alpha_1,\alpha_1,\alpha_1)=  -24(c_\Gamma)^2C_A^2\nn\\
&&\quad\cdot\;\; \Bigg\{
\frac{1}{\eps^2}\Big[ 
\frac{1}{2}L_{\alpha_2}^2+2L_{\alpha_1}L_{\alpha_2}+2L_{\alpha_1}^2+\zeta_2+\frac{11}{6}L_{\alpha_2}+\frac{11}{2}L_{\alpha_1}+\frac{3}{2}\Big]\nn\\
&&\qquad
+\frac{1}{\eps}\Big[
\Big(-2L_{\alpha_2}-6L_{\alpha_1}-2L_{\beta_1}-\frac{11}{2}\Big)\zeta_2-\frac{1}{2}L_{\alpha_2}^3-L_{\alpha_1}L_{\alpha_2}^2\nn\\
&&\qquad\qquad
-L_{\alpha_2}^2L_{\beta_1}-4L_{\beta_1}L_{\alpha_1}L_{\alpha_2}-2L_{\alpha_1}^3-4L_{\alpha_1}^2L_{\beta_1}-\frac{11}{3}L_{\beta_1}L_{\alpha_2}\nn\\
&&\qquad\qquad
-\frac{11}{6}L_{\alpha_1}^2-11L_{\alpha_1}L_{\beta_1}+\frac{67}{18}L_{\alpha_2}+\frac{67}{6}L_{\alpha_1}-3L_{\beta_1}+\frac{5}{2}\zeta_3+\frac{14}{3}
\Big]\nn\\
&&\qquad
+\Big[-\frac{919}{108}L_{\beta_1}-\frac{89}{108}L_{\beta_2}-\frac{11}{18}L_{\alpha_2}^3-\frac{67}{18}L_{\alpha_1}^2-\frac{3}{2}L_{\beta_2}^2+4L_{\beta_1}L_{\alpha_1}L_{\beta_2}L_{\alpha_2}\nn\\
&&\qquad\quad\;\;
-\frac{2}{3}L_{\alpha_1}^3L_{\alpha_2}+\frac{202}{9}L_{\alpha_1}+\frac{202}{27}L_{\alpha_2}+\frac{77}{8}\zeta_4+\frac{11}{3}L_{\beta_1}L_{\beta_2}L_{\alpha_2}
+\frac{193}{24}\nn\\
&&\qquad\quad\;\;
-\frac{67}{9}L_{\beta_1}L_{\alpha_2}-\frac{11}{2}L_{\beta_2}^2L_{\alpha_1}+\frac{7}{6}L_{\alpha_1}^4-\frac{11}{6}L_{\beta_2}L_{\alpha_2}^2-\frac{1}{2}L_{\beta_2}^2L_{\alpha_2}^2\nn\\
&&\qquad\quad\;\;
+\frac{7}{24}L_{\alpha_2}^4-\frac{11}{6}L_{\beta_2}^2L_{\alpha_2}-\frac{11}{6}L_{\beta_2}L_{\alpha_1}^2+\frac{3}{2}L_{\beta_1}^2-\frac{469}{18}L_{\alpha_1}L_{\beta_1}\nn\\
&&\qquad\quad\;\;
+3L_{\beta_2}L_{\beta_1}-2L_{\alpha_1}^2L_{\beta_2}^2
+L_{\beta_1}L_{\alpha_2}^3+2L_{\beta_1}^2L_{\alpha_1}^2+4L_{\beta_1}L_{\alpha_1}^3\nn\\
&&\qquad\quad\;\;
+\Big(-\frac{55}{6}+5L_{\beta_2}-10L_{\beta_1}-12L_{\alpha_1}
-L_{\alpha_2}\Big)\zeta_3+\Big(-\frac{67}{9}+11L_{\beta_1}\nn\\
&&\qquad\quad\;\;
-L_{\beta_2}^2+L_{\beta_1}^2+4L_{\alpha_1}^2+\frac{1}{2}L_{\alpha_2}^2+2L_{\alpha_1}L_{\alpha_2}
+4L_{\beta_1}L_{\alpha_2}+14L_{\alpha_1}L_{\beta_1}\nn\\
&&\qquad\quad\;\;
-2L_{\beta_2}L_{\alpha_1}+2L_{\beta_2}L_{\beta_1}\Big)\zeta_2+\frac{67}{18}L_{\beta_2}L_{\alpha_1}
+\frac{11}{2}L_{\alpha_1}^2L_{\beta_1}+\frac{11}{6}L_{\alpha_2}^2L_{\beta_1}\nn\\
&&\qquad\quad\;\;
+\frac{1}{2}L_{\beta_1}^2L_{\alpha_2}^2+\frac{11}{2}L_{\beta_1}^2L_{\alpha_1}+\frac{11}{6}L_{\beta_1}^2L_{\alpha_2}+\frac{1}{3}L_{\alpha_1}L_{\alpha_2}^3+2L_{\beta_1}^2L_{\alpha_1}L_{\alpha_2}\nn\\
&&\qquad\quad\;\;
-2L_{\alpha_1}^2L_{\alpha_2}L_{\beta_2}-2L_{\alpha_1}L_{\beta_2}^2L_{\alpha_2}-L_{\alpha_1}L_{\alpha_2}^2L_{\beta_2}+4L_{\beta_1}L_{\alpha_1}^2L_{\beta_2}\nn\\
&&\qquad\quad\;\;
+2L_{\beta_1}L_{\alpha_1}^2L_{\alpha_2}+3L_{\beta_1}L_{\alpha_1}L_{\alpha_2}^2
+L_{\alpha_2}^2L_{\beta_1}L_{\beta_2}+11L_{\beta_1}L_{\alpha_1}L_{\beta_2}\Big]\nn\\
&&\qquad
+\order(\eps)\Bigg\}\,.
\eea

\subsection{The poles for the $H\to gggg$ phase space integral}
A simple example which allows us to test the validity of eq. (\ref{eq:ngluonpoles}) is given by the quantity
\beq
\O_{H\to g_1g_2g_3g_4}=\int\d\Phi_{1234}\, |\M_{H\to g_1g_2g_3g_4}|^2\,.
\eeq
We consider the corresponding amplitude in the heavy quark effective theory where the Higgs boson couples to gluons directly via the effective Lagrangian:
\beq
\mathcal{L}=-\frac{C_{eff}}{4}HG^{\mu\nu}_aG_{\mu\nu}^a\,,
\eeq
with $C_{eff}$ a Wilson coefficient, $H$ the Higgs boson field and $G^{\mu\nu}_a$ the gluon field strength tensor. We can evaluate the inclusive quantity $\O_{H\to g_1g_2g_3g_4}$ using IBP reduction and the master integrals presented in \cite{Gehrmann-DeRidder:2003pne} to get (setting $Q^2=1$):
\bea
\O_{H\to g_1g_2g_3g_4}&=&120(c_\Gamma)^2g_s^4(C_A)^2 \O_{H\to g_1g_2}\nn\\
&&
\cdot\Bigg\{
-\frac{1}{\epsilon^4}
-\frac{1}{\epsilon^3}\frac{121}{30}
+\frac{1}{\epsilon^2}\Big[\,\frac{39}{5}\zeta_2-\frac{872}{45}\,\Big]
+\frac{1}{\epsilon}
\Big[\,\frac{123}{5}\zeta_3+\frac{473}{15}\zeta_2-\frac{4691}{54}\,\Big]\nn\\
&&\quad
+\Big[\,-\frac{37}{10}\zeta_4-\frac{304951}{810}+99\zeta_3+\frac{2303}{15}\zeta_2\,\Big]
+\order(\eps)\Bigg\}
\eea
Using eq. (\ref{eq:ngluonpoles}) we can confirm the pole parts of this expression 
independently to obtain (using again the parameter choice of eq. (\ref{eq:paramchoice})):
\bea
\O_{H\to g_1g_2g_3g_4}^{\mathrm{Singular}}
&=&120(c_\Gamma)^2g_s^4(C_A)^2 \O_{H\to g_1g_2}\nn\\
&&
\cdot\Bigg\{
-\frac{1}{\epsilon^4}
-\frac{1}{\epsilon^3}\frac{121}{30}
+\frac{1}{\epsilon^2}\Big[\,\frac{39}{5}\zeta_2-\frac{872}{45}\,\Big]
+\frac{1}{\epsilon}
\Big[\,\frac{123}{5}\zeta_3+\frac{473}{15}\zeta_2-\frac{4691}{54}\,\Big]\nn\\
&&\quad
+\Big[
-\frac{586351}{1620}
+\frac{6788}{45}\zeta_2
+\frac{1496}{15}\zeta_3
-\frac{8}{5}\zeta_4
-\frac{1}{5}L_{\alpha_2}^4
-\frac{17}{3}L_{\alpha_1}^2
-\frac{89}{135}L_{\beta_2}\nn\\
&&\quad\quad\;
-\frac{6}{5}L_{\beta_2}^2
-\frac{22}{15}L_{\beta_2}L_{\alpha_2}^2
-\frac{22}{15}L_{\beta_2}^2L_{\alpha_2}
-\frac{2}{5}L_{\beta_2}^2L_{\alpha_2}^2
-\frac{8}{5}L_{\alpha_1}^2L_{\beta_2}^2
+\frac{4}{5}L_{\alpha_1}^4\nn\\
&&\quad\quad\;
-\frac{44}{15}L_{\alpha_1}^2L_{\beta_1}
-\frac{22}{15}L_{\alpha_2}^2L_{\beta_1}
-\frac{16}{5}L_{\beta_1}L_{\alpha_1}^3
-\frac{22}{15}L_{\beta_2}L_{\alpha_1}^2
-\frac{22}{5}L_{\beta_2}^2L_{\alpha_1}\nn\\
&&\quad\quad\;
-\frac{4}{5}L_{\beta_2}^2\zeta_2
-\frac{16}{5}L_{\alpha_1}\zeta_3
-\frac{8}{5}L_{\alpha_2}\zeta_3
-\frac{44}{15}L_{\alpha_2}\zeta_2
+\frac{22}{15}L_{\alpha_2}^3
+\frac{503}{27}L_{\alpha_1}\nn\\
&&\quad\quad\;
+\frac{187}{18}L_{\beta_1}
+\frac{121}{90}L_{\beta_1}^2
-\frac{44}{15}L_{\alpha_1}\zeta_2
+4\zeta_3L_{\beta_2}
+\frac{8}{5}L_{\beta_2}L_{\beta_1}\zeta_2\nn\\
&&\quad\quad\;
+\frac{16}{5}L_{\beta_1}L_{\alpha_1}^2L_{\beta_2}
+\frac{44}{15}L_{\beta_1}L_{\beta_2}L_{\alpha_2}
+\frac{4}{5}L_{\alpha_2}^2L_{\beta_1}L_{\beta_2}
+\frac{44}{5}L_{\beta_1}L_{\alpha_1}L_{\beta_2}\nn\\
&&\quad\quad\;
-\frac{8}{5}L_{\alpha_1}L_{\beta_1}\zeta_2
+\frac{8}{5}L_{\alpha_1}^2\zeta_2
-\frac{16}{5}L_{\alpha_2}L_{\alpha_1}\zeta_2
-\frac{8}{5}L_{\beta_2}L_{\alpha_1}\zeta_2
+\frac{8}{5}L_{\alpha_2}^2\zeta_2\nn\\
&&\quad\quad\;
+\frac{4}{5}L_{\alpha_2}^2L_{\alpha_1}^2
+\frac{134}{45}L_{\beta_2}L_{\alpha_1}
+\frac{12}{5}L_{\beta_2}L_{\beta_1}
+\frac{8}{5}L_{\alpha_1}L_{\alpha_2}^3
+\frac{644}{45}L_{\alpha_1}L_{\beta_1}\nn\\
&&\quad\quad\;
+\frac{44}{15}L_{\beta_1}^2L_{\alpha_1}
+\frac{8}{5}L_{\beta_1}^2L_{\alpha_1}^2
-\frac{12}{5}L_{\beta_1}L_{\alpha_1}L_{\alpha_2}^2
-\frac{8}{5}L_{\alpha_1}^2L_{\alpha_2}L_{\beta_2}\\
&&\quad\quad\;
-\frac{8}{5}L_{\alpha_1}L_{\beta_2}^2L_{\alpha_2}
-\frac{4}{5}L_{\alpha_1}L_{\alpha_2}^2L_{\beta_2}
+\frac{16}{5}L_{\beta_1}L_{\alpha_1}L_{\beta_2}L_{\alpha_2}
\,\Big]
+\order(\eps)\Bigg\}\nn
\eea
The $\alpha_{i..}$ and $\beta_{ij..}$ parameters thus cancel in the poles and, more importantly, reproduce the correct result. A proper integrand-level implementation of these counter-terms must therefore numerically cancel the finite $\log$-dependent parts which make up the finite part of the integrated counter-term $\O_{H\to g_1g_2g_3g_4}^{\mathrm{Singular}}$. To accomplish this task will be left for future work.

\section{Conclusions}
\label{sec:Conclusion}
In this work we introduced a new scheme for the subtraction of IR divergences in real radiation phase space integrals, which is based on a particular Feynman Diagram dependent slicing observable. We proposed that this slicing scheme can be promoted to a fully local integrand subtraction scheme and illustrated how this can be achieved for a simple example. Based on the geometric properties of the observable we established a subtraction formula which summarises the combinatorics of the various counter-terms for single and double real emissions and conjecture its general form for an arbitrary number of unresolved emissions.

We applied the formalism to final state real radiation at NLO and NNLO in Yang Mills theory and integrated all the required counter-terms. We employed reverse unitarity and IBP reduction to simplify the calculation of the most complicated counter-terms. We showed in particular how the Master integrals required for these counter-terms can be extracted from existing calculations of unrelated quantities. We were thereby able to compute or extract all required counter-terms in terms of $\Gamma$ and ${}_pF_q$ hypergeometric functions to all orders in the dimensional regulator.

We tested the integrated counter-terms by reproducing the poles of the purely double real emission contribution to the gluonic Higgs decay in the heavy quark effective theory.

There exist many possible directions to extend this work in the future. The most  important step will be to show that the scheme can indeed be employed to build local counter-terms at the level of the integrand. The next logical step would be to extend the scheme to include also initial states; this would open up a new path for computing LHC observables. Another step is to extend the scheme to real-virtual corrections; one can foresee that this should be a straight forward application of the techniques presented here for the case of real radiation at NLO. Beyond one can also imagine to use the scheme for N${}^3$LO calculations and/or to include massive quarks into the formalism. 

Given the simplicity of the integrated counter-terms, the potential locality of the counter-terms and their well defined combinatorial properties, we believe that the proposed scheme may well become an important method for performing higher order calculations in perturbative QCD in the future.

\section*{Acknowledgements}
This work has been supported by the European Research Council (ERC) advanced grant 320389 and the NWO Vidi grant 680-47-551  ``Decoding Singularities of Feynman graphs''. I am indebted to S. Badger and T. Melia with whom I collaborated on the project during its early stages and who were influential in establishing some of the basic ideas. I would also like to thank them for their continued enthusiasm and the countless conversations we had on the topic since then. I would further like to thank M. Borrinsky, H. Kissler, D. Kreimer, E.Laenen, and W. Waalewijn for useful discussions and E. Laenen, T. Melia and V. Del Duca for comments on the manuscript.

\newpage
\appendix
\section{Region cancellations}
\label{appendix:Regioncancellation}
In the following we will derive several identities among different sets of overlapping regions.
\begin{itemize}

\item Let us start with the identity
\beq
\lim_{A||B,A||D} \Theta(C_{A B}\cap C_{A D})=\lim_{A||B,A||D} \Theta(C_{A B}\cap C_{A D}\cap C_{A B D})\,,
\eeq
where $A$, $B$ and $D$ are non-intersecting sets of momenta. We consider here the special case where the region $C_{A B}\cap C_{A D}$ corresponds to a collinear momentum configuration only. It is sufficient to show that $s_{ABD}\le b_{ABD}$ given that $s_{AB}\le b_{AB}$ and $s_{AD}\le b_{AD}$. Now since $A||B$ and $A||D$ it follows that $A||B||D$, which in turn implies that 
\beq
s_{ABD}\lesssim b_{AB}\sim b_{AD} \ll b_{ABD}\,,
\eeq
which is in accord with the ordering given in eq. (\ref{eq:aborderl}) and guarantees eq. (\ref{eq:geometriccancellationII}).

\item
We proceed with the identity
\beq
\lim_{A,B\to 0}\Theta(S_{A}\cap S_B)=\lim_{A,B\to 0}\Theta(S_{A}\cap S_{B}\cap S_{A B})\,,
\eeq
where again $A$ and $B$ are two non-intersecting sets of momenta. 
In order to prove this identity we have to specify the hard momenta which enter the constraint of the soft slicing parameter. Two different choices will be relevant for us. The derivation is easiest in the case when the hard momenta of the different slicing parameters are chosen identically as say $k$ and $l$, such that
\bea
&& S_A:\qquad s_{A(kl)}<a_A s_{kl}\,,\\
&& S_B:\qquad s_{B(kl)}<a_B s_{kl}\,,\\
&& S_{A B}:\qquad s_{(AB)(kl)}<a_{AB} s_{kl}\,.
\eea
Applying the limits $\lim_{a_A\to 0}$ and $\lim_{a_B\to 0}$ to the region boundary $S_{A B}$ results depends on the order in which the limits are taken. For instance order 
\beq
\lim_{a_B\to 0} S_{A B}:\qquad s_{A(kl)}<a_{AB} s_{kl}\,,\\
\eeq
which is automatically satisfied given that $a_A\ll a_{AB}$. Taking the limits in the other order leads to a similar 
conclusion and we conclude that $(S_{A}\cap S_{B})\subset S_{AB}$. However there exists another case of interest where the hard momenta appearing in the different soft order parameters are not identical but are nested in the following sense:
\bea
&& S_A:\qquad s_{A(kl)}<a_A s_{kl}\,,\\
&& S_B:\qquad s_{B(il)}<a_B s_{il},\qquad i\in A\,,\\
&& S_{A B}:\qquad s_{(AB)(kl)}<a_{AB} s_{kl}\,.
\eea
Again we obtain different results for $S_{A B}$ depending on whether we take first the limit in $a_A$ or $a_B$. If for instance we take first $B\to0$ the conclusions are as in the case before. If however we first take the limit $A\to 0$ we have to ask whether $s_{B(kl)}<a_{AB} s_{kl}$ is guaranteed by $s_{B(il)}<a_B s_{il}$. For this purpose it is useful to write 
\beq
s_{B(kl)}=s_{B(il)}\frac{s_{B(kl)}}{s_{B(il)}}=
\frac{s_{B(il)}\,s_{ikl}\,y_{B;kl}}{s_{il}}\,,
\eeq
where
\beq
y_{B;kl}=\frac{1-\vec{v}_{B}^{(i,l)}.\vec{v}_{kl}^{(i,l)}}{2}\,,
\eeq
and where we have written out the ratio $\frac{s_{B(kl)}}{s_{B(il)}}$ using energies and angles in the rest-frame of $p_{il}$. 
Here $\vec{v}_{X}^{(i,l)}$ denotes the $D-1$ dimensional space-like velocity vector of the momentum $X$ in the rest-frame of $p_{il}$. We thus obtain the following bound:
\beq
s_{B(kl)}\le a_B\, (s_{i(kl)}+s_{kl})\,y_{B;kl}\lesssim a_B\, s_{kl}\,,
\eeq
where we have used that $s_{i(kl)}$ is much smaller than $s_{kl}$ and $y_{B;kl}\in (0,1]$.
It thus follows that $s_{B(kl)}\ll a_{AB}s_{kl}$, given $a_{B}\ll a_{AB}$. Thus for the cases of interest eq. (\ref{eq:geometriccancellationI}) is fulfilled, given the ordering of eq. (\ref{eq:aborderl}).

Using similar arguments one can show that a more general identity
\beq
\lim_{A,B,C\to 0}\Theta(S_{A C}\cap S_{B C})=\lim_{A,B,C\to 0}\Theta(S_{A C}\cap S_{B C}\cap S_{A B C})\,,
\eeq
is also true.
   
\item   
Let us now consider the following 4-term cancellation identity:
\bea
\label{eq:appgeometriccancellationIII}
0&=&\Theta(C_{A i}\cap C_{A j})\nn\\
&-&\Theta(S_A \cap C_{A i}\cap C_{A j})   \\
&-&\Theta(C_{A i j} \cap C_{A i}\cap C_{A j})  \nn\\ 
&+&\Theta(S_A\cap C_{A i j} \cap C_{A i}\cap C_{A j})\,,   \nn
\eea
where $A$ is a set of momenta not containing the single momenta $i$ and $j$. The identity derives from the fact that the overlap region $C_{A i}\cap C_{B i}$ contains singular momentum configurations of two types: 
\begin{itemize}
   \item [(i)] $A\to0$,
   \item [(ii)] $A||i||j$,
\end{itemize}
or their overlap. While for the second case $A||i||j$ we can relie on the identity given in eq. (\ref{eq:geometriccancellationII}), we must show that it holds also for the case $A\to 0$. To accomplish this it is sufficient to show that
\beq
\lim_{A\to 0} \Theta(C_{A i}\cap C_{B i})-\lim_{A\to 0} \Theta(S_A \cap C_{A i}\cap C_{B i})=0\,,
\eeq
since the other two terms in eq. (\ref{eq:appgeometriccancellationIII}) are contained in a sub-region of these regions and must thus cancel by the same mechanism. The soft region is given by the bound
\beq
S_A: \qquad s_{A(ij)}\le a_As_{ij}
\eeq
Using the constraints $s_{Ai}<b_{Ai}$ and $s_{Aj}<b_{Aj}$ we find 
\beq
S_A\cap C_{A i}\cap C_{Aj}: \qquad s_{A(ij)}\le b_{Ai}+b_{Aj}\le a_As_{ij}
\eeq
Thus, assuming $b_{Ai}\sim b_{Aj}$ we must fulfil the bound:
\beq
\frac{2b_{Ai}}{a_A}\le s_{ij}\,.
\eeq
Now since the momenta $i$ and $j$ are not allowed to be collinear to the momenta in $A$, we can write this as
\beq
b_{Ai}\le \frac{a_Ab_{Aij}}{2}\,.
\eeq
This bound corresponds to the worst scenario, since it may be that $A||i||j$ may not be allowed by the measurement function; nevertheless this inequality is consistent with the ordering suggested in eq. (\ref{eq:aborderl}).

\end{itemize}

\section{Integrated counterterms}
\label{sec:Integrated counterterms}
In the following we provide expressions for the counterterms associated to all the different regions. 
\begin{itemize}
\item $\{C_{ij}\}$:
\bea
&&\lim_{b_{ij}\to0}  \int \mT(C_{ij}) \ast\d\O_{2;1..i..j..n+2} = \I^C_{gg}(Q^2b_{ij})\, \int\d\O_{1;1..\t{ij}..n+2}\nn
\eea
\item $\{C_{ijk}\}$:
\bea
&&\lim_{b_{ijk}\to0}  \int \mT(C_{ijk}) \ast\d\O_{2;1..i..j..k..n+2} =\I^C_{ggg}(Q^2b_{ijk})\, \int\d\O_{0;1..\t{ijk}..n+2} \nn
\eea
\item $\{S_{k}\}$:
\bea
&&\lim_{a_{k}\to0}  \int \mT(S_{k}) \ast\d\O _{2;1..n+2}=-\sum_{i,j=1\neq k}^{n+2}\,\int \d\O^{(i,j)}_{1;1..\not{k}..n+2}\,\I^S_{gg}(s_{ij},a_{k})\nn
\eea
\item $\{S_{kl}\}$: 
\bea
&&\lim_{a_{kl}\to0}  \int \mT(S_{kl}) \ast\d\O_{2;1..n+2}=-\frac{1}{2}C_A\sum_{i,j=1\neq k,l}^{n+2}\,\int\d\O^{(i,j)}_{0;1..\not{k}..\not{l}..n+2}\, 
\I^S_{gg}(s_{ij},a_{kl})\nn
\eea
\item $\{C_{ijk}, C_{ij}\}$:
\bea
&&\lim_{b_{ijk}\to0} \lim_{b_{ij}\to0}  \int \mT(C_{ijk})\mT(C_{ij}) \ast\d\O_{2;1..i..j..k..n+2} =\nn\\&&
\qquad\qquad\qquad\qquad\qquad\qquad  \I^C_{gg}(Q^2,b_{ijk}) \,\I^C_{gg}(Q^2,b_{ij})\,\int\d\O_{0;1..\t{ijk}..n+2} \, \nn
\eea
\item $\{C_{ijk}, S_{ij}\}$ :
\bea
&&\lim_{a_{ij}\to0}  \lim_{b_{ijk}\to0}  \int \mT(C_{ijk}) \mT(S_{ij})  \ast\d\O_{2;1..i..j..k..n+2}=\nn\\&&
\qquad\qquad\qquad\qquad \I^{SC}_{ggg}(Q^2,a_{ij},b_{ijk})\,\int\d\O_{0;1..\t{ijk}..n+2}  \nn
\eea
\item $\{C_{ijk}, S_{k}\}$ :
\bea
&&\lim_{a_{k}\to0}\lim_{b_{ijk}\to0}  \int \mT(C_{ijk})\mT(S_{k}) \ast\d\O_{2;1..i..j..k..n+2}=\nn\\&&
\qquad\qquad\qquad\int\d\O_{0;1..\t{ijk}..n+2}\int \d\I_{g_ig_j}^{C}(Q^2,b_{ijk})\nn\\&&
\qquad\qquad\qquad\qquad  
\cdot\;\frac{1}{2}\Big[
\I_{g}^S(s_{ij},a_{k})+\I_{gg}^{SC}(Q^2b_{ijk}-s_{ij},z_i a_{k})+\I_{gg}^{SC}(Q^2b_{ijk}-s_{ij},z_j a_{k})
\Big]
\nn
\eea
   \item 
$\{C_{ij}, C_{kl}\}$
\bea
&&\lim_{b_{ij}\to0} \lim_{b_{kl}\to0}  \int \mT(C_{ij})\mT(C_{kl}) \ast\d\O_{2;1..i..j..k..l..n+2} =\nn\\&&
\qquad\qquad\qquad\qquad \, \I^C_{gg}(Q^2,b_{ij}) \,\I^C_{gg}(Q^2,b_{kl})\,\int\d\O_{0;1..\t{ijk}..n+2} \nn
\eea
   \item 
$\{C_{kl}, S_{kl}\}$
\bea
&&\lim_{a_{kl}\to0} \lim_{a_{kl}\to0}  \int \mT(S_{kl}) \mT(C_{kl})\ast\d\O_{2;1..k..l..n+2}=\nn\\&&\qquad\qquad
-\,\I^C_{gg}(Q^2b_{kl})\sum_{i,j=1\neq k,l}^{n+2}\,\int\d\O^{(i,j)}_{0;1..\not{k}..\not{l}..n+2}\, \I^S_{g}(s_{ij},a_{kl}) \nn
\eea
\item $\{C_{ij}, S_{i}\}$:
\bea
&&\lim_{a_{i}\to0}\lim_{b_{ij}\to0}  \int \mT(S_{i}) \mT(C_{ij}) \ast\d\O_{2;1..i..j..n+2} =\nn\\&&
\qquad\qquad\qquad\qquad \int\d\O_{1;1..\t{ij}..n+2} \, \I^{SC}_{gg}(Q^2b_{ij},a_i)\nn
\eea
\item $\{C_{ij}, S_{k}\}$ :
\bea
&&\lim_{a_{k}\to0} \lim_{b_{ij}\to0} \int \mT(S_{k}) \mT(C_{ij}) \ast\d\O_{2;1..i..j..k..n+2}=\nn\\&&\qquad\qquad
-\sum_{l,m \in \{1,..,\t{ij},..,\not{k},..n+2\}}\,\int\d\O^{(l,m)}_{0;1..\t{ij}..\not{k}..n+2}\,\I^S_{g}(s_{lm},a_{k})\,
\I^C_{gg}(Q^2b_{ij})\nn
\eea
\item $\{S_{ij}, S_{i}\}$ 
\bea
&& \lim_{a_{kl}\to0}\lim_{a_{k}\to0} \int\mT(S_{k})\mT(S_{kl}) \ast\d\O_{2;1..n+2} =
\nn\\&&\qquad\qquad
-C_A\sum_{i,j\neq k,l}\,\d\O^{(i,j)}_{0;1..\not{k}..\not{l}..n+2}
\nn\\&&\qquad\qquad\qquad\cdot
\Big[
\int \d\I_{g_l}^S(s_{ij},a_{kl})
\big(
\I_{g}^S(s_{il},a_{k})+\I_{g}^S(s_{jl},a_{k})\big)
-\I_{g}^S(s_{ij},a_{kl}) \I_{g}^S(s_{ij},a_{k})
\Big]\nn
\eea

   \item 
$\{\{S_{k}, S_{l}\},\{S_{kl},S_{k}, S_{l}\}\}$
\bea
&&\lim_{a_{kl}\to0} \lim_{(a_k,a_l)\to0}  \int (1-\mT(S_{kl}))\mT(S_{k})\mT(S_{l})\big)\ast\d\O_{2;1..n+2}=\\
&&\qquad\qquad +\frac{1}{2}\sum_{i,j,r,t\neq k,l}\,\int \d\O^{(i,j)(r,t)}_{0;1..\not{k}..\not{l}..n+2}\, 
\I^S_g(s_{ij},a_k)\I^S_g(s_{rt},a_l)\nn
\eea
\item $\{C_{ijk}, C_{ij}, S_{ij}\}$:
\bea
&&\lim_{a_{ij}\to0} \lim_{b_{ijk}\to0} \lim_{b_{ij}\to0}  \int \mT(S_{ij}) \mT(C_{ijk})\mT(C_{ij}) \ast\d\O_{2;1..i..j..k..n+2}=\nn\\&&
\qquad\qquad\qquad\qquad\qquad\qquad \int\d\O_{0;1..\t{ijk}..n+2} \, 
\I^{SC}_{gg}(Q^2b_{ijk},a_{ij}) \,\I^C_{gg}(Q^2b_{ij})\nn
\eea
   \item $\{C_{ijk}, C_{ij}, S_{i}\}$:
\bea
&&\lim_{a_{i}\to0} \lim_{b_{ijk}\to0} \lim_{b_{ij}\to0}  \int \mT(S_{i}) \mT(C_{ijk})\mT(C_{ij}) \ast \d\O_{2;1..i..j..k..n+2}=\nn\\&&
\qquad\qquad\qquad\qquad\qquad\qquad \int\d\O_{0;1..\t{ijk}..n+2} \, 
\I^{C}_{gg}(Q^2b_{ijk}) \,\I^{SC}_{gg}(Q^2b_{ij},a_i)\nn
\eea   
\item $\{C_{ijk}, C_{ij}, S_{k}\}$:
\bea
&&\lim_{a_{k}\to0}\lim_{b_{ijk}\to0} \lim_{b_{ij}\to0} \int \mT(C_{ijk})\mT(C_{ij})\mT(S_{k}) \ast\d\O_{2;1..i..j..k..n+2} =\nn\\&&
\qquad\qquad\qquad\int\d\O_{0;1..\t{ijk}..n+2}
\int \d\I_{g_ig_j}^{C}(Q^2b_{ijk})\nn\\&&
\qquad\qquad\qquad\qquad  
\cdot\;\frac{1}{2}\Big[
\I_{gg}^{SC}(Q^2b_{ijk},z_i a_{k})+\I_{gg}^{SC}(Q^2b_{ijk},z_j a_{k})
\Big]
\nn
\eea
\item $\{C_{ijk}, S_{ik}, S_{k}\}$:
\bea
&&\lim_{a_{ik}\to0}\lim_{a_{k}\to0}\lim_{b_{ijk}\to0}  \int \mT(C_{ijk})\mT(S_{k})\mT(S_{ik}) \ast\d\O_{2;1..i..j..k..n+2} =\nn\\&&
\qquad\qquad\qquad\int \d\O_{0;1..\t{ijk}..n+2}
\int \d\I_{g_ig_j}^{SC}(Q^2b_{ijk},a_{ik})\nn\\&&
\qquad\qquad\qquad\qquad  
\cdot\;\frac{1}{2}\Big[
\I_{g}^S(s_{ij},a_{k})+\I_{gg}^{SC}(Q^2b_{ijk}-s_{ij},z_i a_{k})-\I_{gg}^{SC}(Q^2b_{ijk}-s_{ij},a_{k})
\Big]
\nn
\eea
\item $\{\{C_{ijk}, S_{i}, S_{j}\},\{C_{ijk}, S_{ij}, S_{i}, S_{j}\}\}$:
\bea
&&\lim_{a_{i}\to0}\lim_{a_{j}\to0}\lim_{b_{ijk}\to0}  \int (1-\mT(S_{ij}))\mT(C_{ijk})\mT(S_{i})\mT(S_{j}) \ast\d\O_{2;1..i..j..k..n+2}=\nn\\&&
\qquad\qquad\qquad\int \d\I_{g_ig_k}^{SC}(Q^2b_{ijk},a_{i}) \,\I_{gg}^{SC}(Q^2b_{ijk}-s_{ik},a_{j})\int\d\O_{0;1..\t{ijk}..n+2}
 \nn
\eea
\item 
$\{C_{ij}, C_{kl}, S_{i}\}$:
\bea
&&\lim_{a_{i}\to0}\lim_{b_{ij}\to0} \lim_{b_{kl}\to0}  \int \mT(S_{i})\mT(C_{ij})\mT(C_{kl}) \ast\d\O_{2;1..i..j..k..l..n+2} =\nn\\&&
\qquad\qquad\qquad\qquad \I^{SC}_{gg}(Q^2b_{ij},a_i) \,\I^C_{gg}(Q^2,b_{kl})\,\int\d\O_{0;1..\t{ij}..\t{kl}..n+2}
\, \nn
\eea
\item $\{C_{kl}, S_{kl}, S_{k}\}$:
\bea
&&\lim_{a_{kl}\to0} \lim_{a_{k}\to0} \lim_{b_{kl}\to0}  \int \mT(S_{kl})\mT(C_{kl})\mT(S_{k}) 
\ast \d \O_{2;1..k..l..n+2}=\\&&\qquad\qquad
-\I^{SC}_{gg}(Q^2b_{kl},a_k)\,\sum_{i,j=1\neq k,l}^{n+2}\,\int\d \O^{(i,j)}_{0;1..\not{k}..\not{l}..n+2} \, \I^S_{g}(s_{ij},a_{kl}) \nn
\eea

   \item 
$\{\{C_{ij}, S_{i}, S_{k}\},\{C_{ij},S_{ik}, S_{i}, S_{k}\}\}$
\bea
&&\lim_{a_{k}\to0}\lim_{a_{i}\to0} \lim_{b_{ij}\to0} \int (1-\mT(S_{ik}))\mT(S_{k}) \mT(S_{i}) \mT(C_{ij}) \ast\d\O_{2;1..i..j..k..n+2}=\nn\\&&\qquad\qquad
-\,\I^{SC}_{gg}(Q^2b_{ij},a_i)\,\sum_{l,m \in \{1,..,\t{ij},..,\not{k},..n+2\}}\,\int\d\O^{(l,m)}_{0;1..\t{ij}..\not{k}..n+2}\,
\I^S_{gg}(s_{lm},a_{k})\,
\nn
\eea
   \item 
$\{C_{il}, S_{kl}, S_{k}\}$
\bea
&&\lim_{a_{kl}\to0}\lim_{a_{k}\to0} \lim_{b_{il}\to0}\int \mT(C_{il})\mT(S_{k})\mT(S_{kl}) \ast\d\O_{2;1..n+2}=
\nn\\&&\qquad\; 
 -C_A\sum_{j \in \{1,..,\t{il},..,\not{k},..n+2\}}\,\,\int \d\O^{(\t{il},j)}_{0;1..\not{k}..\t{il}..n+2}
\int \d\I^{SC}_{g_lg_i}(Q^2b_{il},a_{kl})\nn\\&&\qquad\qquad\qquad\qquad\; 
\cdot\;\Big(
\I^S_g(z_ls_{\t{il}j},a_k)-\I^S_g(s_{\t{il}j},a_k)
\Big)\nn
\eea
\item $\{C_{ijk}, C_{ij}, S_{ij}, S_{i}\}$:
\bea
&&\lim_{a_{ij}\to0}\lim_{a_{i}\to0} \lim_{b_{ijk}\to0} \lim_{b_{ij}\to0}  \int \mT(S_{ij})\mT(S_{i}) \mT(C_{ijk})\mT(C_{ij}) \ast\d\O_{2;1..i..j..k..n+2}=\nn\\&&
\qquad\qquad\qquad\qquad\qquad\qquad\, 
\I^{SC}_{gg}(Q^2b_{ijk},a_{ij}) \,\I^{SC}_{gg}(Q^2b_{ij},a_i)\, \int\d\O_{0;1..\t{ijk}..n+2} \nn
\eea
\item 
$\{C_{ijk}, C_{ij}, S_{ik}, S_{k}\}$
\bea
&&\lim_{a_{ik}\to0}\lim_{a_{i}\to0} \lim_{b_{ijk}\to0} \lim_{b_{ij}\to0}  \int \mT(S_{ik})\mT(S_{i}) \mT(C_{ijk})\mT(C_{ij}) \ast\d\O_{2;1..i..j..k..n+2}= \nn\\&&
\qquad\qquad\, 
\frac{1}{2}\int\d\I^{SC}_{g_ig_j}(Q^2b_{ij},a_{ik}) \,\big( \I^{SC}_{gg}(Q^2b_{ijk},z_ia_k)-\I^{SC}_{gg}(Q^2b_{ijk},a_k) \big)\int\d\O_{0;1..\t{ijk}..n+2}\, \nn
\eea
\item 
$\{\{C_{ijk}, C_{ij}, S_{i}, S_{k}\},\{C_{ijk}, C_{ij}, S_{i}, S_{k},S_{ik}\}\}$
\bea
&&\lim_{a_{k}\to0}\lim_{a_{i}\to0} \lim_{b_{ijk}\to0} \lim_{b_{ij}\to0}  \int (1-\mT(S_{ik}))\mT(S_{k})\mT(S_{i}) \mT(C_{ijk})\mT(C_{ij}) \ast\d\O_{2;1..i..j..k..n+2}=\nn\\&&
\qquad\qquad\qquad\qquad\qquad\qquad\, 
\I^{SC}_{gg}(Q^2b_{ijk},a_{k}) \,\I^{SC}_{gg}(Q^2b_{ij},a_i)\, \int\d\O_{0;1..\t{ijk}..n+2} \nn
\eea

\item $\{\{C_{ij}, C_{kl}, S_{i}, S_{k}\},\{C_{ij}, C_{kl}, S_{i}, S_{k},S_{ik}\}\}$:
\bea
&&\lim_{a_{k}\to0}\lim_{a_{i}\to0}\lim_{b_{ij}\to0} \lim_{b_{kl}\to0}  \int (1-\mT(S_{ik}))\mT(S_{i})\mT(S_{k})\mT(C_{ij})\mT(C_{kl}) \ast\d\O_{2;1..i..j..k..l..n+2} =\nn\\&&
\qquad\qquad\qquad\qquad \I^{SC}_{gg}(Q^2b_{ij},a_i) \,\I^{SC}_{gg}(Q^2,b_{kl},a_k)\,\int\d\O_{0;1..\t{ij}..\t{kl}..n+2}
\, \nn
\eea
\end{itemize}
Apart from the counterterms corresponding to the regions $\{\{S_{ij}\},\{C_{ijk}\},\{C_{ijk},S_{ij}\}\}$ all other counterterms are expressible in terms of either factorised or simple iterated NLO limits. This allows to evaluate them straight forwardly in terms of $\Gamma$-functions by employing the paramaterisations given in eqs. (\ref{eq:softparam}), 
(\ref{eq:collparam}) and (\ref{eq:softcollparam}). For the regions $\{\{S_{ij}\},\{C_{ijk}\},\{C_{ijk},S_{ij}\}\}$ the corresponding integrated counterterms can be expressed via IBP reduction in terms of the Master integrals defined in section \ref{sec:masters}:
\beq
\I^S_{gg}(s_{ij},a_{kl})=
-16\frac{(11\eps^2-19\eps+3)(-1+4\eps)(-3+4\eps)}{(-3+2\eps)\eps^3}\MI^{(2;1)}_{S}(s_{ij};a_{kl})
+8\MI^{(2;2)}_{S}(s_{ij};a_{kl})
\eeq
\beq
\I^{SC}_{ggg}(Q^2b_{ijk},a_{ij})=C_A^2\bigg[ -\frac{8(22\eps^3-49\eps^2+25\eps-3)}{\eps^2(-3+2\eps)}
\MI^{(2,2;1)}_{SC}(Q^2b_{ijk},a_{ij})
+8\MI^{(2,2;2)}_{SC}(Q^2b_{ijk},a_{ij}) \bigg]
\eeq
\bea
&&\I^C_{ggg}(Q^2b_{ijk})=C_A^2\bigg[
12\Big(\MI^{(2;3)}_{C}(Q^2b_{ijk})+\MI^{(2;4)}_{C}(Q^2b_{ijk})\Big)\nn\\&&       
\qquad\qquad\qquad -\frac{24(4\eps^3-6\eps^2+3)}{(2\eps-1)(-1+\eps)(-3+2\eps)}
\MI^{(2;2)}_{C}(Q^2b_{ijk})\\&&        
-\frac{12(12\eps^8+232\eps^7-1587\eps^6+3632\eps^5-3677\eps^4+1350\eps^3+354\eps^2-384\eps+72)}{\eps^3(2\eps-1)(-1+\eps)(-3+2\eps)^2}\MI^{(2;1)}_{C}(Q^2b_{ijk}) 
\bigg]\nn
\eea



\bibliographystyle{JHEP}
\bibliography{refs.bib}

\providecommand{\href}[2]{#2}\begingroup\raggedright\begin{thebibliography}{10}

\bibitem{Kinoshita:1962ur}
T.~Kinoshita, \emph{{Mass singularities of Feynman amplitudes}},
  \href{http://dx.doi.org/10.1063/1.1724268}{\emph{J. Math. Phys.} {\bf 3}
  (1962) 650}.

\bibitem{Lee:1964is}
T.D.~Lee and M.~Nauenberg, \emph{{Degenerate Systems and Mass Singularities}},
  \href{http://dx.doi.org/10.1103/PhysRev.133.B1549}{\emph{Phys. Rev.} {\bf
  133} (1964) B1549}.

\bibitem{Politzer:1974fr}
H.D.~Politzer, \emph{{Asymptotic Freedom: An Approach to Strong Interactions}},
  \href{http://dx.doi.org/10.1016/0370-1573(74)90014-3}{\emph{Phys. Rept.} {\bf
  14} (1974) 129}.

\bibitem{Georgi:1951sr}
H.~Georgi and H.D.~Politzer, \emph{{Electroproduction scaling in an
  asymptotically free theory of strong interactions}},
  \href{http://dx.doi.org/10.1103/PhysRevD.9.416}{\emph{Phys. Rev.} {\bf D9}
  (1974) 416}.

\bibitem{Altarelli:1977zs}
G.~Altarelli and G.~Parisi, \emph{{Asymptotic Freedom in Parton Language}},
  \href{http://dx.doi.org/10.1016/0550-3213(77)90384-4}{\emph{Nucl. Phys.} {\bf
  B126} (1977) 298}.

\bibitem{Catani:1996vz}
S.~Catani and M.H.~Seymour, \emph{{A General algorithm for calculating jet
  cross-sections in NLO QCD}},
  \href{http://dx.doi.org/10.1016/S0550-3213(96)00589-5,
  10.1016/S0550-3213(98)81022-5}{\emph{Nucl. Phys.} {\bf B485} (1997) 291}
  [\href{https://arxiv.org/abs/hep-ph/9605323}{{\tt hep-ph/9605323}}].

\bibitem{Catani:1996jh}
S.~Catani and M.H.~Seymour, \emph{{The Dipole formalism for the calculation of
  QCD jet cross-sections at next-to-leading order}},
  \href{http://dx.doi.org/10.1016/0370-2693(96)00425-X}{\emph{Phys. Lett.} {\bf
  B378} (1996) 287} [\href{https://arxiv.org/abs/hep-ph/9602277}{{\tt
  hep-ph/9602277}}].

\bibitem{Frixione:1995ms}
S.~Frixione, Z.~Kunszt and A.~Signer, \emph{{Three jet cross-sections to
  next-to-leading order}},
  \href{http://dx.doi.org/10.1016/0550-3213(96)00110-1}{\emph{Nucl. Phys.} {\bf
  B467} (1996) 399} [\href{https://arxiv.org/abs/hep-ph/9512328}{{\tt
  hep-ph/9512328}}].

\bibitem{Frederix:2009yq}
R.~Frederix, S.~Frixione, F.~Maltoni and T.~Stelzer, \emph{{Automation of
  next-to-leading order computations in QCD: The FKS subtraction}},
  \href{http://dx.doi.org/10.1088/1126-6708/2009/10/003}{\emph{JHEP} {\bf 10}
  (2009) 003} [\href{https://arxiv.org/abs/0908.4272}{{\tt arXiv:0908.4272}}].

\bibitem{Nagy:2003qn}
Z.~Nagy and D.E.~Soper, \emph{{General subtraction method for numerical
  calculation of one loop QCD matrix elements}},
  \href{http://dx.doi.org/10.1088/1126-6708/2003/09/055}{\emph{JHEP} {\bf 09}
  (2003) 055} [\href{https://arxiv.org/abs/hep-ph/0308127}{{\tt
  hep-ph/0308127}}].

\bibitem{Binoth:2004jv}
T.~Binoth and G.~Heinrich, \emph{{Numerical evaluation of phase space integrals
  by sector decomposition}},
  \href{http://dx.doi.org/10.1016/j.nuclphysb.2004.06.005}{\emph{Nucl. Phys.}
  {\bf B693} (2004) 134} [\href{https://arxiv.org/abs/hep-ph/0402265}{{\tt
  hep-ph/0402265}}].

\bibitem{Anastasiou:2003gr}
C.~Anastasiou, K.~Melnikov and F.~Petriello, \emph{{A new method for real
  radiation at NNLO}},
  \href{http://dx.doi.org/10.1103/PhysRevD.69.076010}{\emph{Phys. Rev.} {\bf
  D69} (2004) 076010} [\href{https://arxiv.org/abs/hep-ph/0311311}{{\tt
  hep-ph/0311311}}].

\bibitem{Czakon:2010td}
M.~Czakon, \emph{{A novel subtraction scheme for double-real radiation at
  NNLO}}, \href{http://dx.doi.org/10.1016/j.physletb.2010.08.036}{\emph{Phys.
  Lett.} {\bf B693} (2010) 259} [\href{https://arxiv.org/abs/1005.0274}{{\tt
  arXiv:1005.0274}}].

\bibitem{Boughezal:2011jf}
R.~Boughezal, K.~Melnikov and F.~Petriello, \emph{{A subtraction scheme for
  NNLO computations}},
  \href{http://dx.doi.org/10.1103/PhysRevD.85.034025}{\emph{Phys. Rev.} {\bf
  D85} (2012) 034025} [\href{https://arxiv.org/abs/1111.7041}{{\tt
  arXiv:1111.7041}}].

\bibitem{Caola:2017dug}
F.~Caola, K.~Melnikov and R.~Röntsch, \emph{{Nested soft-collinear
  subtractions in NNLO QCD computations}},
  \href{http://dx.doi.org/10.1140/epjc/s10052-017-4774-0}{\emph{Eur. Phys. J.}
  {\bf C77} (2017) 248} [\href{https://arxiv.org/abs/1702.01352}{{\tt
  arXiv:1702.01352}}].

\bibitem{Czakon:2015owf}
M.~Czakon, D.~Heymes and A.~Mitov, \emph{{High-precision differential
  predictions for top-quark pairs at the LHC}},
  \href{http://dx.doi.org/10.1103/PhysRevLett.116.082003}{\emph{Phys. Rev.
  Lett.} {\bf 116} (2016) 082003} [\href{https://arxiv.org/abs/1511.00549}{{\tt
  arXiv:1511.00549}}].

\bibitem{Caola:2016trd}
F.~Caola, M.~Dowling, K.~Melnikov, R.~Röntsch and L.~Tancredi, \emph{{QCD
  corrections to vector boson pair production in gluon fusion including
  interference effects with off-shell Higgs at the LHC}},
  \href{http://dx.doi.org/10.1007/JHEP07(2016)087}{\emph{JHEP} {\bf 07} (2016)
  087} [\href{https://arxiv.org/abs/1605.04610}{{\tt arXiv:1605.04610}}].

\bibitem{Weinzierl:2003fx}
S.~Weinzierl, \emph{{Subtraction terms at NNLO}},
  \href{http://dx.doi.org/10.1088/1126-6708/2003/03/062}{\emph{JHEP} {\bf 03}
  (2003) 062} [\href{https://arxiv.org/abs/hep-ph/0302180}{{\tt
  hep-ph/0302180}}].

\bibitem{Frixione:2004is}
S.~Frixione and M.~Grazzini, \emph{{Subtraction at NNLO}},
  \href{http://dx.doi.org/10.1088/1126-6708/2005/06/010}{\emph{JHEP} {\bf 06}
  (2005) 010} [\href{https://arxiv.org/abs/hep-ph/0411399}{{\tt
  hep-ph/0411399}}].

\bibitem{Anastasiou:2014nha}
C.~Anastasiou, J.~Cancino, F.~Chavez, C.~Duhr, A.~Lazopoulos, B.~Mistlberger
  et~al., \emph{{NNLO QCD corrections to pp → γ$^{*}$ γ$^{*}$ in the large
  N$_{F}$ limit}}, \href{http://dx.doi.org/10.1007/JHEP02(2015)182}{\emph{JHEP}
  {\bf 02} (2015) 182} [\href{https://arxiv.org/abs/1408.4546}{{\tt
  arXiv:1408.4546}}].

\bibitem{Anastasiou:2010pw}
C.~Anastasiou, F.~Herzog and A.~Lazopoulos, \emph{{On the factorization of
  overlapping singularities at NNLO}},
  \href{http://dx.doi.org/10.1007/JHEP03(2011)038}{\emph{JHEP} {\bf 03} (2011)
  038} [\href{https://arxiv.org/abs/1011.4867}{{\tt arXiv:1011.4867}}].

\bibitem{GehrmannDeRidder:2005cm}
A.~Gehrmann-De~Ridder, T.~Gehrmann and E.W.N.~Glover, \emph{{Antenna
  subtraction at NNLO}},
  \href{http://dx.doi.org/10.1088/1126-6708/2005/09/056}{\emph{JHEP} {\bf 09}
  (2005) 056} [\href{https://arxiv.org/abs/hep-ph/0505111}{{\tt
  hep-ph/0505111}}].

\bibitem{GehrmannDeRidder:2007jk}
A.~Gehrmann-De~Ridder, T.~Gehrmann, E.W.N.~Glover and G.~Heinrich,
  \emph{{Infrared structure of e+ e- ---> 3 jets at NNLO}},
  \href{http://dx.doi.org/10.1088/1126-6708/2007/11/058}{\emph{JHEP} {\bf 11}
  (2007) 058} [\href{https://arxiv.org/abs/0710.0346}{{\tt arXiv:0710.0346}}].

\bibitem{Currie:2013vh}
J.~Currie, E.W.N.~Glover and S.~Wells, \emph{{Infrared Structure at NNLO Using
  Antenna Subtraction}},
  \href{http://dx.doi.org/10.1007/JHEP04(2013)066}{\emph{JHEP} {\bf 04} (2013)
  066} [\href{https://arxiv.org/abs/1301.4693}{{\tt arXiv:1301.4693}}].

\bibitem{Catani:1999ss}
S.~Catani and M.~Grazzini, \emph{{Infrared factorization of tree level QCD
  amplitudes at the next-to-next-to-leading order and beyond}},
  \href{http://dx.doi.org/10.1016/S0550-3213(99)00778-6}{\emph{Nucl. Phys.}
  {\bf B570} (2000) 287} [\href{https://arxiv.org/abs/hep-ph/9908523}{{\tt
  hep-ph/9908523}}].

\bibitem{Currie:2017eqf}
J.~Currie, A.~Gehrmann-De~Ridder, T.~Gehrmann, E.W.N.~Glover, A.~Huss and
  J.~Pires, \emph{{Precise predictions for dijet production at the LHC}},
  \href{http://dx.doi.org/10.1103/PhysRevLett.119.152001}{\emph{Phys. Rev.
  Lett.} {\bf 119} (2017) 152001} [\href{https://arxiv.org/abs/1705.10271}{{\tt
  arXiv:1705.10271}}].

\bibitem{Cruz-Martinez:2018rod}
J.~Cruz-Martinez, T.~Gehrmann, E.W.N.~Glover and A.~Huss, \emph{{Second-order
  QCD effects in Higgs boson production through vector boson fusion}},
  \href{https://arxiv.org/abs/1802.02445}{{\tt arXiv:1802.02445}}.

\bibitem{Somogyi:2005xz}
G.~Somogyi, Z.~Trocsanyi and V.~Del~Duca, \emph{{Matching of singly- and
  doubly-unresolved limits of tree-level QCD squared matrix elements}},
  \href{http://dx.doi.org/10.1088/1126-6708/2005/06/024}{\emph{JHEP} {\bf 06}
  (2005) 024} [\href{https://arxiv.org/abs/hep-ph/0502226}{{\tt
  hep-ph/0502226}}].

\bibitem{Somogyi:2006da}
G.~Somogyi, Z.~Trocsanyi and V.~Del~Duca, \emph{{A Subtraction scheme for
  computing QCD jet cross sections at NNLO: Regularization of doubly-real
  emissions}},
  \href{http://dx.doi.org/10.1088/1126-6708/2007/01/070}{\emph{JHEP} {\bf 01}
  (2007) 070} [\href{https://arxiv.org/abs/hep-ph/0609042}{{\tt
  hep-ph/0609042}}].

\bibitem{Bolzoni:2010bt}
P.~Bolzoni, G.~Somogyi and Z.~Trocsanyi, \emph{{A subtraction scheme for
  computing QCD jet cross sections at NNLO: integrating the iterated
  singly-unresolved subtraction terms}},
  \href{http://dx.doi.org/10.1007/JHEP01(2011)059}{\emph{JHEP} {\bf 01} (2011)
  059} [\href{https://arxiv.org/abs/1011.1909}{{\tt arXiv:1011.1909}}].

\bibitem{DelDuca:2016csb}
V.~Del~Duca, C.~Duhr, A.~Kardos, G.~Somogyi and Z.~Trócsányi,
  \emph{{Three-Jet Production in Electron-Positron Collisions at
  Next-to-Next-to-Leading Order Accuracy}},
  \href{http://dx.doi.org/10.1103/PhysRevLett.117.152004}{\emph{Phys. Rev.
  Lett.} {\bf 117} (2016) 152004} [\href{https://arxiv.org/abs/1603.08927}{{\tt
  arXiv:1603.08927}}].

\bibitem{Giele:1991vf}
W.T.~Giele and E.W.N.~Glover, \emph{{Higher order corrections to jet
  cross-sections in e+ e- annihilation}},
  \href{http://dx.doi.org/10.1103/PhysRevD.46.1980}{\emph{Phys. Rev.} {\bf D46}
  (1992) 1980}.

\bibitem{Fabricius:1981sx}
K.~Fabricius, I.~Schmitt, G.~Kramer and G.~Schierholz, \emph{{Higher Order
  Perturbative QCD Calculation of Jet Cross-Sections in e+ e- Annihilation}},
  \href{http://dx.doi.org/10.1007/BF01578281}{\emph{Z. Phys.} {\bf C11} (1981)
  315}.

\bibitem{GehrmannDeRidder:1997gf}
A.~Gehrmann-De~Ridder and E.W.N.~Glover, \emph{{A Complete O (alpha alpha-s)
  calculation of the photon + 1 jet rate in e+ e- annihilation}},
  \href{http://dx.doi.org/10.1016/S0550-3213(97)00818-3}{\emph{Nucl. Phys.}
  {\bf B517} (1998) 269} [\href{https://arxiv.org/abs/hep-ph/9707224}{{\tt
  hep-ph/9707224}}].

\bibitem{Catani:2007vq}
S.~Catani and M.~Grazzini, \emph{{An NNLO subtraction formalism in hadron
  collisions and its application to Higgs boson production at the LHC}},
  \href{http://dx.doi.org/10.1103/PhysRevLett.98.222002}{\emph{Phys. Rev.
  Lett.} {\bf 98} (2007) 222002}
  [\href{https://arxiv.org/abs/hep-ph/0703012}{{\tt hep-ph/0703012}}].

\bibitem{Boughezal:2015aha}
R.~Boughezal, C.~Focke, W.~Giele, X.~Liu and F.~Petriello, \emph{{Higgs boson
  production in association with a jet at NNLO using jettiness subtraction}},
  \href{http://dx.doi.org/10.1016/j.physletb.2015.06.055}{\emph{Phys. Lett.}
  {\bf B748} (2015) 5} [\href{https://arxiv.org/abs/1505.03893}{{\tt
  arXiv:1505.03893}}].

\bibitem{Gaunt:2015pea}
J.~Gaunt, M.~Stahlhofen, F.J.~Tackmann and J.R.~Walsh, \emph{{N-jettiness
  Subtractions for NNLO QCD Calculations}},
  \href{http://dx.doi.org/10.1007/JHEP09(2015)058}{\emph{JHEP} {\bf 09} (2015)
  058} [\href{https://arxiv.org/abs/1505.04794}{{\tt arXiv:1505.04794}}].

\bibitem{Catani:2011qz}
S.~Catani, L.~Cieri, D.~de~Florian, G.~Ferrera and M.~Grazzini, \emph{{Diphoton
  production at hadron colliders: a fully-differential QCD calculation at
  NNLO}}, \href{http://dx.doi.org/10.1103/PhysRevLett.108.072001,
  10.1103/PhysRevLett.117.089901}{\emph{Phys. Rev. Lett.} {\bf 108} (2012)
  072001} [\href{https://arxiv.org/abs/1110.2375}{{\tt arXiv:1110.2375}}].

\bibitem{Boughezal:2016dtm}
R.~Boughezal, X.~Liu and F.~Petriello, \emph{{W-boson plus jet differential
  distributions at NNLO in QCD}},
  \href{http://dx.doi.org/10.1103/PhysRevD.94.113009}{\emph{Phys. Rev.} {\bf
  D94} (2016) 113009} [\href{https://arxiv.org/abs/1602.06965}{{\tt
  arXiv:1602.06965}}].

\bibitem{Moult:2017jsg}
I.~Moult, L.~Rothen, I.W.~Stewart, F.J.~Tackmann and H.X.~Zhu, \emph{{N
  -jettiness subtractions for $gg\to H$ at subleading power}},
  \href{http://dx.doi.org/10.1103/PhysRevD.97.014013}{\emph{Phys. Rev.} {\bf
  D97} (2018) 014013} [\href{https://arxiv.org/abs/1710.03227}{{\tt
  arXiv:1710.03227}}].

\bibitem{Boughezal:2018mvf}
R.~Boughezal, A.~Isgrò and F.~Petriello, \emph{{Next-to-leading-logarithmic
  power corrections for $N$-jettiness subtraction in color-singlet
  production}}, \href{http://dx.doi.org/10.1103/PhysRevD.97.076006}{\emph{Phys.
  Rev.} {\bf D97} (2018) 076006} [\href{https://arxiv.org/abs/1802.00456}{{\tt
  arXiv:1802.00456}}].

\bibitem{Cacciari:2015jma}
M.~Cacciari, F.A.~Dreyer, A.~Karlberg, G.P.~Salam and G.~Zanderighi,
  \emph{{Fully Differential Vector-Boson-Fusion Higgs Production at
  Next-to-Next-to-Leading Order}},
  \href{http://dx.doi.org/10.1103/PhysRevLett.115.082002,
  10.1103/PhysRevLett.120.139901}{\emph{Phys. Rev. Lett.} {\bf 115} (2015)
  082002} [\href{https://arxiv.org/abs/1506.02660}{{\tt arXiv:1506.02660}}].

\bibitem{Currie:2018fgr}
J.~Currie, T.~Gehrmann, E.W.N.~Glover, A.~Huss, J.~Niehues and A.~Vogt,
  \emph{{N3LO Corrections to Jet Production in Deep Inelastic Scattering using
  the Projection-to-Born Method}},
  \href{https://arxiv.org/abs/1803.09973}{{\tt arXiv:1803.09973}}.

\bibitem{Dulat:2017prg}
F.~Dulat, B.~Mistlberger and A.~Pelloni, \emph{{Differential Higgs production
  at N$^{3}$LO beyond threshold}},
  \href{http://dx.doi.org/10.1007/JHEP01(2018)145}{\emph{JHEP} {\bf 01} (2018)
  145} [\href{https://arxiv.org/abs/1710.03016}{{\tt arXiv:1710.03016}}].

\bibitem{Anastasiou:2002yz}
C.~Anastasiou and K.~Melnikov, \emph{{Higgs boson production at hadron
  colliders in NNLO QCD}},
  \href{http://dx.doi.org/10.1016/S0550-3213(02)00837-4}{\emph{Nucl. Phys.}
  {\bf B646} (2002) 220} [\href{https://arxiv.org/abs/hep-ph/0207004}{{\tt
  hep-ph/0207004}}].

\bibitem{Eynck:2001en}
T.O.~Eynck, E.~Laenen, L.~Phaf and S.~Weinzierl, \emph{{Comparison of phase
  space slicing and dipole subtraction methods for gamma* ---> anti-Q}},
  \href{http://dx.doi.org/10.1007/s100520100868}{\emph{Eur. Phys. J.} {\bf C23}
  (2002) 259} [\href{https://arxiv.org/abs/hep-ph/0109246}{{\tt
  hep-ph/0109246}}].

\bibitem{Bloch:2008jk}
S.~Bloch and D.~Kreimer, \emph{{Mixed Hodge Structures and Renormalization in
  Physics}}, \href{http://dx.doi.org/10.4310/CNTP.2008.v2.n4.a1}{\emph{Commun.
  Num. Theor. Phys.} {\bf 2} (2008) 637}
  [\href{https://arxiv.org/abs/0804.4399}{{\tt arXiv:0804.4399}}].

\bibitem{Brown:2015fyf}
F.~Brown, \emph{{Feynman amplitudes, coaction principle, and cosmic Galois
  group}}, \href{http://dx.doi.org/10.4310/CNTP.2017.v11.n3.a1}{\emph{Commun.
  Num. Theor. Phys.} {\bf 11} (2017) 453}
  [\href{https://arxiv.org/abs/1512.06409}{{\tt arXiv:1512.06409}}].

\bibitem{Sterman:1978bi}
G.F.~Sterman, \emph{{Mass Divergences in Annihilation Processes. 1. Origin and
  Nature of Divergences in Cut Vacuum Polarization Diagrams}},
  \href{http://dx.doi.org/10.1103/PhysRevD.17.2773}{\emph{Phys. Rev.} {\bf D17}
  (1978) 2773}.

\bibitem{Sterman:1978bj}
G.F.~Sterman, \emph{{Mass Divergences in Annihilation Processes. 2.
  Cancellation of Divergences in Cut Vacuum Polarization Diagrams}},
  \href{http://dx.doi.org/10.1103/PhysRevD.17.2789}{\emph{Phys. Rev.} {\bf D17}
  (1978) 2789}.

\bibitem{Collins:2011zzd}
J.~Collins, \emph{{Foundations of perturbative QCD}}, Cambridge University
  Press (2013).

\bibitem{Hahn:2004fe}
T.~Hahn, \emph{{CUBA: A Library for multidimensional numerical integration}},
  \href{http://dx.doi.org/10.1016/j.cpc.2005.01.010}{\emph{Comput. Phys.
  Commun.} {\bf 168} (2005) 78}
  [\href{https://arxiv.org/abs/hep-ph/0404043}{{\tt hep-ph/0404043}}].

\bibitem{Zimmermann:1969jj}
W.~Zimmermann, \emph{{Convergence of Bogolyubov's method of renormalization in
  momentum space}}, \href{http://dx.doi.org/10.1007/BF01645676}{\emph{Commun.
  Math. Phys.} {\bf 15} (1969) 208}.

\bibitem{Humpert:1980xj}
B.~Humpert and W.L.~van~Neerven, \emph{{GRAPHICAL MASS FACTORIZATION}},
  \href{http://dx.doi.org/10.1016/0370-2693(81)91246-6}{\emph{Phys. Lett.} {\bf
  102B} (1981) 426}.

\bibitem{Smirnov:1986me}
V.A.~Smirnov and K.G.~Chetyrkin, \emph{{R* Operation in the Minimal Subtraction
  Scheme}}, \href{http://dx.doi.org/10.1007/BF01017902}{\emph{Theor. Math.
  Phys.} {\bf 63} (1985) 462}.

\bibitem{Herzog:2017jgk}
F.~Herzog, \emph{{Zimmermann's forest formula, infrared divergences and the QCD
  beta function}},
  \href{http://dx.doi.org/10.1016/j.nuclphysb.2017.11.011}{\emph{Nucl. Phys.}
  {\bf B926} (2018) 370} [\href{https://arxiv.org/abs/1711.06121}{{\tt
  arXiv:1711.06121}}].

\bibitem{Collins:1981uk}
J.C.~Collins and D.E.~Soper, \emph{{Back-To-Back Jets in QCD}},
  \href{http://dx.doi.org/10.1016/0550-3213(81)90339-4}{\emph{Nucl. Phys.} {\bf
  B193} (1981) 381}.

\bibitem{Anastasiou:2013srw}
C.~Anastasiou, C.~Duhr, F.~Dulat and B.~Mistlberger, \emph{{Soft triple-real
  radiation for Higgs production at N3LO}},
  \href{http://dx.doi.org/10.1007/JHEP07(2013)003}{\emph{JHEP} {\bf 07} (2013)
  003} [\href{https://arxiv.org/abs/1302.4379}{{\tt arXiv:1302.4379}}].

\bibitem{Smirnov:2008iw}
A.V.~Smirnov, \emph{{Algorithm FIRE -- Feynman Integral REduction}},
  \href{http://dx.doi.org/10.1088/1126-6708/2008/10/107}{\emph{JHEP} {\bf 10}
  (2008) 107} [\href{https://arxiv.org/abs/0807.3243}{{\tt arXiv:0807.3243}}].

\bibitem{Smirnov:2014hma}
A.V.~Smirnov, \emph{{FIRE5: a C++ implementation of Feynman Integral
  REduction}}, \href{http://dx.doi.org/10.1016/j.cpc.2014.11.024}{\emph{Comput.
  Phys. Commun.} {\bf 189} (2015) 182}
  [\href{https://arxiv.org/abs/1408.2372}{{\tt arXiv:1408.2372}}].

\bibitem{Anastasiou:2004vj}
C.~Anastasiou and A.~Lazopoulos, \emph{{Automatic integral reduction for higher
  order perturbative calculations}},
  \href{http://dx.doi.org/10.1088/1126-6708/2004/07/046}{\emph{JHEP} {\bf 07}
  (2004) 046} [\href{https://arxiv.org/abs/hep-ph/0404258}{{\tt
  hep-ph/0404258}}].

\bibitem{Anastasiou:2012kq}
C.~Anastasiou, S.~Buehler, C.~Duhr and F.~Herzog, \emph{{NNLO phase space
  master integrals for two-to-one inclusive cross sections in dimensional
  regularization}},
  \href{http://dx.doi.org/10.1007/JHEP11(2012)062}{\emph{JHEP} {\bf 11} (2012)
  062} [\href{https://arxiv.org/abs/1208.3130}{{\tt arXiv:1208.3130}}].

\bibitem{Anastasiou:2015yha}
C.~Anastasiou, C.~Duhr, F.~Dulat, E.~Furlan, F.~Herzog and B.~Mistlberger,
  \emph{{Soft expansion of double-real-virtual corrections to Higgs production
  at N$^{3}$LO}}, \href{http://dx.doi.org/10.1007/JHEP08(2015)051}{\emph{JHEP}
  {\bf 08} (2015) 051} [\href{https://arxiv.org/abs/1505.04110}{{\tt
  arXiv:1505.04110}}].

\bibitem{Ritzmann:2014mka}
M.~Ritzmann and W.J.~Waalewijn, \emph{{Fragmentation in Jets at NNLO}},
  \href{http://dx.doi.org/10.1103/PhysRevD.90.054029}{\emph{Phys. Rev.} {\bf
  D90} (2014) 054029} [\href{https://arxiv.org/abs/1407.3272}{{\tt
  arXiv:1407.3272}}].

\bibitem{Huber:2005yg}
T.~Huber and D.~Maitre, \emph{{HypExp: A Mathematica package for expanding
  hypergeometric functions around integer-valued parameters}},
  \href{http://dx.doi.org/10.1016/j.cpc.2006.01.007}{\emph{Comput. Phys.
  Commun.} {\bf 175} (2006) 122}
  [\href{https://arxiv.org/abs/hep-ph/0507094}{{\tt hep-ph/0507094}}].

\bibitem{Gehrmann-DeRidder:2003pne}
A.~Gehrmann-De~Ridder, T.~Gehrmann and G.~Heinrich, \emph{{Four particle phase
  space integrals in massless QCD}},
  \href{http://dx.doi.org/10.1016/j.nuclphysb.2004.01.023}{\emph{Nucl. Phys.}
  {\bf B682} (2004) 265} [\href{https://arxiv.org/abs/hep-ph/0311276}{{\tt
  hep-ph/0311276}}].

\bibitem{Catani:1998bh}
S.~Catani, \emph{{The Singular behavior of QCD amplitudes at two loop order}},
  \href{http://dx.doi.org/10.1016/S0370-2693(98)00332-3}{\emph{Phys. Lett.}
  {\bf B427} (1998) 161} [\href{https://arxiv.org/abs/hep-ph/9802439}{{\tt
  hep-ph/9802439}}].

\end{thebibliography}\endgroup

\end{document}